\documentstyle[epsfig]{article}

\newcommand{\lra}{\mbox{$\leftrightarrow$}}

\newcommand{\bone}{\mbox{\bf 1}}
\newcommand{\tr}{\mbox{tr$\,$}}
\newcommand{\str}{\mbox{str$\,$}}

\newcommand{\e}{\mbox{$\epsilon$}}
\newcommand{\half}{\mbox{$1\over2$}}
\newcommand{\third}{\mbox{$1\over3$}}
\newcommand{\quarter}{\mbox{$1\over4$}}
\newcommand{\eigth}{\mbox{$1\over8$}}

\newcommand{\NN}{\mbox{$\cal N$}}
\newcommand{\OO}{\mbox{$\cal O$}}
\newcommand{\PP}{\mbox{$\cal P$}}

\newcommand{\KK}{\mbox{$\cal K$}}

\newcommand{\YY}{\mbox{$\cal Y$}}

\begin{document}

\begin{titlepage}
\begin{center}

{\hbox to\hsize{
hep-th/0109065
\hfill 
UCLA/01/TEP/22
}}
{\hbox to\hsize{
\hfill 
September 2001
}}

\vskip 1in
 
{\Large \bf 
Three-Point Functions \\ 
$\phantom{_|}$ of $\phantom{_|}$ \\ 
Quarter BPS Operators in \NN=4 SYM 
}

\vskip 0.45in

\renewcommand{\thefootnote}{\fnsymbol{footnote}}
{\bf 
Eric D'Hoker%
\footnote{
	dhoker@physics.ucla.edu
	}
and 
Anton V. Ryzhov%
\footnote{
	ryzhovav@physics.ucla.edu
	}
}
\setcounter{footnote}{0}

\vskip 0.15in

{\em Department of Physics and Astronomy, \\
University of California, Los Angeles, \\ 
LA, CA 90095-1547 \\ }

\vskip 0.1in

\end{center}

\vskip .3in

\begin{abstract}
In a recent paper \cite{Ryzhov}, 
\quarter-BPS chiral primaries 
were constructed 
in the fully interacting 
four dimensional $\NN=4$ Super-Yang-Mills theory 
with gauge group $SU(N)$. 
These operators are annihilated by four supercharges, 
and at order $g^2$ have protected scaling dimension 
and normalization. 
Here, we compute three-point functions involving 
these \quarter-BPS operators along with \half-BPS operators. 
The combinatorics of the problem is rather involved, and 
we consider the following special cases: 
(1) correlators 
$\langle \OO_{\tiny \! \half} \OO_{\tiny \! \half} \OO_{\rm BPS} 
\rangle$ 
of two \half-BPS primaries with an arbitrary chiral primary; 
(2) certain classes of 
$\langle \OO_{\tiny \! \half} \OO_{\tiny \! \quarter} \OO_{\tiny \! \quarter} 
\rangle$ and 
$\langle \OO_{\tiny \! \quarter} \OO_{\tiny \! \quarter} 
\OO_{\tiny \! \quarter} \rangle$ 
three-point functions; 
(3) three-point functions involving the $\Delta \le 7$ operators 
found in \cite{Ryzhov}; 
(4) 
$\langle \OO_{\tiny \! \half} \OO_{\tiny \! \quarter} 
\OO_{\tiny \! \quarter} \rangle$ 
correlators with the 
special $\OO_{\tiny \! \quarter}$ made of single and double 
trace operators only. 
The analysis in cases (1)-(3) is valid for general $N$, 
while (4) is a 
large $N$ approximation. 
Order $g^2$ corrections to all 
three-point functions 
considered in this paper are found to vanish. 

In the AdS/CFT correspondence, 
\quarter-BPS chiral primaries 
are dual to threshold bound states 
of elementary supergravity excitations. 
We present a supergravity discussion 
of two- and three-point correlators 
involving these bound states.

\end{abstract}

\end{titlepage}

\newpage

\section{Introduction}

The AdS/CFT correspondence \cite{Maldacena, GKP, Witten} 
provides a powerful tool for deriving
dynamical information in \NN=4 superconformal YM theory outside 
the regime of weak coupling perturbation theory. Comparison of
weak and strong coupling behaviors has given rise to a number of
surprising new conjectures 
\cite{LMRS}, \cite{DFS}, \cite{G-rKP}, 
\cite{Skiba}, \cite{ZANON}, \cite{BKRS}
that 2- and 3-point functions, as well
as certain ``extremal" n-point correlators of \half-BPS operators
are ``unrenormalized", \cite{extremal:refs}, 
i.e. when properly defined, their form is
independent of the gauge coupling $g$. 
\NN=2 superspace methods 
have since further confirmed the validity of these results 
\cite{Intriligator}, \cite{HSW}, \cite{AF}.
Extensions to ``near-extremal correlators" have unveiled a 
hierarchical pattern of $g$-dependence that further generalizes 
the non-renormalization conjectures in exciting ways 
\cite{near-extremal:refs}.

The AdS/CFT correspondence relies on a direct duality between
the \half-BPS operators and the canonical fields and states 
of Type IIB supergravity on $AdS_5 \times S^5$, since they both
belong to the shortest multiplets of the superconformal group 
$SU(2,2|4)$, \cite{Maldacena, GKP, Witten, AGMOO}. 
However, in addition the \half-BPS operators, 
also \quarter- and \eigth-BPS operators enjoy certain non-renormalization 
properties, such as the fact that their scaling dimension is
fixed entirely by their internal quantum numbers \cite{MS,DP,AFSZ,Ryzhov}. 
The \quarter- and \eigth-BPS operators 
are dual on the AdS side to threshold bound states 
of elementary supergravity excitations, typically consisting of 
at least two and three supergravity states respectively, which 
have not been explored from the AdS point of view. 

Through the study of four point functions of \half-BPS
operators, certain couplings, such as those of two \half-BPS operators
and one \quarter-BPS operator have been analyzed in SYM theory. In 
weak-coupling perturbation theory this was done in \cite{BKRS}, while 
arguments based on $\NN=2$ superfield methods were presented in 
\cite{ADS:4-pt:N=2 superspace}.
Furthermore, using the $\NN=4$ superfield approach, a general study of
3-point functions and their non-renormalization properties was initiated
in \cite{ADS:4-pt:N=4 superspace}. 
$\NN=4$ superfield methods, however, require on-shell superfields
whose use in the study of off-shell correlators is not fully understood.

In a previous paper \cite{Ryzhov}, a construction was presented 
for \quarter-BPS chiral primaries in the fully interacting N=4 SYM theory. 
In general, these $[p,q,p]$ operators are linear combinations of all 
local, polynomial, gauge invariant, scalar composite operators 
with the correct $SU(4)$ labels. 
Besides the double trace operators from the classification of 
\cite{DP,AFSZ}, single trace and other multiple trace 
operators made of the same scalar fields also have to be taken 
into account. 
The coefficients with which they all combine into 
operators with a well defined scaling dimension are quite involved. 
However, in the large $N$ limit, 
\quarter-BPS primaries of a special form 
(those made of the single trace operator 
and the double trace operator from the classification of \cite{AFSZ}) 
become surprisingly simple. 
The \quarter-BPS chiral primaries, like the 
\half-BPS operators extensively studied in the literature, 
also have protected 
two-point functions, at least at order $g^2$ \cite{Ryzhov}.

Presently, 
we investigate the (non-) renormalization 
properties of three-point correlators involving \quarter-BPS operators 
along with \half-BPS operators. 
Given the elaborate combinatorics of the problem, 
we concentrate on the following special cases. 
First, we discuss several group theoretic simplifications 
of the combinatorial factors multiplying the Feynman 
graphs that contribute to three-point functions of 
chiral primaries. 
Based on $SU(4)$ group theory and conformal invariance only, 
we argue that certain classes of such correlators 
are protected at order $g^2$, for 
all $N$. 
In particular, 
this allows us to compute $\OO(g^2)$ corrections 
to correlators of the form 
$\langle \OO_{\tiny \! \half} \OO_{\tiny \! \half} \OO_{\rm BPS} 
\rangle$, 
where $\OO_{\tiny \! \half}$ are two \half-BPS operators, and 
$\OO_{\rm BPS}$ is an arbitrary (\half-, \quarter-, or \eigth-BPS) 
chiral primary. 
Next, 
we look at the three-point functions 
$\langle \OO_{\tiny \! \half} \OO_{\tiny \! \quarter} \OO_{\tiny \! \quarter} 
\rangle$ and  
$\langle \OO_{\tiny \! \quarter} 
\OO_{\tiny \! \quarter} \OO_{\tiny \! \quarter} 
\rangle$, 
also for general $N$, 
where $\OO_{\tiny \! \quarter}$ are the $\Delta \le 7$ 
\quarter-BPS primaries found in \cite{Ryzhov}. 
Then, we carry out a large $N$ analysis 
of 
$\langle \OO_{\tiny \! \half} \OO_{\tiny \! \quarter} \OO_{\tiny \! \quarter} 
\rangle$ 
correlators involving the special \quarter-BPS operators 
(mixtures of single and double trace scalar composite operators), 
for arbitrary $\Delta$. 
Also in the large $N$ limit, 
we identify the corresponding objects 
in the supergravity description, 
and compute the correlators on the AdS side of the correspondence. 

Finally, we make some speculations. 
Based on the broad range of special cases studied in this 
paper and in \cite{Ryzhov}, 
we conjecture that two- and three-point functions 
of \half- and \quarter-operators receive no quantum corrections, 
for arbitrary $N$. 
Additionally, a set of group theoretic considerations of this 
paper extends straightforwardly from three-point functions 
to extremal correlators. 
Therefore, we suggest 
that extremal correlators involving 
\half- and \quarter-operators 
are protected as well.

\section{The operators}
\label{section:operators}

We begin setting the stage for computing three-point functions, 
by describing the operators we will deal with. 
The construction of gauge invariant scalar composite 
operators was explained 
in \cite{Ryzhov}, so 
here we will briefly review the main points of 
that discussion, as well as some well established facts.

Four dimensional 
\NN=4 SYM is a superconformal theory, 
and has a global 
$SU(2,2|4)$ superconformal symmetry group. 
Operators in the theory fall into multiplets of 
$SU(2,2|4)$ \cite{MS,DP}, and chiral primary operators 
are classified by its maximal bosonic subgroup \cite{AFSZ}, 
which includes the $R$-symmetry group $SU(4) \sim SO(6)$. 
The most widely studied operators in the theory 
are the \half-BPS primaries, 
i.e. chiral primary operators annihilated by 8 
out of 16 Poincar\'e supercharges. 
These are totally symmetric rank $q$ tensors
of the flavor $SO(6)$, 
with highest weight operators of the form 
$\tr (\phi^1)^q$, minus $SO(6)$ traces. 
The $SU(4)$ labels of these representations are 
$[0,q,0]$, and the $SO(6)$ Young tableau%
\footnote{
	See for example \cite{Hamermesh} for a 
	discussion of irreducible tensors of $SO(n)$. 
	}
corresponding 
to totally symmetric rank $q$ tensor is 
$\hspace{-1ex}\mbox{
\setlength{\unitlength}{0.7em}
\begin{picture}(4.5,1)
\put(0,0){\framebox (1,1){}}
\put(1,0){\framebox (2,1){\scriptsize $...$}}
\put(3,0){\framebox (1,1){}}
\end{picture}}$, 
namely one row of length $q$. 
The \half-BPS chiral primaries are Lorentz scalars, 
and their conformal dimension 
is related to their flavor quantum numbers by $\Delta = q$. 

\quarter-BPS chiral primaries 
belong to $[p,q,p]$ representations with $p \ge 2$.%
\footnote{
	They are \half-BPS for $p=0$; 
	and vanish for $p=1$ 
	after we take the $SU(N)$ traces.
	} 
In the $SO(6)$ notation, the highest weight state 
of $[p,q,p]$ 
corresponds to the 
\begin{equation}
\label{pqp:representation} 
\mbox{
\setlength{\unitlength}{1em}
\begin{picture}(7.5,2)
\put(0,1){\framebox (1,1){\scriptsize $1$}}
\put(1,1){\framebox (2,1){\scriptsize $...$}}
\put(3,1){\framebox (1,1){\scriptsize $1$}}
\put(4,1){\framebox (1,1){\scriptsize $1$}}
\put(5,1){\framebox (1,1){\scriptsize $...$}}
\put(6,1){\framebox (1,1){\scriptsize $1$}}
\put(0,0){\framebox (1,1){\scriptsize $2$}}
\put(1,0){\framebox (2,1){\scriptsize $...$}}
\put(3,0){\framebox (1,1){\scriptsize $2$}}
\put(1.8,-.7){\scriptsize $p$}
\put(5.4,.3){\scriptsize $q$}
\end{picture}}
\end{equation}
Young tableau. 
In the free theory, \quarter-BPS primaries with 
the highest $SU(4)$ weights are of the form 
$\tr (\phi^1)^{p+q} \, \tr (\phi^2)^{p}$ 
(modulo $(\phi^1,\phi^2)$ antisymmetrizations, 
and subtraction of the $SO(6)$ traces). 
However, there are many other ways to 
partition a given Young tableau. 
Each partition may result in 
a different operator after we take the $SU(N)$ traces. 
To be more explicit, consider the simplest example of 
$[p,q,p]$ scalar composite operators, namely [2,0,2]. 
The ways to partition the Young tableaux corresponding 
to this representation are 
\begin{equation}
\label{202:partitions} 
\left(\mbox{
\setlength{\unitlength}{0.7em}
\begin{picture}(2.5,1.5)
\put(0,0.5){\framebox (1,1){}}
\put(1,0.5){\framebox (1,1){}}
\put(0,-.5){\framebox (1,1){}}
\put(1,-.5){\framebox (1,1){}}
\end{picture}}\right)
, 
\left(\mbox{
\setlength{\unitlength}{0.7em}
\begin{picture}(2.5,1.5)
\put(0,0.7){\framebox (1,1){}}
\put(1,0.7){\framebox (1,1){}}
\put(0,-.8){\framebox (1,1){}}
\put(1,-.8){\framebox (1,1){}}
\end{picture}}\right)
, 
\left(\mbox{
\setlength{\unitlength}{0.7em}
\begin{picture}(2.5,1.5)
\put(-.4,0.5){\framebox (1,1){}}
\put(1,0.5){\framebox (1,1){}}
\put(-.4,-.5){\framebox (1,1){}}
\put(1,-.5){\framebox (1,1){}}
\end{picture}}\right)
,
\left(\mbox{
\setlength{\unitlength}{0.7em}
\begin{picture}(2.5,1.5)
\put(-.4,0.7){\framebox (1,1){}}
\put(1,0.2){\framebox (1,1){}}
\put(0,-.8){\framebox (1,1){}}
\put(1,-.8){\framebox (1,1){}}
\end{picture}}\right)
, 
\left(\mbox{
\setlength{\unitlength}{0.7em}
\begin{picture}(2.5,1.5)
\put(-.4,0.7){\framebox (1,1){}}
\put(1,0.5){\framebox (1,1){}}
\put(-.4,-.7){\framebox (1,1){}}
\put(1,-.5){\framebox (1,1){}}
\end{picture}}\right)
, 
\left(\mbox{
\setlength{\unitlength}{0.7em}
\begin{picture}(2.5,1.5)
\put(-.4,0.7){\framebox (1,1){}}
\put(1,0.7){\framebox (1,1){}}
\put(-.4,-.7){\framebox (1,1){}}
\put(1,-.7){\framebox (1,1){}}
\end{picture}}\right)
\end{equation}
where each continuous group of boxes corresponds 
to a single $SU(N)$ trace. After taking traces, 
the last three partitions vanish identically 
(since $\tr \phi^I = 0$); and the ``4=2+2'' partitions 
turn out to give the same operator. 
For a general $[p,q,p]$ representation the 
arguments are similar, although they become 
progressively more tedious 
as $2p+q$ gets larger. 
Scalar composite operators with $2p+q \le 7$ 
are listed in Appendix \ref{operators explicitly}.

Sometimes, the way we take the $SU(N)$ traces 
to obtain a $[p,q,p]$ scalar composite operator 
is not important, and the only relevant information 
is what fields are used, and that 
the operator is actually gauge invariant. 
In such cases, we shall use the notation ``$[...]$'' 
to denote gauge invariant combinations of the fields in brackets. 
For example, operators corresponding to 
the highest weight state of representations $[p,q,p]$ 
will be written as 
$[ (\phi^1)^{(p+q)} (\phi^2)^p ]$ $- SO(6)$ traces, etc.

In the interacting theory, none of the operators 
obtained by simply partitioning Young tableaux 
are eigenstates of the dilatation operator 
(or pure) \cite{Ryzhov}.  
Instead, they are mixtures of operators with different scaling dimensions. 
Proper \quarter-BPS primaries are 
certain linear combinations of these mixtures. 
To find the correct linear combinations, 
one can look at two-point functions, and 
The \quarter-BPS primaries are identified as 
the operators 
which receive no $\OO(g^2)$ corrections to 
two-point functions among themselves or with 
other scalar composite operators. 
The protected scaling dimension 
of a $[p,q,p]$ chiral primary 
is $\Delta = 2 p + q$.

The classical (Euclidean theory) Lagrangian can be written as \cite{DFS} 
\begin	{eqnarray}
\label	{lagrangian:for Feynman rules}
L &\!\!=\!\!& \tr \left\{ 
\quarter F_{\mu\nu} F^{\mu\nu} 
+ \half \bar \lambda \gamma^\mu D_\mu \lambda 
+ \overline{D_\mu z_j} D^\mu z_j 
+ \half \bar \psi^j \gamma^\mu D_\mu \psi^j 
\right\} 
\nonumber\\ 
&&
+ i \sqrt{2} g f^{abc} \left( 
\bar \lambda_a \bar z_b^j L \psi_c^j - 
\bar \psi_a^j R z_b^j \lambda_c 
\right) 
- \half Y f^{abc} \e_{ijk} \left( 
\bar \psi_a^i z_b^j L \psi_c^k - 
\bar \psi_a^i R \bar z_b^j \psi_c^k 
\right) 
\nonumber\\ 
&&
- \half g^2 (f^{abc} \bar z_b^j z_c^j ) (f^{ade} \bar z_d^k z_e^k ) 
+ \quarter Y^2 f^{abc} f^{ade} \e_{ijk} \e_{ilm} 
z_b^j z_c^k \bar z_d^l \bar z_e^m 
\end	{eqnarray}
where $L$ and $R$ are chirality projectors. 
Separate coupling constants $g$ and $Y$ are used 
to distinguish the terms coming from the gauge and 
superpotential sectors. 
The terms in the Lagrangian 
(\ref{lagrangian:for Feynman rules}) 
proportional to $g$ and $g^2$ come from the $D$-terms; 
while the ones appearing with $Y$ and $Y^2$, 
from the $F$-terms. 

The theory defined by (\ref{lagrangian:for Feynman rules}) 
has a global symmetry group $SU(3) \times U(1)$. 
When $Y = g \sqrt{2}$, SUSY is enhanced from \NN=1 to \NN=4; 
and the $R$-symmetry group becomes $SU(4)$, 
although its manifestly realized part is still $SU(3) \times U(1)$. 
Elementary fields in the Lagrangian (\ref{lagrangian:for Feynman rules}) 
have good quantum numbers under 
this subgroup of $SU(4)$. 
The scalars $z_i$ and $\bar z_i$ are related to the 
fields $\phi^I$, $I = 1 , ... , 6$ of a manifestly 
$SO(6)$ invariant formulation by a projection 
$\phi_a = {1\over\sqrt{2}}(z_a + \bar z_a)$, 
$\phi_{a+3} = {1\over i \sqrt{2}}(z_a - \bar z_a)$, $a=1,2,3$. 
Order $g^2$ calculations 
are simpler in terms of the fields 
$\{ 
\lambda, z_i, \bar z_i, \psi_i, \bar \psi_i
\, | \, 
i=1,2,3 
\}$,
rather than 
$\{
\phi^I, \lambda^j
\, | \, 
I=1,...,6, \, j=1,...,4
\}$, so we 
will use the Lagrangian (\ref{lagrangian:for Feynman rules}) 
throughout this paper.

\section{Contributing diagrams}
\label{section:contributing diagrams}

Next, we to sort out 
the diagrams which contribute to 
correlators of these operators 
at order $g^0$ and $g^2$. 

\begin{figure}[t!]
{\begin{center}
\epsfig{width=4.75in, file=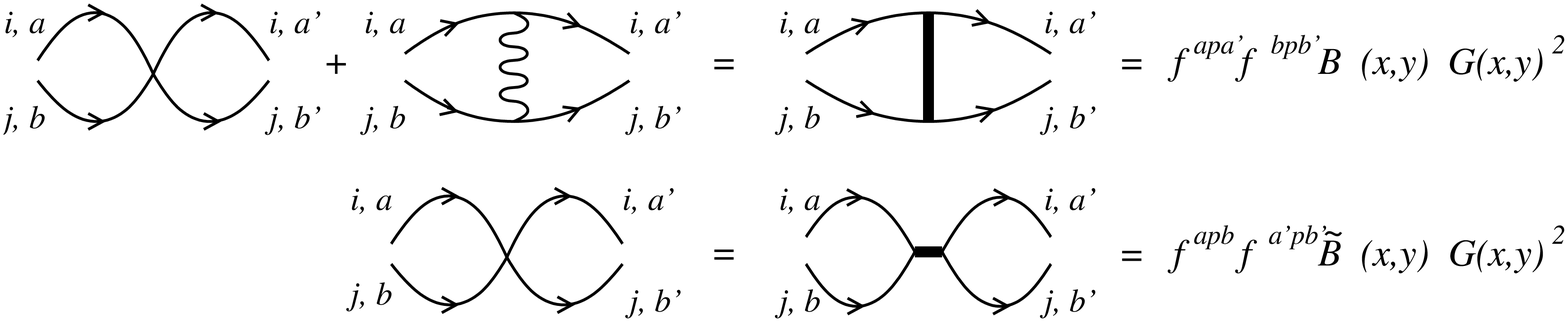, angle=0}
\vskip 0.15in
\epsfig{width=4.6in, file=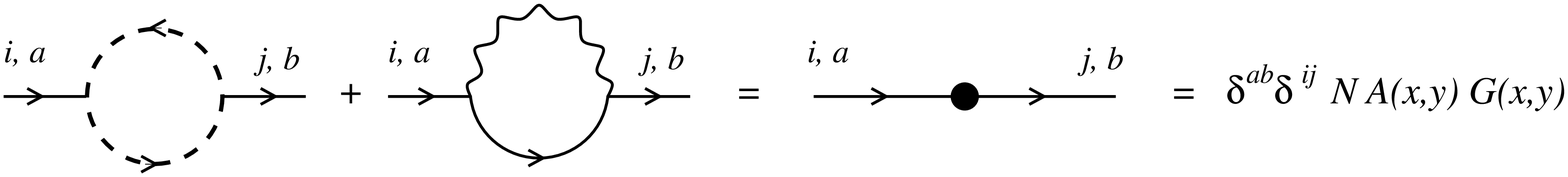, angle=0}\quad\quad\quad
\end{center}}
\vskip -0.2in
\caption{%
Structures contributing to two-point functions of scalars at order $g^2$ 
through 
four-scalar blocks 
and 
the propagator. 
Thick lines correspond to exchanges of the gauge boson, 
and of the auxiliary fields $F_i$ and $D$
(in the \NN=1 formulation; 
after integrating out $F_i$ and $D$, 
the $z z \bar z \bar z$ vertex). 
The scalar propagator remains diagonal in both color and 
flavor indices at order $g^2$. 
\label {fig:four-scalar and propagator}
}%
\end {figure}

The two-point functions of the 
scalar composites discussed in \cite{Ryzhov}, 
have the schematic form 
\begin{equation}
\label{eq:2-pt generic}
\langle 
\left[ {z_1}^{(p+q)} {z_2}^p \right](x) 
\left[ {\bar z_1}^{(p+q)} {\bar z_2}^p \right](y) \rangle 
\end{equation}
where $[...]$ are some gauge invariant combinations. 
The free field part of (\ref{eq:2-pt generic}) 
is given by a power of the free
correlator $[G(x,y)]^{(2p+q)}$, times a combinatoric factor. 
From the Lagrangian (\ref{lagrangian:for Feynman rules}) 
we can read off the structures
for the leading
correction to the propagator, and the four-scalar blocks. 
These are shown in Figure \ref{fig:four-scalar and propagator}, 
where they are categorized according to their 
gauge group (color) index structure 
(we use the same notation as in \cite{DFS}). 
Gauge fixing
and ghost terms in the Lagrangian do not contribute at this order
(as we are only looking at operators made up of scalars).

\begin{figure}[t!]
{\begin{center}
\epsfig{width=3.6in, file=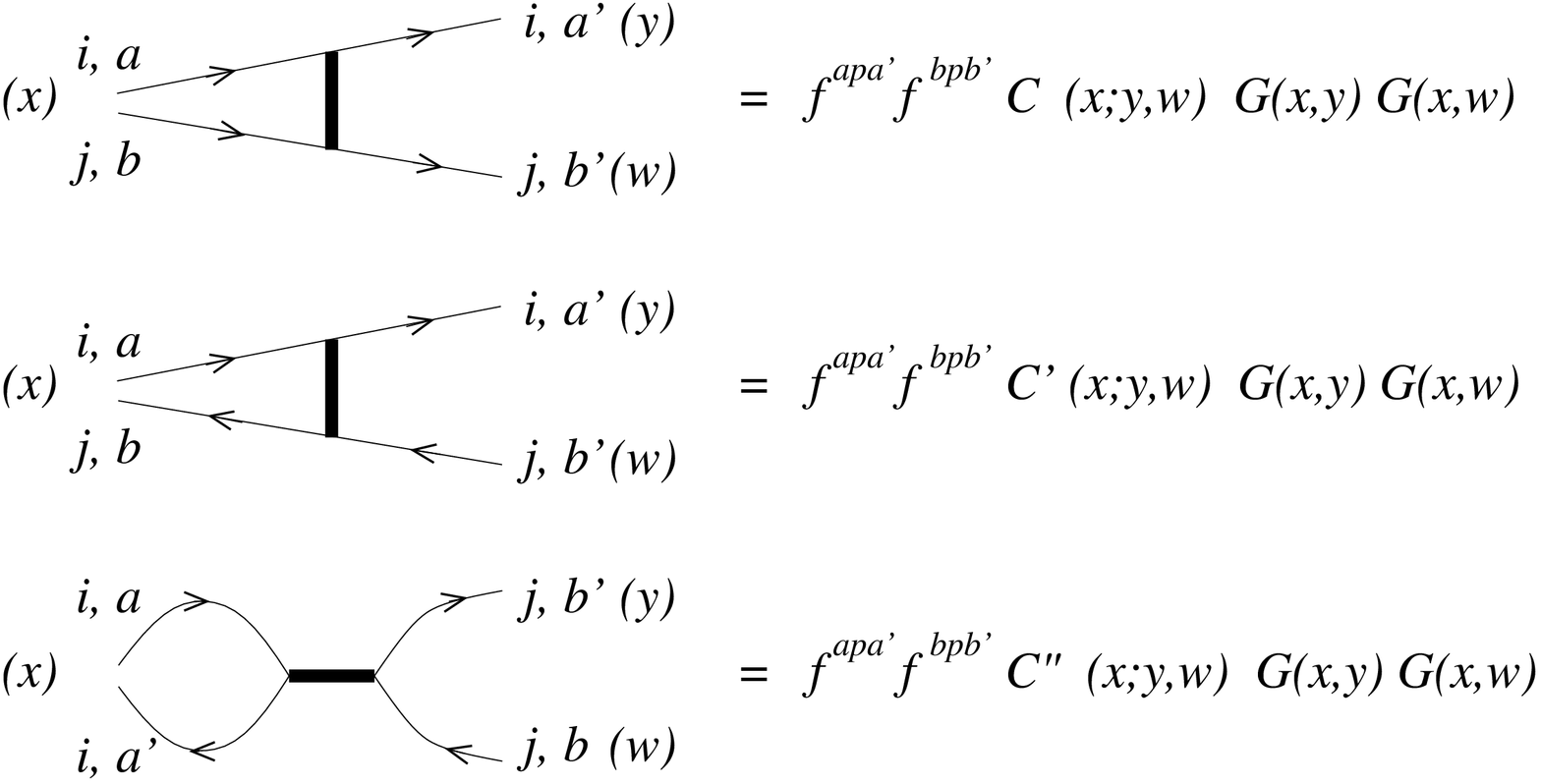, angle=-90}
\vskip 0.2in
\epsfig{width=3.6in, file=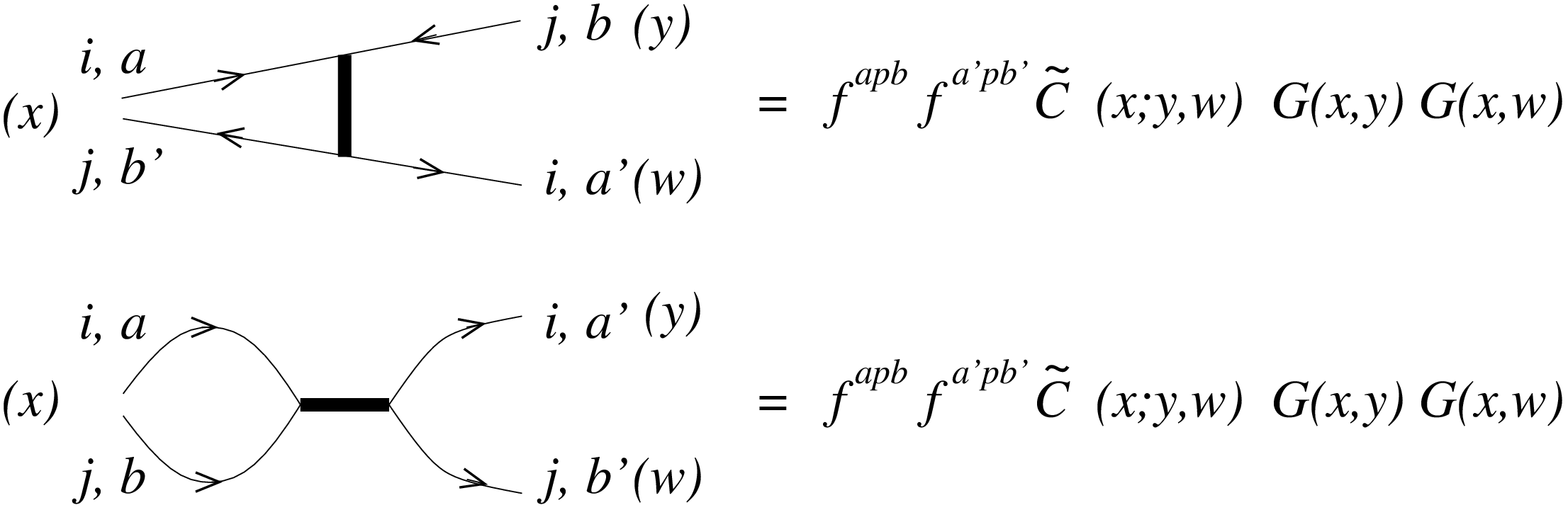, angle=-90}
\end{center}}
\vskip -0.2in
\caption{%
Building blocks for $g^2$ corrections to three-point functions. 
The three-points are $x$ (with two legs attached) and $y$ and $w$ 
(with a single leg each). 
\label {fig:three-pt-blocks}
}%
\end {figure}

Three-point functions to be considered in this paper
are of the form 
\begin{equation}
\label{eq:3-pt generic}
\langle 
\left[ z^{k+l} \right](x) 
\left[ \bar z^{k+m} \right](y) 
\left[ \bar z^l z^m \right](w) 
\rangle 
\end{equation}
The free result is just 
the product of appropriate powers of 
free correlators $[G(x,y)]^k [G(x,w)]^l [G(y,w)]^m$. 
The same structures that contribute to the two-point functions 
at order $g^2$ 
(see Figure \ref{fig:four-scalar and propagator}), 
also contribute to the three-point functions (\ref{eq:3-pt generic}). 
Apart from these, 
there are new building blocks, shown in Figure \ref{fig:three-pt-blocks}. 
They have the same index structure, 
but are now functions of three 
space-time coordinates rather than two. 

Notice that the $F$-term corrections proportional to $\tilde B(x,y)$ 
in Figure \ref{fig:four-scalar and propagator}, 
and the last graph (proportional to $\tilde C(x;y,w)$)
in Figure \ref{fig:three-pt-blocks}, 
are antisymmetric in $i$ and $j$, hence they are absent when the 
scalars in the four legs have pairwise the same flavor. 
For the same reason, 
these corrections are also absent when 
the operator at point $x$ 
is symmetric in all of its flavor indices. 
In particular, this is the case when the operator 
at $x$ is \half-BPS.

\section{Restrictions from \NN=4 SUSY and gauge invariance}
\label{section:gauge-dependent}

The form of quantum corrections to two and three-point functions 
is known \cite{DFS}. 
Space-time coordinate dependence 
of the Feynman diagrams contributing to these 
correlators at order $\OO(g^2)$ 
is constrained, since all exchanged fields are massless. 
We know the parametric form of 
the functions $A(x,y)$, $B(x,y)$, $\tilde B(x,y)$; and 
$C(x;y,w)$, $C'(x;y,w)$, $C''(x;y,w)$, and $\tilde C(x;y,w)$, 
without having to perform integrals explicitly. 
Functions which depend on two space-time points, 
are of the form $A(x_1, x_2) = a \log x_{12}^2 \mu^2 + b$ 
with $x_{ij} \equiv x_i-x_j$; 
three-point contributions look like 
$C(x_1;x_2,x_3) = a' \log x_{12}^2 x_{13}^2 \mu^4 - a'' \log x_{23}^2 + b'$ 
(making use of the $x_2 \lra x_3$ symmetry of these building blocks).

\NN=4 SUSY tells us more. 
From non-renormalization of two and three-point functions 
of operators in the stress tensor multiplet, one can see 
\cite{DFS} that 
$B(x,y) = - 2 A(x,y)$, 
and 
$C'(0;x,y) + \tilde C(0;x,y) = -C(0;x,y)$; 
the authors of \cite{DFS} chose to combine these and call it just $C'$.%
\footnote{
	One way to see this is to consider the protected 
	correlators of [0,2,0] scalar composite operators 
	$\langle \tr z_1 z_2 (x) \; \tr \bar z_1 \bar z_2 (y) \rangle$, 
	and 
	$\langle [z^2](x) [\bar z^2](y) [z \bar z](w) \rangle$
	and
	$\langle [z \bar z](x) [z \bar z](y) [z \bar z](w) \rangle$. 
	}
The coefficients $a'$, 
$a''$ and $b'$ are determined%
\footnote{
	This follows from 
	$C(x;y,w) + C(y;x,w) + C(w;x,y) + A(x,y) + A(y,w) + A(x,w)  =0$. 
	}
in terms of $a$ and $b$: 
\begin{eqnarray}
A (x,0) ~=~ 
-\half B (x,0) 
&=& a \log {x^2 \mu^2} + b
\nonumber\\
- C (0;x,y) &=& a \log {x^2 y^2 \mu^2\over (x-y)^2} + b
\end{eqnarray}
Therefore, the net contribution 
to the three-point function (\ref{eq:3-pt generic}) 
of the $\OO(g^2)$ diagrams involving a gauge boson exchange 
(the ones proportional to $A$, $B$, and $C$), 
is 
\begin{eqnarray}
\label{three-point:D-term}
\langle 
[z^k](x_1) [\bar z^l](x_2) [z^m \bar z^n](x_3) 
\rangle |_{(A+B+C)} 
\hspace{-15em}
\nonumber\\
&=& 
a 
(
c_g^{12} \log x_{12}^2 \mu^2 + 
c_g^{13} \log x_{13}^2 \mu^2 + 
c_g^{23} \log x_{23}^2 \mu^2 
)
+
b c_g^{123}
\end{eqnarray}
where and $c_g^{ij}$ and $c_g^{123}$ are 
some combinatorial coefficients.

Now we use gauge invariance of the theory. 
On the one hand, 
we observe that 
the coefficients $a$ and $b$ are gauge dependent 
\cite{Ryzhov}, 
\begin 	{eqnarray}
A(x,0) &=& 
\half \pi^2 g^2 \xi 
\left[ 
\log x^2 \mu^2 + \log 4\pi - \gamma 
\right]
+ (\mbox{$\xi$-independent}) 
\end 	{eqnarray}
where $\xi$ is the gauge fixing parameter. 
On the other hand, 
a correlator of gauge invariant operators 
can not depend on $\xi$. 
Therefore, the combinatorial coefficients 
multiplying $a$ and $b$ in equation (\ref{three-point:D-term}) 
must vanish, 
$c_g^{ij} = c_g^{123} = 0$. 
Hence, the $D$-term diagrams 
proportional to $A$, $B$, and $C$ all cancel; 
their net contribution 
to the three-point functions (\ref{eq:3-pt generic}) is zero. 

So just like in the case of two-point functions, 
we only have to consider the $F$-term graphs. 
They are proportional to $\tilde B$ and $\tilde C$, 
the only gauge independent diagrams around 
($C'=-(C+\tilde C)$ and $C''=C-\tilde C$ do not have to 
be treated separately as they are linear combinations 
of the other ones).

In the $\OO(g^2)$ calculations of correlators of 
\half-BPS operators \cite{DFS} and \cite{Skiba}, 
there were no other contributions to 
three-point functions except for 
those proportional to $A$ and $B$. 
Thus, gauge invariance together with \NN=4 SUSY 
(which is needed to relate $C$ and $B$ to $A$) 
guarantees that the correlators of \cite{DFS} and \cite{Skiba} 
receive no order $g^2$ corrections.

\section{Position dependence of $\tilde B$ and $\tilde C$}
\label{section:c-tilde}

Having shown that $D$-term corrections to three-point functions 
(\ref{eq:3-pt generic}) are absent, it remains to 
consider the $F$-term interactions. 
In this Section we derive a relation between 
functions $\tilde B$ and $\tilde C$, 
which will play a key role in the analysis of 
three-point functions of \quarter-BPS chiral primaries, 
see Section \ref{section:cat diagrams}.

Space-time position dependence of 
$\tilde B$ and $\tilde C$ 
(shown Figures \ref{fig:four-scalar and propagator} and 
\ref{fig:three-pt-blocks}) 
is parametrically determined to be 
$\tilde B(x,0) = \tilde a \log (x^2 \mu^2) + \tilde b$ and 
$\tilde C(0;x,y) = \tilde a' \log (x^2 y^2 \mu^4) 
- \tilde a'' \log ((x-y)^2 \mu^2) + \tilde c$; furthermore, 
the leading divergent behavior can be read off from the 
integrals unambiguously and so from the limit 
$\tilde C(0;x,y \to x)$ we infer $\tilde a' = \half \tilde a$.

To evaluate the remaining coefficients 
$\tilde a$, $\tilde a''$, and $\tilde b$, 
we use differential regularization \cite{FJMV},%
\footnote{
	This is the fastest way to calculate the 
	integral for $\tilde C$, 
	but one can obtain the same results 
	using dimensional regularization. 
	}
or a simpler equivalent prescription:
replace 
$1/x^2 \to 1/(x^2 + \e^2)$ for scalar propagators inside 
integrals. 
With this prescription 
\begin{eqnarray}
\label{b-tilde}
\tilde B (x,0) &\!\!=\!\!& 
- \quarter Y^2  
\int { (d^4 z) \left[ 4 \pi^2 x^2 \right]^2 \over 
\left[ 4 \pi^2 ((z-x)^2 + \e^2)\right]^2 \left[ 4 \pi^2 (z^2 + \e^2)\right]^2}
\nonumber\\&\!\!=\!\!& 
-Y^2  
{1 \over 32 \pi^2 } 
\left[ \log (x^2/\e^2) - 1\right] 
\\
\noalign{\noindent
is the regularized two-point function, 
while the three-point function becomes 
}
\label{c-tilde}
\tilde C (x;y,0) &\!\!=\!\!& 
- \quarter Y^2  
\int { (d^4 z) \left[ 4 \pi^2 x^2 \right] \left[ 4 \pi^2 (x-y)^2 \right] \over 
\left[ 4 \pi^2 ((z-x)^2 + \e^2)\right]^2 
\left[ 4 \pi^2 ((z-y)^2 + \e^2)\right]
\left[ 4 \pi^2 (z^2 + \e^2)\right] }
\nonumber\\&\!\!=\!\!& 
-Y^2  
{1 \over 64 \pi^2 } 
\left[ \log {x^2 (x-y)^2 \over y^2 \e^2}\right] 
\end{eqnarray}
(The numerators inside the integrals come about 
because of the powers of free scalar propagator 
in the definitions of $\tilde B$ and $\tilde C$, 
see Figures \ref{fig:four-scalar and propagator} and 
\ref{fig:three-pt-blocks}.) 
Therefore, 
\begin{eqnarray}
\label{eq:cat diagrams}
\tilde C(x;y,0) + 
\tilde C(y;x,0) - 
\tilde B(x,y) = - Y^2 \times {1 \over 32 \pi^2 } 
\end{eqnarray}
is a nonzero constant 
(for \NN=4 SUSY, $Y^2 = 2 g^2$).

The value of the constant $-Y^2/32 \pi^2$ 
in equation (\ref{eq:cat diagrams}) 
does not depend on the regulator $\e$. 
Also note that with the ``point splitting regularization'' 
one would get the incorrect result of vanishing constant 
in (\ref{eq:cat diagrams}).

\section{Structure of the three-point functions}
\label{section:3-pt structure}

With the results of Section \ref{section:gauge-dependent} 
at hand, we can write down the form of a general 
three-point function of scalar composite operators 
(\ref{eq:3-pt generic}) to order $g^2$: 
\begin{eqnarray}
\label{eq:3-pt g^2}
\langle 
\left[ z^{k+l} \right] (x) 
\left[ \bar z^{k+m} \right] (y) 
\left[ \bar z^l z^m \right] (w) 
\rangle 
&\!\!=\!\!& 
G(x,y)^k 
G(x,w)^l 
G(w,y)^m 
\nonumber\\&&\hspace{-5.55em} \times
\Big( \Big. 
\alpha_{\rm free} 
+ \tilde \beta_{xy} \tilde B(x,y) 
+ \tilde \beta_{xw} \tilde B(x,w) 
+ \tilde \beta_{yw} \tilde B(y,w) 
\nonumber\\&&\hspace{-2em}
+ \tilde \gamma_{x} \tilde C(x;y,w) 
+ \tilde \gamma_{y} \tilde C(y;x,w) 
+ \tilde \gamma_{w} \tilde C(w;x,y) 
\nonumber\\&&\hspace{-2em}
+ \OO(g^4) 
\Big. \Big) 
\end{eqnarray}
where $\alpha_{\rm free}$, $\tilde \beta$-s and $\tilde \gamma$-s 
are some combinatorial coefficients. 
Using the expressions 
(\ref{b-tilde}) and (\ref{c-tilde}) 
from Section \ref{section:c-tilde}, 
we can determine the $\OO(g^2)$ position dependence of 
(\ref{eq:3-pt g^2}) completely 
--- if we know these combinatorial coefficients. 
Together with conformal invariance, 
and the $SU(4)$ symmetry properties 
of the operators in (\ref{eq:3-pt g^2}), 
we can often go a long way to figuring out 
which of the combinatorial coefficients must 
vanish, without doing any actual calculations.

\subsection{Space-time coordinate dependence}

Like in the case of two-point functions, 
conformal invariance restricts the position 
dependence of three-point correlators of
pure operators (i.e. ones which have a well 
defined scaling dimension). 
Consider three (gauge invariant Lorentz scalar) 
operators $\OO_1$, $\OO_2$, and $\OO_3$, 
of dimensions 
$\Delta_i = k_i + \delta_i$, 
inserted at corresponding points $x_i$. 
Let $k_i$ be integers and $\delta_i$ 
the order $g^2$ corrections to the scaling dimensions 
(which may or may not be zero). 
The three-point function 
$\langle \OO_1 \OO_2 \OO_3 \rangle$
is completely determined up to 
a multiplicative constant $C_{123} = C^0_{123} + C^1_{123}$ 
(where again $C^0_{123}$ is the free field result 
and $C^1_{123} \sim g^2$), 
\begin{eqnarray}
\label{eq:3-pt constraint}
\langle
\OO_1 (x_1) \OO_2 (x_2) \OO_3 (x_3) 
\rangle
&=& 
{
C_{123} 
\over 
x_{12}^{\Delta_1 + \Delta_2 - \Delta_3}
x_{13}^{\Delta_1 + \Delta_3 - \Delta_2}
x_{23}^{\Delta_2 + \Delta_3 - \Delta_1}
}
\nonumber\\
&=& 
\langle
\OO_1 \OO_2 \OO_3 
\rangle_{\rm free} 
\Big( \Big. 
1 
+ C^1_{123} / C^0_{123} 
\nonumber\\&&\quad 
- \delta_1 \log {x_{12}^2 x_{13}^2 \over x_{23}^2 \e^2}
- \delta_2 \log {x_{21}^2 x_{23}^2 \over x_{13}^2 \e^2}
- \delta_3 \log {x_{31}^2 x_{32}^2 \over x_{12}^2 \e^2}
\nonumber\\&&\quad 
+ \OO(g^4) 
\Big. \Big) 
\end{eqnarray}
with $x_{ij} = x_i - x_j$ as usual.

Suppose that all three operators have 
protected scaling dimensions, $\delta_i = 0$. 
Then (\ref{eq:3-pt constraint}) reduces to 
\begin{eqnarray}
\label{eq:3-pt constraint:all protected}
\langle
\OO_1 (x_1) \OO_2 (x_2) \OO_3 (x_3) 
\rangle
= 
\langle
\OO_1 \OO_2 \OO_3 
\rangle_{\rm free} 
\Big( \Big. 
1 
+ C^1_{123} / C^0_{123} 
\Big. \Big) 
\end{eqnarray}
and no logs arise. In this case, the 
combinatorial factors in (\ref{eq:3-pt g^2}) 
satisfy 
\begin{eqnarray}
\label{eq:coeffs - all three:beg}
\tilde \gamma_x 
&=& 
- \left( 
\tilde \beta_{xy} + \tilde \beta_{xw} 
\right) 
\\
\tilde \gamma_y 
&=& 
- \left( 
\tilde \beta_{xy} + \tilde \beta_{yw} 
\right) 
\\
\tilde \gamma_w 
&=& 
- \left( 
\tilde \beta_{xw} + \tilde \beta_{yw} 
\right) 
\label{eq:coeffs - all three:end}
\end{eqnarray}
and we need to calculate only three coefficients 
(the $\tilde \beta$-s, for example), 
to find all the $\OO(g^2)$ corrections 
to this correlator. 
In fact, the only allowed correction 
is the constant 
$
\tilde \beta \equiv 
\tilde \beta_{xy} + \tilde \beta_{xw} + \tilde \beta_{yw}$
times $(- Y^2 / 32 \pi^2)$. 
To show that a three-point function of BPS operators is protected, 
we have to demonstrate that $\tilde \beta = 0$.

\subsection{Group theory simplifications}
\label{section:simplifications}

There are several simplifications which set some of the 
combinatorial coefficients in (\ref{eq:3-pt g^2}) to zero. 
These considerations are based on the underlying 
$SU(4) \sim SO(6)$ symmetry of the theory only, 
and are applicable for general $N$. 
We will leave aside the trivial case when the 
correlator is forced to vanish by group theory, 
and assume that the Born level coefficient 
in (\ref{eq:3-pt g^2}) $\alpha_{\rm free} \ne 0$.

The simplest BPS operators are \half-BPS chiral primaries, 
gauge invariant scalar composites in $[0,q,0]$ 
representations of $SO(6)$. 
These are 
totally symmetric tensors of $SO(6)$, so if 
for example the operator $\OO_x$ is \half-BPS, 
the coefficients 
$\tilde \beta_{xy} = \tilde \beta_{xw} = \tilde \gamma_x = 0$
since the diagrams they multiply are antisymmetric 
in the flavor indices of $\OO_x$.%
\footnote{
	If we chose $\OO_w$ as such a \half-BPS operator, 
	we would not be able to conclude that $\tilde \gamma_w = 0$
	just from the symmetries of $\OO_w$: 
	the fourth diagram of Figure \ref{fig:three-pt-blocks}, 
	is not antisymmetric in flavor indices 
	at the vertex where the operator is made 
	of both $z$-s and $\bar z$-s. 
	However, using equation (\ref{eq:coeffs - all three:end})
	we find $\tilde \gamma_w = \tilde \beta_{xw} + \tilde \beta_{yw}= 0$
	since $\tilde \beta_{xw} = \tilde \beta_{yw}= 0$ 
	when all three operators have protected scaling dimensions. 
	}
Similarly, if both $\OO_x$ and $\OO_y$ are \half-BPS and 
$\OO_w$ is any BPS operator, we have 
$\tilde \beta_{xy} = \tilde \beta_{xw} = \tilde \gamma_x = 
\tilde \beta_{yw} = \tilde \gamma_y = 0$, 
and hence 
$\tilde \beta = 
\tilde \beta_{xy} + \tilde \beta_{xw} + \tilde \beta_{yw} = 0$; 
there are no $\OO(g^2)$ corrections in this case. 
In particular, this reproduces the result 
of \cite{DFS} when all three $\OO_{x,y,w}$ are \half-BPS chiral primaries.

\begin{figure}[t!]
{\begin{center}
\epsfig{height=1.0in, file=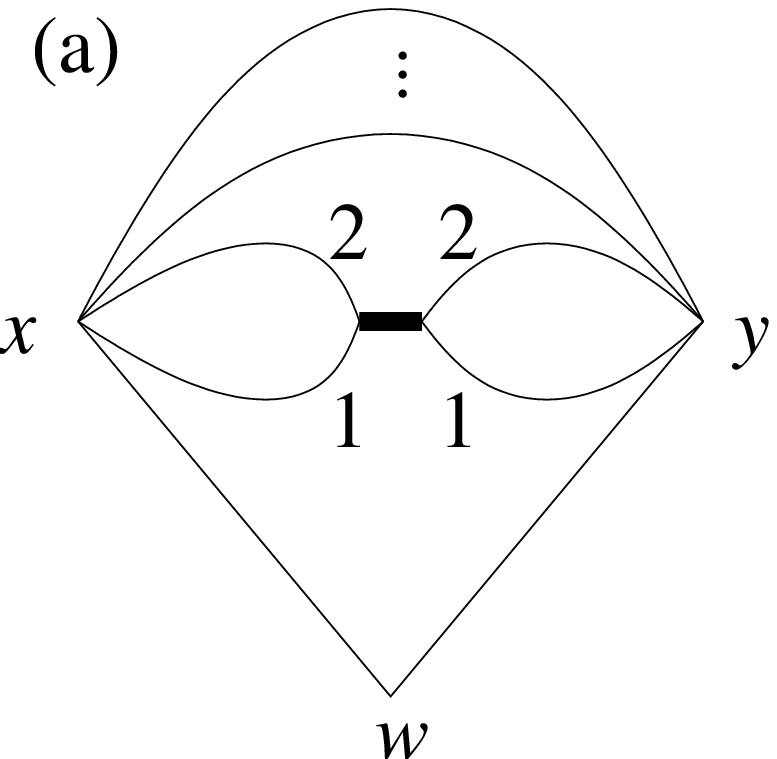, angle=0}
\hspace{1.6em}
\epsfig{height=1.0in, file=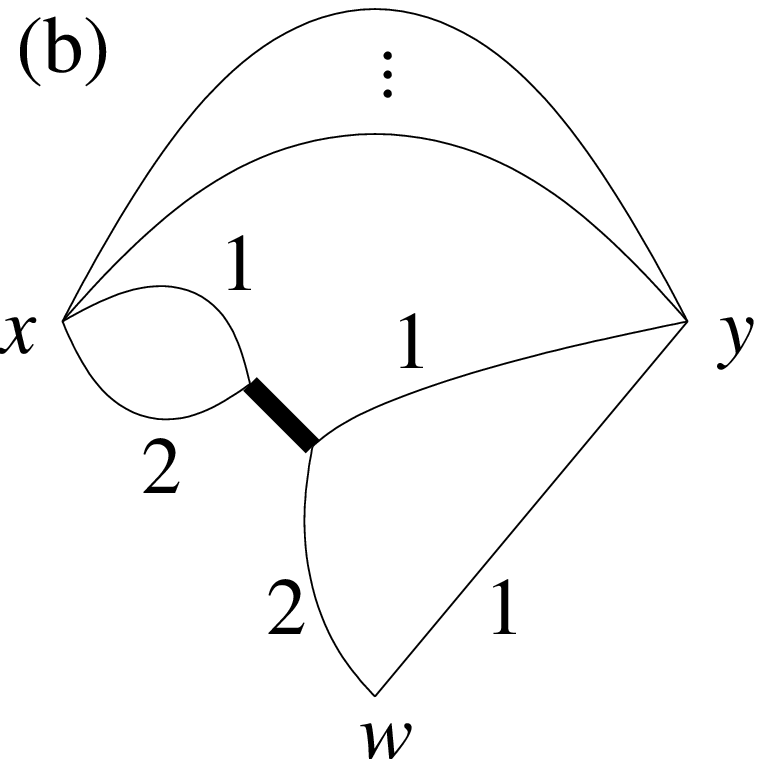, angle=0}
\hspace{1.6em}
\epsfig{height=1.0in, file=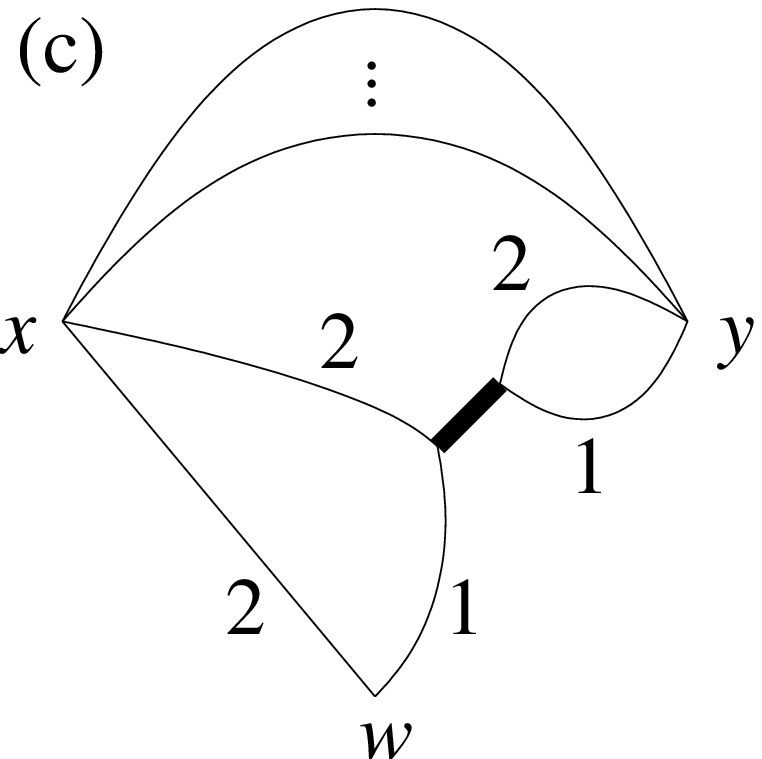, angle=0}
\end{center}}
\vskip -0.2in
\caption{%
$F$-term contributions to 
$\langle \OO_{\rm BPS}(x) \OO'_{\rm BPS}(y) \OO_{\tiny \half}(w)
\rangle_{g^2}$ 
in the case when the correlator can be ``partitioned into two flavors:'' 
(a) proportional to $\tilde B(x,y)$; 
(b) proportional to $\tilde C(x;y,w)$; 
(c) proportional to $\tilde C(y;x,w)$. 
\label {fig:3-point-z1-z2}
}%
\end {figure}

Now suppose that $\OO_w$ is \half-BPS, and furthermore we can 
``partition the correlator into two flavors,''
i.e. choose the operators such that 
$\OO_w = [\bar z_1^m z_2^n]$ 
while 
$\OO_x = [z_1^k z_2^l]$ 
and $\OO_y = [\bar z_1^{(k-m)} \bar z_2^{(l-n)}]$. 
Consider a diagram proportional to $\tilde C(x;y,w)$, 
see Figure \ref{fig:3-point-z1-z2}. 
The sum of all such diagrams 
is symmetric in the flavors at $y$ (since they are the same), 
and symmetric in flavors at $w$ (since $\OO_w$ is \half-BPS). 
But it must be antisymmetric in the flavors $z_1$ and $z_2$ 
leaving the interaction vertex, so 
all such diagrams cancel, and so $\tilde \gamma_x = 0$. 
In the same fashion, we conclude that $\tilde \gamma_y = 0$ as well, 
and together with 
$\tilde \beta_{yw} = \tilde \beta_{xw} = 0$
(as $\OO_w$ is \half-BPS), we find that $\beta = 0$ 
when $\OO_x$ and $\OO_y$ are any operators with 
protected scaling dimensions.

\begin{figure}[t!]
\vskip 0.2in
{\begin{center}
\epsfig{width=4.5in, file=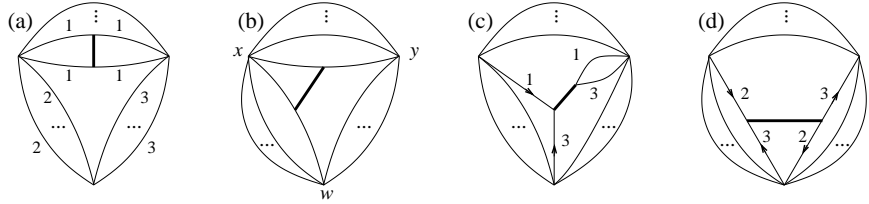, angle=0}
\end{center}}
\vskip -0.2in
\caption{%
Order $g^2$ corrections to correlators of the form 
(\ref{disjoint-flavors}): 
(a) and (b) includes a gauge boson exchange;
(c) and (d) $F$-terms.
Self energy contributions (not shown) also 
include a gauge boson exchange. 
\label {fig:disjoint-flavors}
}%
\end {figure}

There is another type of three-point functions 
of BPS chiral primaries which receive no 
$\OO(g^2)$ corrections by similar considerations. 
Consider a correlator 
(\ref{eq:3-pt g^2}) such that $\OO_x$ 
is made of $z_1$ and $z_2$; 
$\OO_y$ made of $\bar z_1$ and $\bar z_3$; 
and $\OO_w$, made of $\bar z_2$ and $z_3$, 
i.e. a correlator of the form  
\begin	{eqnarray}
\langle 
\label{disjoint-flavors}
[z_1^m z_2^n] (x) \; [\bar z_1^m \bar z_3^k] (y) \; [z_3^k \bar z_2^n] (w) 
\rangle 
\end	{eqnarray}
This correlator is ``partitioned into three disjoint flavors.'' 
Order $g^2$ contributions to this three-point function are shown in 
Figure \ref{fig:disjoint-flavors}. 
There are no corrections proportional to any $\tilde B$-s 
since all lines within any rainbow carry the same flavor, 
so immediately 
$\tilde \beta_{xy} = \tilde \beta_{yw} = \tilde \beta_{xw} = 0$ 
and hence
there are no $\OO(g^2)$ corrections here, as well.

\begin{figure}[t!]
\vskip 0.2in
{\begin{center}
\epsfig{width=4.7in, file=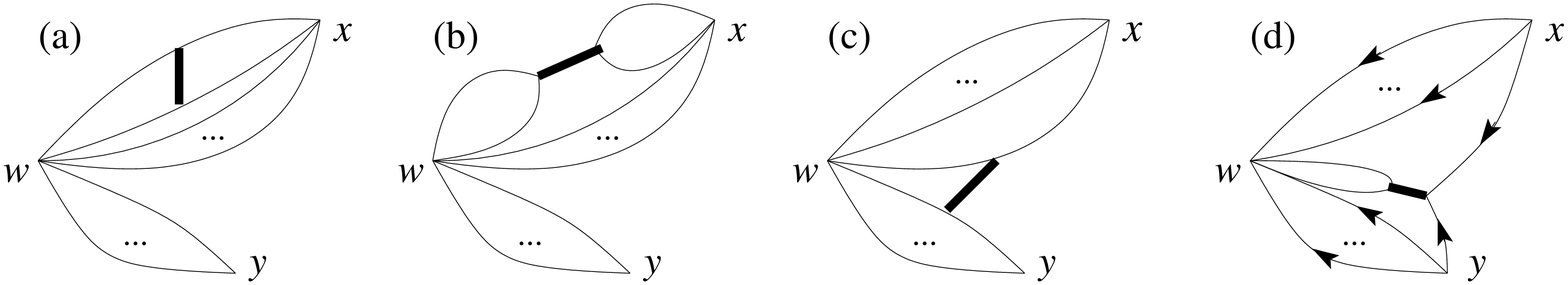, angle=0}
\end{center}}
\vskip -0.2in
\caption{%
Order $g^2$ corrections to extremal correlators:
(a) and (b) within a single peddle;
(c) and (d) between the two peddles. 
Self energy contributions (not shown) and diagrams (a) and (c) 
are gauge dependent, while (b) and (d) diagrams arise from 
the $F$-terms. 
\label {fig:extremal}
}%
\end {figure}

Finally, extremal three-point functions 
can be analyzed in a simple way. 
Here, the scaling dimension of one of the operators 
is equal to the sum of scaling dimensions of the other two.%
\footnote{
	In general, 
	$(n+1)$-point functions 
	$\langle \OO_0 (x_0) \OO_1(x_1) ... \OO_n(x_n) \rangle$ 
	are called extremal if one of the scaling dimensions 
	is the sum of all the others, 
	$\Delta_0 = \Delta_1 + ... + \Delta_n$.
	}
Suppose that $\Delta_x + \Delta_y = \Delta_w$ in 
(\ref{eq:3-pt g^2}). 
At Born level, there are no $G(x,w)$ propagators, 
and so there are no corrections proportional 
to $\tilde B(x,y)$, $\tilde C(x;y,w)$, or $\tilde C(y;x,w)$, 
see Figure \ref{fig:extremal}. 
Together with the constraints 
(\ref{eq:coeffs - all three:beg}-\ref{eq:coeffs - all three:end}), 
this determines $\tilde \beta = 0$ 
when the three (Lorentz scalar) operators inserted $x$, $y$, and $w$ 
are arbitrary operators with protected scaling dimensions.

Another remark about extremal correlators is in order. 
As it is easy to see, one of the above group theory 
simplifications 
generalizes straightforwardly to 
extremal correlators of chiral primaries. 
Namely, if all operators except for one are \half-BPS 
(and the remaining one is an arbitrary chiral primary), 
extremal correlators 
receive no order $g^2$ corrections.

\section{Three-point functions of BPS operators}

We are now ready to discuss correlators of 
three BPS chiral primaries. 
The simplest correlators 
$\langle 
\OO_{\tiny \! \half} \OO_{\tiny \! \half} \OO_{\tiny \! \half} 
\rangle$ 
(where each $\OO_{\tiny \! \half}$ stands for a \half-BPS operator), 
were considered by 
\cite{DFS}, who found that three-point functions
of \half-BPS operators do not get corrected at order $g^2$, 
for any $N$. 
These are a special case of 
correlators of the form 
$\langle \OO_{\tiny \! \half} \OO_{\tiny \! \half} \OO_{\rm BPS} 
\rangle$, 
which we discussed in Section \ref{section:simplifications} 
(here $\OO_{\rm BPS}$ is an arbitrary BPS operator). 
These three-point functions receive no $\OO(g^2)$ corrections 
by group theory reasoning.

\subsection
{Correlators $\langle 
\OO_{\tiny \! \quarter} \OO_{\tiny \! \quarter} \OO_{\tiny \! \half} 
\rangle$}

Not all three-point functions of chiral primaries can be simplified 
using the results of Section \ref{section:simplifications}, 
so occasionally we will have to actually 
compute some of the combinatorial coefficients. 
In this Section we will look at correlators 
of two \quarter-BPS operators with one \half-BPS operator. 

\subsubsection{$\langle \OO_{[p,q,p]}(x) \bar{\OO}_{[p,q,p]}(y) (\tr \bar z t z) (w) \rangle$}
\label{section:cat diagrams}

The simplest 
$\langle 
\OO_{\tiny \! \quarter} \OO_{\tiny \! \quarter} \OO_{\tiny \! \half} 
\rangle$ 
three-point functions 
are of the form 
\begin{eqnarray}
\label{o-o-x-square}
\langle \OO (x) \bar{\OO}' (y) (\tr X^2) (w) \rangle 
\end{eqnarray}
where the \half-BPS primary $\tr X^2$ is a 
scalar composite operator in the [0,2,0] of $SU(4)$. 
Group theory restricts the quantum numbers of 
operators which can have nontrivial tree point functions. 
Tensoring $[p,q,p] \otimes [0,2,0]$ 
using Young diagrams of $SO(6)$ gives 
\begin{eqnarray}
\label{pqp-times-020}
\mbox{
\setlength{\unitlength}{0.6em}
\begin{picture}(7.5,2)
\put(0,1){\framebox (4,1){\scriptsize $p$}}
\put(4,1){\framebox (3,1){\scriptsize $q$}}
\put(0,0){\framebox (4,1){\scriptsize $p$}}
\end{picture}}
\otimes
\mbox{
\setlength{\unitlength}{0.6em}
\begin{picture}(2,2)
\put(-0.5,1){\framebox (1,1){$~$}}
\put(0.5,1){\framebox (1,1){$~$}}
\end{picture}}
&=& 
\mbox{
\setlength{\unitlength}{0.6em}
\begin{picture}(9.5,2)
\put(0,1){\framebox (4,1){$~$}}
\put(4,1){\framebox (3,1){$~$}}
\put(0,0){\framebox (4,1){$~$}}
\put(7,1){\framebox (1,1){$~$}}
\put(8,1){\framebox (1,1){$~$}}
\end{picture}}
\oplus
\mbox{
\setlength{\unitlength}{0.6em}
\begin{picture}(8.5,2)
\put(0,1){\framebox (4,1){$~$}}
\put(4,1){\framebox (3,1){$~$}}
\put(0,0){\framebox (4,1){$~$}}
\put(7,1){\framebox (1,1){$~$}}
\put(4,0){\framebox (1,1){$~$}}
\end{picture}}
\oplus
\mbox{
\setlength{\unitlength}{0.6em}
\begin{picture}(7.5,2)
\put(0,1){\framebox (4,1){$~$}}
\put(4,1){\framebox (3,1){$~$}}
\put(0,0){\framebox (4,1){$~$}}
\put(4,0){\framebox (1,1){$~$}}
\put(5,0){\framebox (1,1){$~$}}
\end{picture}}
\nonumber\\
&\oplus&
\mbox{
\setlength{\unitlength}{0.6em}
\begin{picture}(9.5,2)
\put(0,1){\framebox (4,1){$~$}}
\put(4,1){\framebox (3,1){$~$}}
\put(0,0){\framebox (4,1){$~$}}
\put(7,1){\framebox (1,1){X}}
\put(8,1){\framebox (1,1){}}
\put(6,1){\framebox (1,1){X}}
\end{picture}}
\oplus
\mbox{
\setlength{\unitlength}{0.6em}
\begin{picture}(8.5,2)
\put(0,1){\framebox (4,1){$~$}}
\put(4,1){\framebox (3,1){$~$}}
\put(0,0){\framebox (4,1){$~$}}
\put(7,1){\framebox (1,1){X}}
\put(4,0){\framebox (1,1){}}
\put(6,1){\framebox (1,1){X}}
\end{picture}}
\oplus
\mbox{
\setlength{\unitlength}{0.6em}
\begin{picture}(11.5,2)
\put(0,1){\framebox (4,1){$~$}}
\put(4,1){\framebox (3,1){$~$}}
\put(0,0){\framebox (4,1){$~$}}
\put(7,1){\framebox (1,1){X}}
\put(8,1){\framebox (1,1){}}
\put(3,0){\framebox (1,1){X}}
\end{picture}}
\nonumber\\
&\oplus&
\mbox{
\setlength{\unitlength}{0.6em}
\begin{picture}(9.5,2)
\put(0,1){\framebox (4,1){$~$}}
\put(4,1){\framebox (3,1){$~$}}
\put(0,0){\framebox (4,1){$~$}}
\put(7,1){\framebox (1,1){X}}
\put(8,1){\framebox (1,1){X}}
\put(5,1){\framebox (1,1){X}}
\put(6,1){\framebox (1,1){X}}
\end{picture}}
\oplus
\mbox{
\setlength{\unitlength}{0.6em}
\begin{picture}(8.5,2)
\put(0,1){\framebox (4,1){$~$}}
\put(4,1){\framebox (3,1){$~$}}
\put(0,0){\framebox (4,1){$~$}}
\put(7,1){\framebox (1,1){X}}
\put(4,0){\framebox (1,1){X}}
\put(3,0){\framebox (1,1){X}}
\put(6,1){\framebox (1,1){X}}
\end{picture}}
\oplus
\mbox{
\setlength{\unitlength}{0.6em}
\begin{picture}(7.5,2)
\put(0,1){\framebox (4,1){$~$}}
\put(4,1){\framebox (3,1){$~$}}
\put(0,0){\framebox (4,1){$~$}}
\put(5,0){\framebox (1,1){X}}
\put(4,0){\framebox (1,1){X}}
\put(3,0){\framebox (1,1){X}}
\put(2,0){\framebox (1,1){X}}
\end{picture}}
\nonumber\\
&\oplus& 
...
\end{eqnarray}
where in the first row there are no contractions (i.e. $SO(6)$ traces), 
only symmetrizations and antisymmetrizations; in the 
second row, one contraction; 
and in the third row, two contractions;
the ``...'' stands for tableaux with more than two rows.%
\footnote{
	Tensoring representations in the manner of 
	equation (\ref{pqp-times-020}) gets messy 
	for larger representations. 
	Another 
	method 
	(of Berenstein-Zelevinsky triangles) 
	is discussed in Appendix \ref{BZ-triangles}. 
	} 
In terms of Dynkin labels, equation (\ref{pqp-times-020})  reads 
\begin{eqnarray}
\label{pqp-times-020:dynkin}
[p,q,p] \otimes [0,2,0] &=& 
[p,q+2,p] \oplus [p+1,q,p+1] \oplus [p+2,q-2,p+2]
\nonumber\\
&\oplus&
[p,q,p] \oplus [p+1,q-2,p+1] \oplus [p-1,q+2,p-1]
\nonumber\\
&\oplus&
[p,q-2,p] \oplus [p-1,q,p-1] \oplus [p-2,q+2,p-2]
\nonumber\\
&\oplus&
...
\end{eqnarray}
Now, the ``...'' stands for representations with 
$[r,s,r+2k]$ Dynkin labels with $k \ne 0$. 
Thus the only three-point functions 
of the form (\ref{o-o-x-square}) 
which can possibly have a nonzero value 
are the extremal correlators 
\begin{eqnarray}
\label{eq:extremal X^2}
\langle\OO_{[p,q,p]}(x)\bar{\OO}_{[p,q-2,p]}(y) (\tr\bar z_1^2)(w) \rangle 
\\
\langle\OO_{[p,q,p]}(x)\bar{\OO}_{[p-2,q+2,p-2]}(y)(\tr\bar z_2^2)(w) \rangle 
\\
\langle\OO_{[p,q,p]}(x)\bar{\OO}_{[p-1,q,p-1]}(y)(\tr\bar z_1\bar z_2)(w) 
\rangle 
\end{eqnarray}
which correspond to those diagrams in (\ref{pqp-times-020}) 
with zero or maximal number of contractions; 
and non-extremal correlators 
\begin{eqnarray}
\label{eq1:non-extremal X^2}
\langle \OO_{[p,q,p]}(x) \bar{\OO}_{[p,q,p]}(y) (\tr \bar z t z)(w) \rangle 
\\
\langle\OO_{[p,q,p]}(x)\bar{\OO}_{[p-1,q+2,p-1]}(y)(\tr\bar z_2 z_1)(w)\rangle 
\label{eq2:non-extremal X^2}
\end{eqnarray}
where $t$ is a diagonal $SU(3)$ generator. 
All other correlators of the form (\ref{o-o-x-square})
either vanish because the tensor product of 
irreps $[p,q,p]$ and $[0,2,0]$ does not contain $[r,s,r]$, 
or are related to the ones 
in (\ref{eq:extremal X^2}-\ref{eq2:non-extremal X^2}).

\begin{figure}[t!]
{\begin{center}
\quad \epsfig{width=3.8in, file=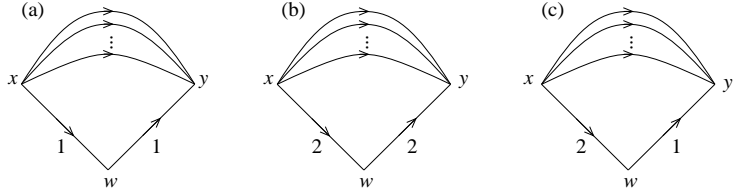, angle=-90}
\vskip -0.2in
\end{center}}
\caption{%
Nonvanishing Born diagrams for 
non-extremal three-point functions 
$\langle \OO_{[p,q,p]} \bar{\OO}_{[p,q,p]} (\tr \bar z t z) \rangle$ 
(a-b); 
$\langle \OO_{[p,q,p]} \bar{\OO}_{[p-1,q+2,p-1]} (\tr \bar z_2 z_1) 
\rangle$ 
(c). 
\label {fig:born-3-point-x-square}
}%
\end {figure}

Extremal three-point functions were discussed 
in Section \ref{section:simplifications}, and were found 
to be protected at order $g^2$. 
The only correlators of the form 
$\langle \OO \bar {\OO}' \tr X^2 \rangle$
we need to consider are those given by 
(\ref{eq1:non-extremal X^2}) and (\ref{eq2:non-extremal X^2}). 
However, 
the three-point functions 
of Figure \ref{fig:born-3-point-x-square}(c), 
must in fact vanish: 
$\tr \bar z_2 z_1 = \half \bar z_2^a z_1^a$ 
is diagonal in color indices, and hence 
the combinatorial factors for the Born graph 
of 
$\langle \OO_{[p,q,p]}(x) \bar{\OO}_{[p-1,q+2,p-1]}(y) 
(\tr \bar z_2 z_1)(w) \rangle$ 
are 
proportional to the ones for the two-point function 
$\langle \OO_{[p,q,p]} \bar{\OO}_{[p-1,q+2,p-1]} \rangle = 0$.
The same thing happens at order $g^2$, etc.%
\footnote{
	Explicitly, in Section \ref{section:simplifications} 
	we saw that the $\OO(g^2)$ part of this 
	three-point function vanishes. 
	} 
So correlators (\ref{eq2:non-extremal X^2}), 
although allowed by 
(\ref{pqp-times-020:dynkin}), 
are in fact forbidden by a combination of 
$SU(N)$ and $SU(4)$ group theory.

\begin{figure}[t!]
{\begin{center}
\epsfig{width=1.4in, file=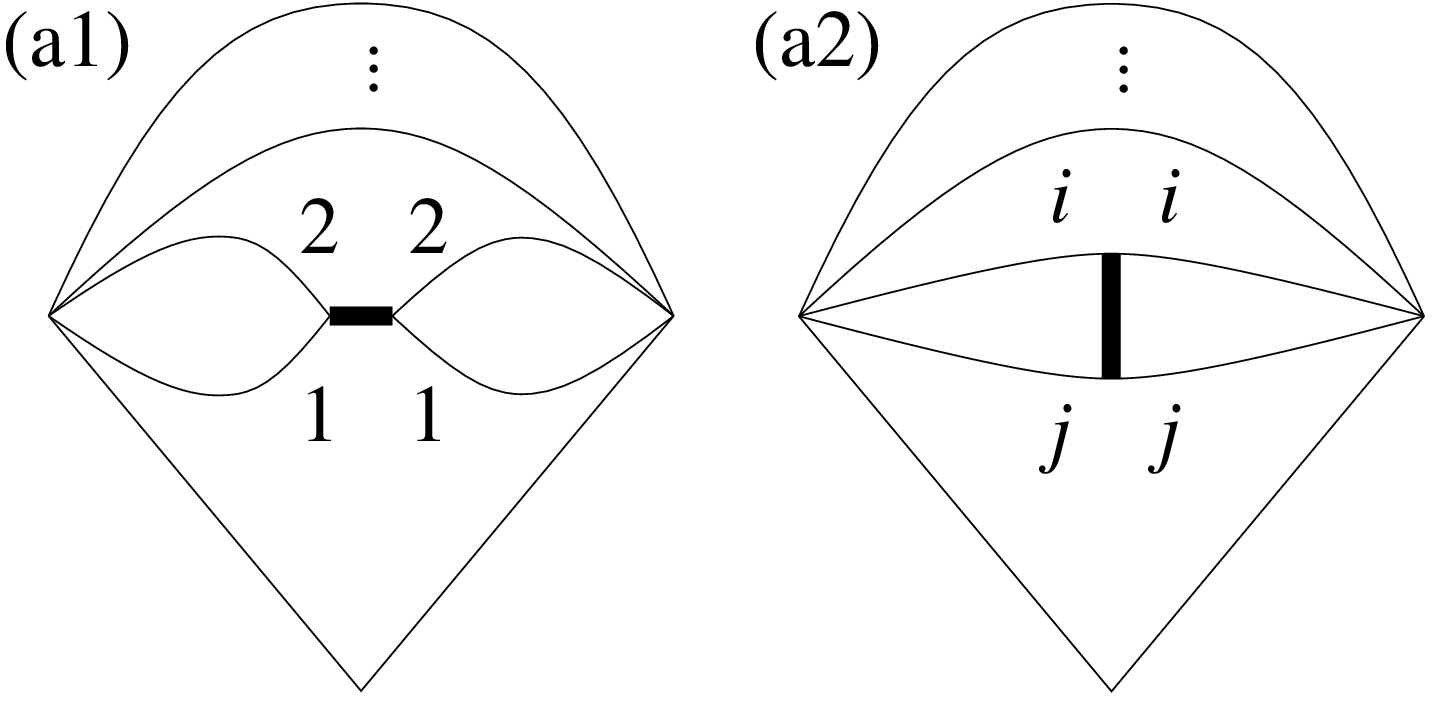, angle=-90}
\quad
\epsfig{width=1.4in, file=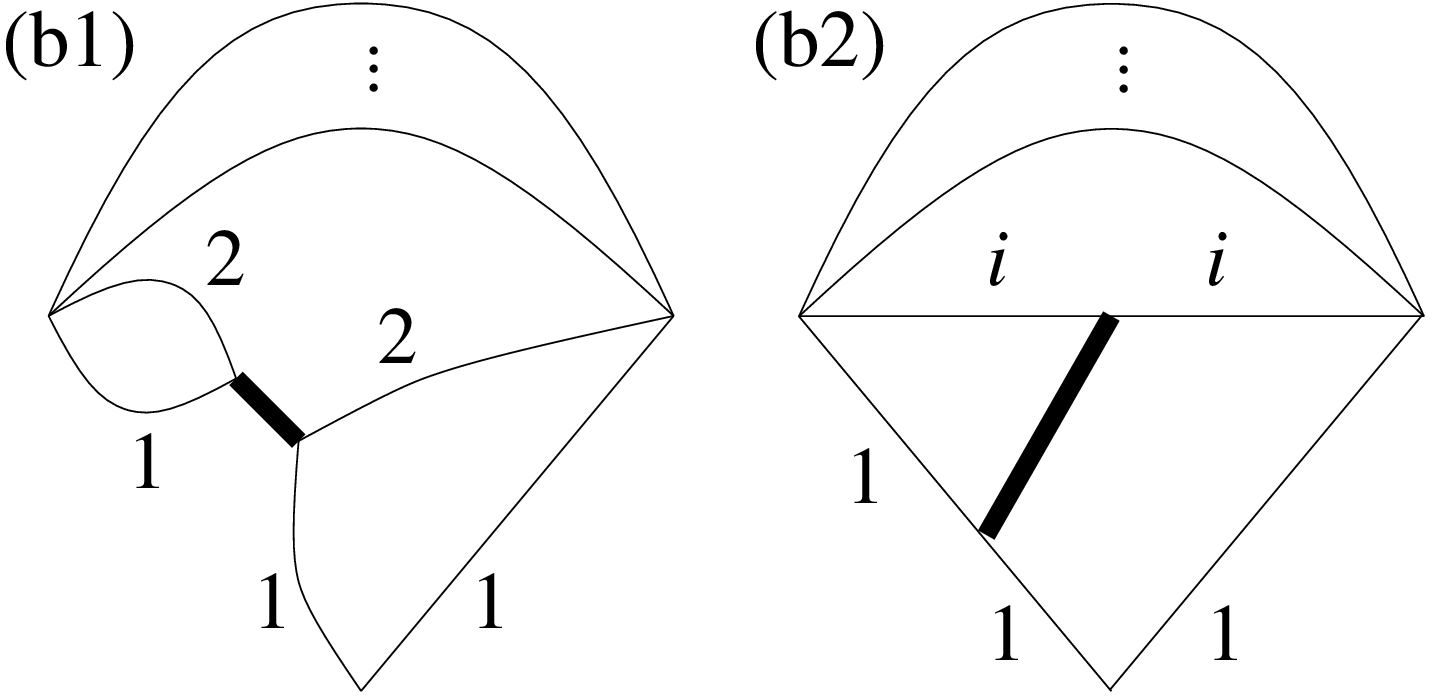, angle=-90}
\quad
\epsfig{width=1.4in, file=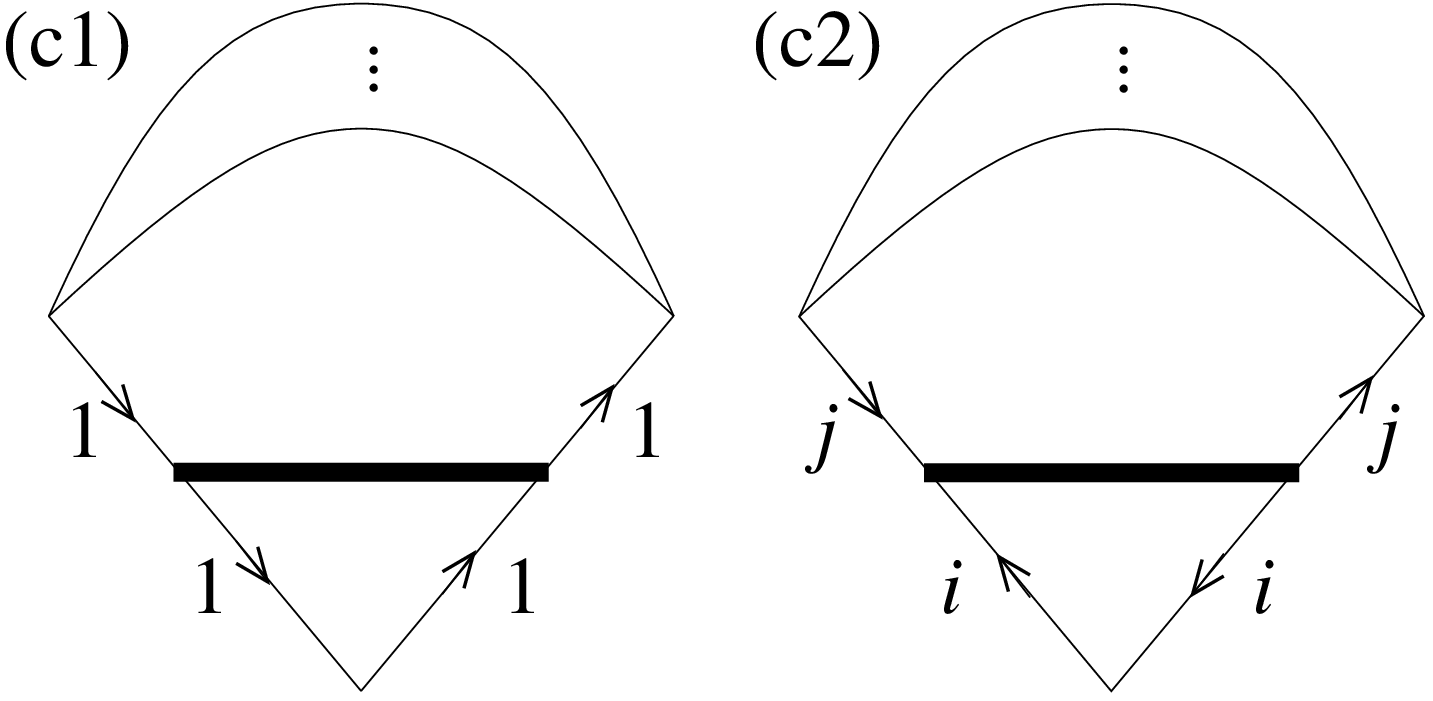, angle=-90}
\end{center}}
\vskip -0.2in
\caption{%
Order $g^2$ corrections to 
$\langle \OO_{[p,q,p]}(x) \bar{\OO}_{[p,q,p]}(y) (\tr \bar z t z) (w) \rangle$ 
with $t_{22} = 0$: 
(a) within the rainbow ($i,j=1,2$ in the second diagram); 
(b) from the rainbow to $X^2$ 
(there are similar ones with the other leg of $X^2$ uncorrected). 
\label {fig:3-point-x-square}
}%
\end {figure}

Correlators 
$\langle \OO_{[p,q,p]}(x) \bar{\OO}_{[p,q,p]}(y) (\tr \bar z t z)(w)\rangle$ 
are the only ones that remain to be considered. 
The contributing 
Born level diagrams are shown in 
Figure \ref{fig:born-3-point-x-square}(a,b), 
and the $\OO(g^2)$ graphs 
appear 
in Figure \ref{fig:3-point-x-square}
(corrections to the scalar propagator are not shown, 
but are also present). 
Repeating the 
arguments of \cite{DFS} from the 
\half-BPS calculations, we see that the combinatorial structure 
of this three-point function 
$\langle \OO \bar{\OO} (\tr \bar z t z) \rangle$
is the same as that of the 
two-point function $\langle \OO \bar{\OO} \rangle$. 
At Born level, we find that 
\begin{eqnarray}
\label{eq:o-o-tr-ztz Born}
\langle \OO_{[p,q,p]}(x) \bar{\OO}_{[p,q,p]}(y) (\tr \bar z t z) (w) \rangle 
|_{\rm free} && \nonumber\\
&&
\hspace{-14em}
= 
\half [(p+q) t_{11} + p t_{22} ] 
[{G(x,w) G(y,w) \over G(x,y)}]
\langle \OO_{[p,q,p]}(x) \bar{\OO}_{[p,q,p]}(y) \rangle 
|_{\rm free} 
\quad\quad 
\end{eqnarray}
At order $g^2$, the contributions 
proportional to $\tilde B$ and $\tilde C$
(diagrams (a1) and (b1) in Figure \ref{fig:3-point-x-square})
have the same index structure, 
which in turn is identical to that of the two-point functions 
$\langle \OO_{[p,q,p]}(x) \bar{\OO}_{[p,q,p]}(y) \rangle$. 
Because $\tr \bar z_1 z_1$ is diagonal in color indices, 
its only effect on the combinatorics 
is to distinguish the pair of indices which go 
to $\OO_w$ rather than stretch directly between 
$\OO_x$ and $\OO_y$.

There is a curious relation between the functions
$\tilde B(x,y)$ and $\tilde C(x;y,w)$, 
which can be graphically expressed as%
\footnote{
	The fact that $\tilde C(x;y,w) + \tilde C(y;x,w) - \tilde B(x,y)$ 
	is just a constant was established in 
	Section \ref{section:c-tilde} by an explicit calculation. 
	The value of this constant was also found there. 
	}
\begin{equation}
\label{cat:diagrams}
\epsfig{width=4.2in, file=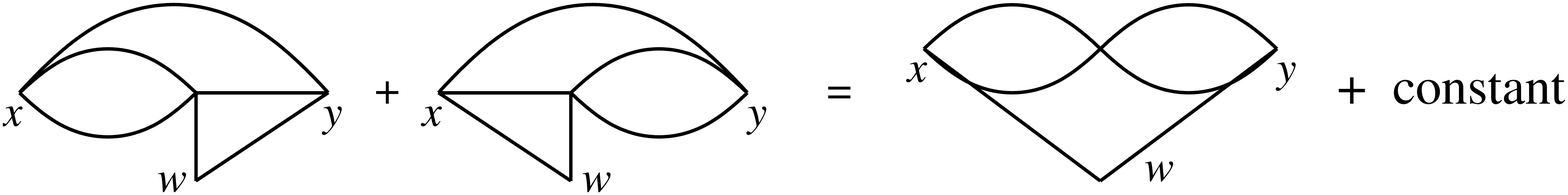, angle=0}
\end{equation}
This is a consequence of conformal invariance and 
nonrenormalization of the scaling dimension of $\tr X^2$. 
To see how this comes about, let $\OO_{1,2}$ 
be 
arbitrary (not necessarily BPS) 
scalar operators of the 
same scaling dimension $\Delta$; 
then 
\begin{eqnarray}
\label{eq:o-o-conformal}
\langle \OO_1(x) \OO_2(y) \rangle &=& 
{C_{12} \over (x-y)^{2 \Delta} }
\\
\langle \OO_1(x) \OO_2(y) \, \tr X^2(w) \rangle &=& 
{\tilde C_{12} \over (x-y)^{2 \Delta} } 
{G(x,w) G(y,w) \over G(x,y) }
\label{eq:o-o-X^2-conformal}
\end{eqnarray}
in $d=4$. 
In other words, coordinate dependence 
(modulo the ratio of free scalar correlators)
is the same, 
and the difference is just a constant factor. 
Assume now that $\OO_1$ is constructed of only $z_i$-s; 
and $\OO_2$, $\bar z_j$-s.  
Then, the only contributions to the two-point functions 
$\langle \OO_1(x) \OO_2(y) \rangle$ are proportional to $\tilde B(x,y)$; 
similarly, the correlators 
$\langle \OO_1(x) \OO_2(y) \, \tr X^2(w) \rangle$ 
are proportional to 
$\tilde B(x,y) + \tilde C(x;y,w) + \tilde C(y;x,w)$. 
Index structure of these building blocks 
is the same (as discussed after equation \ref{eq:o-o-tr-ztz Born}), so 
\begin{eqnarray}
\label{eq:c-c-b-1}
{ \langle \OO_1(x) \OO_2(y) \rangle|_{g^2 ~~} 
\over \langle \OO_1(x) \OO_2(y) \rangle|_{\rm free}} &=& 
\zeta_{12} \; \tilde B(x,y) 
\quad\quad\\
{\langle \OO_1(x) \OO_2(y) \, \tr X^2(w) \rangle|_{g^2 ~~} 
\over \langle \OO_1(x) \OO_2(y) \, \tr X^2(w) \rangle|_{\rm free}} &=& 
\half \zeta_{12} \left[ \tilde C(x;w,y) + \tilde C(y;w,x) + \tilde B(x,y) \right] 
\quad\quad 
\label{eq:c-c-b-2}
\end{eqnarray}
with the same 
$\zeta_{12}$.%
\footnote{
	In particular, if 
	$\Delta = \Delta_0$ is not corrected, then neither are 
	$\langle \OO_1(x) \OO_2(y) \rangle$
	and 
	$\langle \OO_1(x) \OO_2(y) \, \tr X^2(w) \rangle$, 
	at least at one loop. 
	}
As was discussed in Section \ref{section:c-tilde}, 
$\tilde B$ and $\tilde C$ 
have the form 
$\tilde B(x,y) = \tilde a \log {(x-y)^2 \over \e^2} + \tilde b$, 
$\tilde C(x;y,w) 
= \tilde a' \log {(x-y)^2 (x-w)^2 \over \e^4} 
- \tilde a'' \log {(y-w)^2 \over \e^2} + \tilde b'$. 
By comparing 
(\ref{eq:o-o-X^2-conformal}) and (\ref{eq:o-o-conformal}), 
we see that 
expression (\ref{eq:c-c-b-2}) 
must have the same coordinate dependence as (\ref{eq:c-c-b-1}). 
This restricts 
$\tilde a' = \tilde a''  = \half \tilde a$, 
which reproduces the ``winking cat'' identity (\ref{cat:diagrams}).

Finally, we can relate 
$\langle \OO_1(x) \OO_2(y) \, \tr X^2(w) \rangle$ 
to 
$\langle \OO_1(x) \OO_2(y) \rangle$ 
by a Ward identity. 
As shown in \cite{OSBORN}, the ratio 
\begin{equation}
{\langle \OO(x) \OO(y) T_{\mu\nu}(0) \rangle \over
\langle \OO(x) \OO(y) \rangle} = 
{2 \Delta \over 3 \pi^2} 
{t_{\mu\nu}(\gamma) (x-y)^4 \over x^4 y^4}
\end{equation}
depends on the scaling dimension 
$\Delta$ of the operator $\OO$
(here, $\gamma = {x \over x^2} - {y \over y^2}$ and 
$t_{\mu\nu}(\gamma) =
{\gamma_\mu \gamma_\nu \over \gamma^2} - \quarter \eta_{\mu\nu}$). 
Since the energy momentum tensor 
$T_{\mu \nu}$ is in the same $\NN=4$ multiplet with 
$\tr X^2$, there is also nothing 
peculiar about the fact that $\tilde C_{12} / C_{12}$ 
can in general receive $\OO(g^2)$ correction. 
This ratio also depends on $\Delta$.

\subsubsection{General 
$\langle 
\OO_{\tiny \! \quarter} \OO_{\tiny \! \quarter} \OO_{\tiny \! \half} 
\rangle$ 
correlators}
\label{section:quarter-quarter-half}

Three-point functions of two \quarter-BPS operator and 
one \half-BPS operator are similar to the ones described 
in Section \ref{section:cat diagrams}. 
It suffices to consider a single three-point function 
(such that the 
Clebsch-Gordon coefficient%
\footnote{
	By Wigner-Eckart theorem, for any three representations 
	we only need to calculate one (nonvanishing) correlator
	of any representatives from these irreps. 
	} for 
these three vectors in the given 
irreps of $SU(4)$ 
is nonzero) for each set of three representations.
Without loss of generality, we can choose 
a $[p,q,p]$ scalar composite $\OO(x)$ to be made of only $z$-s; 
a $[r,s,r]$ scalar composite $\OO'(y)$ to be made of only $\bar z$-s; 
and a $[0,k,0]$ scalar composite $\tr X^{\alpha_1+\alpha_2}$ at $w$ 
of the form 
\begin{eqnarray}
t_{i_1 ... i_{\alpha_1} ; j_1 ... j_{\alpha_2}} 
\str z_{i_1} ... z_{i_{\alpha_1}} \bar z_{j_1} ... \bar z_{j_{\alpha_2}} 
\end{eqnarray}
where $\alpha_1+\alpha_2=k$, and 
$t_{i_1 ... i_{\alpha_1} ; j_1 ... j_{\alpha_2}}$ 
is the appropriate irreducible $SU(3)$ tensor 
(like in \cite{DFS}). 
The correlators we are after are 
\begin{eqnarray}
\label{o-o-X^k}
\langle \OO (x) \OO' (y) (\tr X^{\alpha_1+\alpha_2}) (w) \rangle 
\end{eqnarray}
Position dependence of (\ref{o-o-X^k}) is 
\begin{eqnarray}
\label{eq:o-o-X^k position}
[G(x,y)^{(2p+r)+(2r+s)-k} 
G(x,w)^{k+(2p+r)-(2r+s)} 
G(w,y)^{k-(2p+r)+(2r+s)}]^{1/2}
\end{eqnarray}
at Born level. 
The contributing free diagrams are similar to the ones 
shown in Figure \ref{fig:born-3-point-x-square}; 
and $\OO(g^2)$ diagrams, to those of 
Figure \ref{fig:3-point-x-square}, 
but now there can be a 
different number of lines stretching 
between $x$ and $w$ and between $w$ and $y$. 
Apart from the 
factor (\ref{eq:o-o-X^k position}), 
the general 
$\langle 
\OO_{\tiny \! \quarter} \OO_{\tiny \! \quarter} \OO_{\tiny \! \half} 
\rangle$ 
correlator (\ref{o-o-X^k}) is given by 
\begin{eqnarray}
\label{eq1:o-o-X^k corrections}
\alpha_{\rm free} + 
\tilde \beta_{xy} \tilde B (x,y) + 
\tilde \gamma_x \tilde C (x;y,w) + 
\tilde \gamma_y \tilde C (y;x,w) 
+ \OO(g^4) 
.
\end{eqnarray}
According to the discussion of Section \ref{section:3-pt structure}, 
the remaining combinatorial coefficients vanish, 
$\beta_{xw} = \beta_{yw} = \gamma_w = 0$. 
Moreover, 
$\tilde \gamma_x = \tilde \gamma_y = - \tilde \beta_{xy}$ 
as follows from equations 
(\ref{eq:coeffs - all three:beg}-\ref{eq:coeffs - all three:end}), 
so the $\OO(g^2)$ corrections in (\ref{eq1:o-o-X^k corrections}) 
add up to 
\begin{eqnarray}
\label{eq2:o-o-X^k corrections}
\alpha_{\rm free} - 
\tilde \beta_{xy} 
\left( 
\tilde C (x;y,w) + \tilde C (y;x,w) - \tilde B (x,y) 
\right) 
+ \OO(g^4) 
\end{eqnarray}
and we only need to verify that $\tilde \beta_{xy} = 0$.%
	\footnote{
	The expression multiplying $\tilde \beta_{xy}$ 
	in (\ref{eq2:o-o-X^k corrections}) 
	is a nonzero, renormalization scale independent 
	constant. In Section \ref{section:c-tilde}, 
	its value was computed to be $-\left( Y^2 / 32 \pi^2 \right)$. 
	}

The simplifications we can use to deduce that 
$\tilde \beta_{xy} = 0$ without doing calculations, 
are discussed in Section \ref{section:simplifications}. 
Extremal three-point functions are always 
easy to identify, and 
with the BPS primaries in representations 
$[p,q,p]$, $[r,s,r]$, and $[0,k,0]$, 
the restrictions on the scaling dimension translate into 
\begin{equation}
\label{restriction:extremal}
2r+s = 2p+q + k, 
\quad 
2p+q = 2r+s + k, 
\quad\mbox{or}\quad 
2p+q + 2r+s = k, 
\end{equation}
depending on which scaling dimension is the 
sum of the other two.

The ``three flavor partition'' boils down to being able to 
choose a single flavor (at Born level) for the lines between 
the two \quarter-BPS operators, when the third operator is \half-BPS. 
This is possible whenever 
\begin{equation}
\label{restriction:disjoint-flavors}
2 r + s \le k+q 
\quad\mbox{and}\quad 
2 p + q \le k+s. 
\end{equation}

Alternatively, suppose we can choose 
the \half-BPS operator $\OO_w$ 
to be made of only $\bar z_1$-s and $z_2$-s; 
and the \quarter-BPS operators as 
$\OO_x$ of $z_1$-s and $z_2$-s, 
$\OO_y$ of $\bar z_1$-s and $\bar z_2$-s. 
This is the ``two flavor partition'' 
of Section \ref{section:simplifications}. 
Flavors can be chosen this way if 
\begin{equation}
\label{restriction:two-flavors}
k \le q + s. 
\end{equation}

In all three cases 
(\ref{restriction:extremal}), 
(\ref{restriction:disjoint-flavors}), 
and (\ref{restriction:two-flavors}) 
there are no $\OO(g^2)$ corrections, 
as established in Section \ref{section:simplifications} 
using only $SU(4)$ group theory and conformal invariance 
arguments. 
However, there are allowed three-point functions 
of the form 
$\langle 
\OO_{\tiny \! \quarter} \OO_{\tiny \! \quarter} \OO_{\tiny \! \half} 
\rangle$ 
where we can not choose irrep representatives 
in such a nice way.

Throughout the rest of this Section, we will concentrate 
on the \quarter-BPS operators with dimensions 7 and smaller, 
constructed in \cite{Ryzhov}. 
In particular, we will consider scalar composite 
operators in $SU(4)$ representations of the form $[p,q,p]$, 
with $2p+q \le 7$. 
These are [2,0,2], [2,1,2], [2,2,2], [3,1,3], and [2,3,2]. 
We will take \half-BPS (single trace) operators as whichever ones are 
allowed by group theory. 
Of the triple products of the form 
$[p,q,p] \otimes [r,s,r] \otimes [0,k,0]$ 
containing the singlet, 
most satisfy at least one of the 
simplifying constraints 
(\ref{restriction:extremal}), 
(\ref{restriction:disjoint-flavors}), 
or (\ref{restriction:two-flavors}).%
\footnote{
	We omit the tedious details here. 
	In order to find the allowed triple products, 
	we used the method of BZ triangles, 
	see Appendix \ref{BZ-triangles}. 
	Then we just went through the list 
	and checked if any of the conditions 
	(\ref{restriction:extremal}-\ref{restriction:two-flavors}) 
	apply. 
	}
The exceptions are  
$[2,0,2] \otimes [2,0,2] \otimes [0,2,0]$
and 
$[3,1,3] \otimes [3,1,3] \otimes [0,4,0]$. 

Correlators of the form 
$
\langle\OO_{[p,q,p]}(x)\bar{\OO}_{[p,q,p]}(y) \, \tr X^2 (w)\rangle 
$
were considered in Section \ref{section:cat diagrams}, 
so the only three-point function 
we actually have to calculate is 
$\langle 
\OO_{[3,1,3]} \bar{\OO}_{[3,1,3]} \, \tr X^{2+2} 
\rangle$. 
Explicitly, we can take 
\begin{eqnarray}
&&
\OO_x = \sum_{j=1}^{4} C_x^j \OO_j
,\quad
\OO_y = \sum_{j=1}^{4} C_y^j \bar{\OO}_j
\quad\mbox{with}\quad
\OO_j \sim [z_1^4 z_2^2 z_3] , \\ 
&& 
\OO_w \sim [z_2^2 \bar z_2^2] - \mbox{$SO(6)$ traces}, 
\end{eqnarray}
to be the scalar composite operators%
\footnote{
	$\OO_{1,...,4}$ were studied in \cite{Ryzhov}, 
	and the results are summarized here in 
	Appendix \ref{operators explicitly}. 
	} 
in the [3,1,3] of $SU(4)$. 
With this choice of flavors, the 
free combinatorial factor for this three-point function is 
\begin{eqnarray}
\alpha_{\rm free} &=& 
{(N^2-1) (N^2-2) \over 41472 N^2} 
      (189540 C_x^2 C_y^2 - 4860 C_x^2 C_y^1 N - 
\nonumber\\&& \hspace{3em} 
        4860 C_x^1 C_y^2 N - 
	131220 C_x^4 C_y^2 N - 131220 C_x^2 C_y^4 N + 
\nonumber\\&& \hspace{3em} 
        360 C_x^1 C_y^1 N^2 + 
        13500 C_x^4 C_y^1 N^2 - 79380 C_x^2 C_y^2 N^2 - 
\nonumber\\&& \hspace{3em} 
        22680 C_x^3 C_y^2 N^2 - 
	22680 C_x^2 C_y^3 N^2 + 5184 C_x^3 C_y^3 N^2 + 
\nonumber\\&& \hspace{3em} 
        13500 C_x^1 C_y^4 N^2 - 
        30780 C_x^4 C_y^4 N^2 + 2700 C_x^2 C_y^1 N^3 - 
\nonumber\\&& \hspace{3em} 
        270 C_x^3 C_y^1 N^3 + 
	2700 C_x^1 C_y^2 N^3 + 43740 C_x^4 C_y^2 N^3 - 
\nonumber\\&& \hspace{3em} 
        270 C_x^1 C_y^3 N^3 - 
        9720 C_x^4 C_y^3 N^3 + 43740 C_x^2 C_y^4 N^3 - 
\nonumber\\&& \hspace{3em} 
        9720 C_x^3 C_y^4 N^3 - 
	115 C_x^1 C_y^1 N^4 - 2760 C_x^4 C_y^1 N^4 + 
\nonumber\\&& \hspace{3em} 
        13500 C_x^2 C_y^2 N^4 + 
        4410 C_x^3 C_y^2 N^4 + 4410 C_x^2 C_y^3 N^4 - 
\nonumber\\&& \hspace{3em} 
        1332 C_x^3 C_y^3 N^4 - 
	2760 C_x^1 C_y^4 N^4 + 13680 C_x^4 C_y^4 N^4 - 
\nonumber\\&& \hspace{3em} 
        450 C_x^2 C_y^1 N^5 + 
        240 C_x^3 C_y^1 N^5 - 450 C_x^1 C_y^2 N^5 - 
\nonumber\\&& \hspace{3em} 
        4500 C_x^4 C_y^2 N^5 + 
	240 C_x^1 C_y^3 N^5 + 2340 C_x^4 C_y^3 N^5 - 
\nonumber\\&& \hspace{3em} 
        4500 C_x^2 C_y^4 N^5 + 
        2340 C_x^3 C_y^4 N^5 - 15 C_x^1 C_y^1 N^6 - 
\nonumber\\&& \hspace{3em} 
        990 C_x^2 C_y^2 N^6 - 126 C_x^3 C_y^3 N^6 - 
        1980 C_x^4 C_y^4 N^6)
\end{eqnarray}
so 
$\langle \OO_x \OO_y \OO_w \rangle \ne 0$ 
in general (or when $\OO_x$ and $\OO_y$ are \quarter-BPS, in particular). 
We have also explicitly calculated%
\footnote{
	This calculation 
	was done using {\it Mathematica} 
	and took about 200 hours.
	}
the $\OO(g^2)$ combinatorial factor 
in (\ref{eq2:o-o-X^k corrections}): 
\begin{eqnarray}
\tilde \beta_{xy} &=& 
{(N^2-1) (N^2-4) \over 13824} 
     (-10800 C_x^2 C_y^1 + 4320 C_x^3 C_y^1 - 
\nonumber\\&& \hspace{2em} 
       10800 C_x^1 C_y^2 - 
	259200 C_x^4 C_y^2 + 4320 C_x^1 C_y^3 + 
\nonumber\\&& \hspace{2em} 
       103680 C_x^4 C_y^3 - 
       259200 C_x^2 C_y^4 + 103680 C_x^3 C_y^4 - 
\nonumber\\&& \hspace{2em} 
       2025 C_x^1 C_y^1 N - 
	27000 C_x^4 C_y^1 N - 32400 C_x^2 C_y^2 N + 
\nonumber\\&& \hspace{2em} 
       38880 C_x^3 C_y^2 N + 
       38880 C_x^2 C_y^3 N - 25920 C_x^3 C_y^3 N - 
\nonumber\\&& \hspace{2em} 
       27000 C_x^1 C_y^4 N - 
	129600 C_x^4 C_y^4 N - 600 C_x^2 C_y^1 N^2 + 
\nonumber\\&& \hspace{2em} 
       2940 C_x^3 C_y^1 N^2 - 
       600 C_x^1 C_y^2 N^2 + 50400 C_x^4 C_y^2 N^2 + 
\nonumber\\&& \hspace{2em} 
       2940 C_x^1 C_y^3 N^2 - 
	7200 C_x^4 C_y^3 N^2 + 50400 C_x^2 C_y^4 N^2 - 
\nonumber\\&& \hspace{2em} 
       7200 C_x^3 C_y^4 N^2 + 
       175 C_x^1 C_y^1 N^3 + 5400 C_x^4 C_y^1 N^3 + 
\nonumber\\&& \hspace{2em} 
       7200 C_x^2 C_y^2 N^3 - 
	7920 C_x^3 C_y^2 N^3 - 7920 C_x^2 C_y^3 N^3 + 
\nonumber\\&& \hspace{2em} 
       3888 C_x^3 C_y^3 N^3 + 
       5400 C_x^1 C_y^4 N^3 + 28800 C_x^4 C_y^4 N^3 + 
\nonumber\\&& \hspace{2em} 
       600 C_x^2 C_y^1 N^4 - 
	780 C_x^3 C_y^1 N^4 + 600 C_x^1 C_y^2 N^4 - 
\nonumber\\&& \hspace{2em} 
       780 C_x^1 C_y^3 N^4 - 
       2880 C_x^4 C_y^3 N^4 - 2880 C_x^3 C_y^4 N^4 + 
\nonumber\\&& \hspace{2em} 
       50 C_x^1 C_y^1 N^5 + 288 C_x^3 C_y^3 N^5)
\end{eqnarray}
If we choose the coefficients 
$( C_x^1, C_x^2, C_x^3, C_x^4 )$ 
and 
$( C_y^1, C_y^2, C_y^3, C_y^4 )$ 
independently from the set 
$
\{
(
- {\frac{12 N}{{N^2}-2}} , 1 ,  - {\frac{5}{{N^2}-2}} , 0
)
, 
(
{\frac{96}{{N^2}-4}} , - {\frac{4 N}{{N^2}-4}} , 
{\frac{10 N}{{N^2}-4}} , 1
)
\}
$, 
we recover $\tilde \beta_{xy} = 0$. 
This corresponds to taking 
$\OO_x$ and $\OO_y$ as the \quarter-BPS chiral primaries 
found in \cite{Ryzhov}, 
so there are no 
$\OO(g^2)$ corrections in this case either.

\subsection{Three-point functions of \quarter-BPS operators}
\label{section: all BPS}

When all three operators are \quarter-BPS, 
the arguments get progressively more tedious. 
We will chose $2l+k \le 2p+q \le 2r+s$. 
The simplifications 
discussed in Section \ref{section:simplifications} 
applicable to correlators 
$\langle \OO_{[p,q,p]} \OO_{[r,s,r]} \OO_{[l,k,l]} \rangle$
are: 
the extremality condition 
\begin{equation}
\label{restriction:extremal QQQ}
2r+s = 2p+q + 2l+k, 
\end{equation}
and the ``partition into three disjoint flavors'' condition,%
\footnote{
	Which just says that the number 
	of scalars exchanged between each pair of $\OO$-s 
	is no larger than the length of the first column in 
	the corresponding Young tableaux. 
}
\begin{eqnarray}
\label{restriction1:disjoint QQQ}
2r+s &\le& 2l+k + q, \\
2r+s &\le& 2p+q + k, \\
2p+q &\le& 2l+k + s, 
\label{restriction2:disjoint QQQ}
\end{eqnarray}
which have to be satisfied simultaneously 
(there are three more, but they are satisfied trivially 
since we took $2l+k \le 2p+q \le 2r+s$). 
For example, 
(\ref{restriction1:disjoint QQQ}-\ref{restriction2:disjoint QQQ}) 
are true 
when all the \quarter-BPS operators are in the $84 = [2,0,2]$ of $SU(4)$; 
for example, we can take%
\footnote{
	As shown in Appendix \ref{partition-into-two}, 
	such operators are in fact in the 84 of $SU(4)$. 
	}
\begin	{eqnarray}
\label{202:different wts-1}
{\YY} (x) = 
\left\{ 
(\tr z_1^2) (\tr z_2^2) - (\tr z_1 z_2) (\tr z_1 z_2) 
\right\}
+ {1\over N} 
\left\{ 
\tr [z_1 , z_2]^2 
\right\}
,
\\
{\YY} (y) = 
\left\{ 
(\tr \bar z_1^2) (\tr \bar z_3^2) - 
(\tr \bar z_1 \bar z_3) (\tr \bar z_1 \bar z_3) 
\right\}
+ {1\over N} 
\left\{ 
\tr [\bar z_1 , \bar z_3]^2 
\right\}
,
\\
{\YY} (w) = 
\left\{ 
(\tr z_3^2) (\tr \bar z_2^2) - (\tr z_3 \bar z_2) (\tr z_3 \bar z_2) 
\right\}
+ {1\over N} 
\left\{ 
\tr [z_3 , \bar z_2]^2 
\right\}
.
\label{202:different wts-2}
\end	{eqnarray}
The Born amplitude 
\begin	{eqnarray}
\label{84-84-84-born}
\langle 
{\YY}(x) \; {\YY}(y) \; {\YY}(w) 
\rangle_{\rm free}
&\propto& 
(N^2-1)(N^2-4)(2 N^2-15)
\end	{eqnarray}
does not vanish%
\footnote{
	$N>1$ in general since the gauge group is $SU(N)$, 
	while for $N=2$ the single and double trace operators 
	are proportional,
	and there are no \quarter-BPS operators in the 84 of $SU(4)$.
	}
for $N > 2$, so we can't blame the lack of corrections 
on group theory, 
and since the correlator 
$\langle {\YY}(x) \; {\YY}(y) \; {\YY}(w) \rangle$
is of the form (\ref{disjoint-flavors}), 
it receives no radiative corrections at order $g^2$.

Of the allowed 
$\langle \OO_{\tiny\quarter}(x) \OO_{\tiny\quarter}(y) 
\OO_{\tiny\quarter}(w) \rangle$
three-point functions 
where each $\OO_{\tiny\quarter}$ is a scalar composite in a 
$[p,q,p]$ of $SU(4)$ with $2p+q \le 7$, 
ten more satisfy 
(\ref{restriction:extremal QQQ}) or 
(\ref{restriction1:disjoint QQQ}-\ref{restriction2:disjoint QQQ}).%
\footnote{
	We used the method of BZ triangles 
	(see Appendix \ref{BZ-triangles}) 
	to find the allowed triple products. 
	}
For the remaining five correlators 
\begin{eqnarray}
\label{quarter-all: cases}
&& 
\langle \OO_{[2,0,2]} (x) \OO_{[2,0,2]} (y) \OO_{[2,2,2]} (w) \rangle 
\nonumber\\&& 
\langle \OO_{[2,0,2]} (x) \OO_{[2,1,2]} (y) \OO_{[2,3,2]} (w) \rangle 
\nonumber\\&& 
\langle \OO_{[2,0,2]} (x) \OO_{[2,1,2]} (y) \OO_{[3,1,3]} (w) \rangle 
\nonumber\\&& 
\langle \OO_{[2,0,2]} (x) \OO_{[3,1,3]} (y) \OO_{[2,3,2]} (w) \rangle 
\nonumber\\&& 
\langle \OO_{[2,0,2]} (x) \OO_{[3,1,3]} (y) \OO_{[3,1,3]} (w) \rangle 
\end{eqnarray}
we have to verify that there are no contributions 
proportional to any of the functions 
$\tilde B(x,y)$, $\tilde B(x,w)$, or $\tilde B(y,w)$. 
In fact, with $2l+k \le 2p+q \le 2r+s$, we 
automatically have 
\begin{eqnarray}
\label{restriction3:disjoint QQQ}
2l+k \le 2r+s + q \quad\mbox{and}\quad 
2p+q \le 2r+s + k 
\label{restriction4:disjoint QQQ}
\end{eqnarray}
so we can always choose the operators as 
\begin{eqnarray}
\label{QQQ:operators1}
\OO_{[l,k,l]} (x) &\sim& [\bar z_1^a \bar z_2^b z_3^e] \\
\OO_{[p,q,p]} (y) &\sim& [\bar z_1^c \bar z_2^d \bar z_3^e] \\
\OO_{[r,s,r]} (w) &\sim& [z_1^{r+s} z_2^r] 
\label{QQQ:operators2}
\end{eqnarray}
where $e \equiv \half [(2l+k) + (2p+q) - (2r+s)] 
\le l+k, p+q$; 
and integers $a$, $b$, $c$, $d$ partition 
$r+s = a+c$, $r = b+d$. 
Then 
$\beta_{xy}=0$ since 
the operators exchanged between $\OO_x$ and $\OO_y$ 
all have the same flavor, and we only 
need to calculate $\beta_{xw}$ and $\beta_{yw}$. 
Details of these calculations are given 
in Appendix \ref{tedious details}, 
and here we just quote the result: 
as in the cases considered so far, 
$\beta_{xy} = \beta_{xw} = \beta_{yw} = 0$, 
and none of the three-point functions (\ref{quarter-all: cases}) 
receive any $\OO(g^2)$ corrections.

\section{
$\langle \OO_{\tiny \quarter} \OO_{\tiny \quarter} \OO_{\tiny \half} \rangle$
correlators in the large $N$ limit}
\label{section:large N}

Like the two-point functions studied in \cite{Ryzhov}, 
$\langle \OO_x \OO_y \OO_w \rangle$
calculations get progressively more cumbersome as the 
representations of the $\OO$-s become larger. 
In this Section we will calculate correlators of 
two \quarter-BPS operators with one \half-BPS operator, 
in the large $N$ limit. 
The situation when all three operators are 
\quarter-BPS is even less tractable, 
and we avoid it here.

\subsection{Large $N$ operators}

We will use the \quarter-BPS operators found in \cite{Ryzhov}. 
Schematically, the special double and single trace operators 
can be written as 
\begin{eqnarray}
\label{def:O and K}
\OO_{[p,q,p]} \sim 
\left(
\mbox{
\setlength{\unitlength}{1em}
\begin{picture}(7.5,1.6)
\put(0,.5){\framebox (1,1){}}
\put(1,.5){\framebox (2,1){\scriptsize $...$}}
\put(3,.5){\framebox (1,1){}}
\put(4,.5){\framebox (1,1){}}
\put(5,.5){\framebox (1,1){\scriptsize $...$}}
\put(6,.5){\framebox (1,1){}}
\put(0,-1){\framebox (1,1){}}
\put(1,-1){\framebox (2,1){\scriptsize $...$}}
\put(3,-1){\framebox (1,1){}}
\put(1.8,-1.7){\scriptsize $p$}
\put(5.4,-0.7){\scriptsize $q$}
\end{picture}}
\right) 
, \quad 
\KK_{[p,q,p]} \sim 
\left(
\mbox{
\setlength{\unitlength}{1em}
\begin{picture}(7.5,1.6)
\put(0,.2){\framebox (1,1){}}
\put(1,.2){\framebox (2,1){\scriptsize $...$}}
\put(3,.2){\framebox (1,1){}}
\put(4,.2){\framebox (1,1){}}
\put(5,.2){\framebox (1,1){\scriptsize $...$}}
\put(6,.2){\framebox (1,1){}}
\put(0,-.8){\framebox (1,1){}}
\put(1,-.8){\framebox (2,1){\scriptsize $...$}}
\put(3,-.8){\framebox (1,1){}}
\put(1.8,-1.5){\scriptsize $p$}
\put(5.4,-0.5){\scriptsize $q$}
\end{picture}}
\right) 
\end{eqnarray}
(each continuous group of boxes stands for an $SU(N)$ trace); 
explicit formulae for highest $SU(4)$ weight operators of this form 
are given in Appendix \ref{appendix:large N}. 
In the large $N$ limit the linear combinations 
\begin{eqnarray}
\label{K-hat:large N}
\tilde{\YY}_{[p,q,p]} &=& 
\KK_{[p,q,p]}   + \OO(N^{-2}) 
\\
\label{O-hat:large N}
\YY_{[p,q,p]} &=& 
\OO_{[p,q,p]} - {p(p+q)\over N} \KK_{[p,q,p]}  + \OO(N^{-2}) 
\end{eqnarray}
are eigenstates of the dilatations operator. 
$\YY_{[p,q,p]}$ 
have protected normalization and 
scaling dimension ($\Delta_{\tiny \YY} = 2 p + q$) 
at order $g^2$, 
and were argued to be \quarter-BPS.

We did not specify the $SU(4)$ weights of operators 
$\OO_{[p,q,p]}$ and $\KK_{[p,q,p]}$
in (\ref{def:O and K}). 
The choice of weights will depend on 
the representations in the triple product 
$[p,q,p] \otimes [r,s,r] \otimes [0,k,0]$ in the following way. 
Assume $p \le r$; then it is convenient to choose 
\begin{eqnarray}
\label{QQH1:large N}
\YY_{[p,q,p]} (x) &\sim& [\bar z_1^l z_2^n z_3^p] \\
\YY_{[r,s,r]} (y) &\sim& [\bar z_1^m \bar z_2^n \bar z_3^p] \\
\OO_{[0,k,0]} (w) &\sim& [z_1^k] 
\label{QQH2:large N}
\end{eqnarray}
with 
$l \equiv \half [(2p+q)+k-(2r+s)]$, 
$m \equiv k-l$ and $n \equiv p+q-l$. 
We will also assume that none of the simplifications 
(\ref{restriction:extremal}-\ref{restriction:two-flavors}) 
apply, since those cases were already discussed in 
Section \ref{section:quarter-quarter-half}.

\subsection{
$\langle \KK \KK \OO_{\tiny \half} \rangle_{\rm free}$, 
$\langle \KK \OO \OO_{\tiny \half} \rangle_{\rm free}$, 
$\langle \OO \KK \OO_{\tiny \half} \rangle_{\rm free}$, 
and 
$\langle \OO \OO \OO_{\tiny \half} \rangle_{\rm free}$ 
}
\label{section: Born level - large N}

We can estimate the leading large $N$ behavior of 
the combinatorial factors 
$\alpha_{\rm free}$ and $\tilde \beta_{xy}$ 
using the ``trace merging formula'' 
\begin{equation}
\label{merging traces}
2 \left( \tr A t^c \right) \left( \tr B t^c \right) 
= 
\tr A B - 
\mbox{$1\over N$} \left( \tr A \right) \left( \tr B \right) 
\end{equation}
where $A$ and $B$ are arbitrary $N \times N$ matrices 
and $t^c$ are $SU(N)$ generators in the fundamental 
(sums on repeated indices are implied). 
With (\ref{merging traces}) and the expression for the 
quadratic Casimir%
\footnote{
	In particular, the adjoint and fundamental 
	representations will be of interested, and for these 
	$C_2(\mbox{adj}) = N$, 
	$C_2(\mbox{fund}) = (\mbox{$N^2 - 1 \over 2N$})$, 
	$C(\mbox{adj}) = N$, $C(\mbox{fund}) = \half$. 
	}
\begin{equation}
\label{casimir}
T^c_r T^c_r = C_2(r) \; \bone 
\end{equation}
of $SU(N)$, we find for example that for $k \ge 2$,%
\footnote{
	One can derive (\ref{leading traces}) from a 
	one-term 
	recursion relation defined by 
	(\ref{merging traces}), (\ref{casimir}), and 
	$\tr t^a = 0$. 
	}
\begin{eqnarray}
\label{leading traces}
(\tr t^{a_1} t^{a_2} ... t^{a_k}) 
(\tr t^{a_k} ... t^{a_2} t^{a_1}) 
&=& 
\left(
{N^2 - 1 \over 2 N} 
\right)^k
-
\left(
{N^2 - 1 \over 2 N} 
\right) 
\left(
{- 1 \over 2 N} 
\right)^{k-1} 
\nonumber\\
&=& 
\left(
{N \over 2} 
\right)^k
\left[
1 + \OO(1/N^2) 
\right]
\end{eqnarray}
To have this large $N$ behavior, generators in 
the two traces should appear in opposite order. 
When the generators are taken in any other order 
(except cyclic permutations inside the traces), 
such products are suppressed 
by at least a power of $N^2$.

Calculations proceed along the same lines as in \cite{Ryzhov}. 
We begin by considering correlators of the form 
$\langle \OO \OO \OO_{\tiny \half} \rangle$ with 
the two $\OO$-s in the same representation $[p,q,p]$. 
In this case the leading contribution to $\alpha_{\rm free}$ 
comes from terms like 
\begin{eqnarray}
\label{OOH:leading-born:equal}
&& 
( \tr t^{a_1} ... t^{a_l} t^{b_1} ... t^{b_n} )
( \tr t^{c_1} ... t^{c_p} ) 
~ 
( \tr t^{d_1} ... t^{d_l} t^{b_n} ... t^{b_1} )
( \tr t^{c_p} ... t^{c_1} ) 
\nonumber\\ &&\times 
( \tr t^{a_l} ... t^{a_1} t^{d_l} ... t^{d_1} ) 
\nonumber\\
&&\hspace{4em}
\sim 
\left( {1\over2} \right)^2 
N \left( {N\over2} \right)^{2l-2 + n} 
\times 
\left( {N\over2} \right)^p 
= \left( {1\over2} \right) \left( {N\over2} \right)^{2l + n + p - 1}
\hspace{3.5em}
\end{eqnarray}
The factor of $(\half)^2$ comes about because 
we merge traces twice; the exponent $2l + n - 2$ counts 
how many generators collapse using 
$t^c t^c \sim \half N \bone$; 
the extra factor of $N$ is due to $\tr \bone = N$; 
and finally $(N/2)^p$ is from contracting the 
traces of equal length containing the $t^{c_i}$-s. 
All remaining calculations of this and next Sections are 
analogous, and we won't spell things out as much.

If the representations of $[p,q,p]$ and $[r,s,r]$ are 
different, a similar situation occurs when for example 
$p = r+s$, i.e. $\OO_{[p,q,p]}$ and $\OO_{[r,s,r]}$ contain 
traces of equal length. Then we merge traces twice, 
and one set of traces collapses completely 
as in (\ref{leading traces}). 
Otherwise, we have to merge traces three times, so 
the leading contributions to 
$\langle \OO \OO \OO_{\tiny \half} \rangle_{\rm free}$ 
are of the form 
\begin{eqnarray}
\label{OOH:leading-born}
&& 
( \tr t^{a_1} ... t^{a_l} t^{b_1} ... t^{b_n} )
( \tr t^{c_1} ... t^{c_p} ) 
~ 
( \tr t^{d_1} ... t^{d_m} t^{b_n} ... )
( \tr ... t^{b_1} t^{c_p} ... t^{c_1} ) 
\nonumber\\ &&\times
( \tr t^{a_l} ... t^{a_1} t^{d_m} ... t^{d_1} ) 
\nonumber\\
&&\hspace{4em}
\sim 
\left( {1\over2} \right) 
\left( {N\over2} \right)^{p+l+m+n-1} 
~\quad 
\mbox{\parbox{10em}{if $p = r$ or $p+q = r+s$ \par
or $p = r+s$ or $r = p+q$}} 
\nonumber\\&&\hspace{4em}
\sim 
\left( {1\over2} \right)^3 
\left( {N\over2} \right)^{p+l+m+n-3} 
\quad 
\mbox{otherwise}
\hspace{11em}
\end{eqnarray}

For the other three types of correlators, 
no pair of traces ever collapses completely, 
so the answers are more uniform. We find 
that the large $N$ behavior of 
$\langle \KK \KK \OO_{\tiny \half} \rangle_{\rm free}$ 
is defined by the terms like 
\begin{eqnarray}
\label{KKH:leading-born}
( \tr t^{a_1} ... t^{a_l} t^{b_1} ... t^{b_n} t^{c_1} ... t^{c_p} ) 
~ 
( \tr t^{d_1} ... t^{d_m} t^{b_n} ... t^{b_1} t^{c_p} ... t^{c_1} ) 
~ 
( \tr t^{a_l} ... t^{a_1} t^{d_m} ... t^{d_1} ) 
\hspace{-28em}&&\nonumber\\
&&\sim 
\left( {1\over2} \right)^2 
\left( {N\over2} \right)^{p+l+m+n-2} N 
= \left( {1\over2} \right) 
\left( {N\over2} \right)^{p+l+m+n-1} 
\hspace{4em}
\end{eqnarray}
as we merge traces twice. 
Similarly, 
$\langle \OO \KK \OO_{\tiny \half} \rangle_{\rm free}$ 
scales as the terms 
\begin{eqnarray}
\label{OKH:leading-born}
( \tr t^{a_1} ... t^{a_l} t^{b_1} ... t^{b_n} )
( \tr t^{c_1} ... t^{c_p} ) 
~ 
( \tr t^{d_1} ... t^{d_m} t^{b_n} ... t^{b_1} t^{c_p} ... t^{c_1} ) 
~ 
( \tr t^{a_l} ... t^{a_1} t^{d_m} ... t^{d_1} ) 
\hspace{-30em}&&\nonumber\\
&&\sim 
\left( {1\over2} \right)^2 
\left( {N\over2} \right)^{p+l+m+n-2} 
\hspace{16em}
\end{eqnarray}
since traces have to be merged three times now. 
The three-point functions 
$\langle \KK \OO \OO_{\tiny \half} \rangle_{\rm free}$ 
also have the leading large $N$ dependence
(\ref{OKH:leading-born}).

\subsection{
$\langle \KK \KK \OO_{\tiny \half} \rangle_{g^2}$, 
$\langle \KK \OO \OO_{\tiny \half} \rangle_{g^2}$, 
$\langle \OO \KK \OO_{\tiny \half} \rangle_{g^2}$, 
and 
$\langle \OO \OO \OO_{\tiny \half} \rangle_{g^2}$ 
}
\label{section: g^2 - large N}

Here there are no special cases to consider. 
We have to merge traces 
twice for $\langle \KK \KK \OO_{\tiny \half} \rangle_{g^2}$, 
three times for 
$\langle \KK \OO \OO_{\tiny \half} \rangle_{g^2}$ or 
$\langle \OO \KK \OO_{\tiny \half} \rangle_{g^2}$, and 
four times for 
$\langle \OO \OO \OO_{\tiny \half} \rangle_{g^2}$. 
The leading behavior of the $\tilde \beta_{xy}$ 
combinatorial coefficient 
for the three-point functions 
$\langle \KK \KK \OO_{\tiny \half} \rangle_{g^2}$ 
is the same is for terms of the form 
\begin{eqnarray}
\label{KKH:leading-g^2}
&& 
( \tr t^{a_1} ... t^{a_l} t^{b_1} ... t^{b_{n-1}} [t^{c_1},t^s] 
t^{c_1} ... t^{c_p} ) 
~ 
( \tr t^{d_1} ... t^{d_m} t^{c_p} ... t^{c_2} [t^{b_n},t^s] 
t^{b_n} ... t^{b_1} ) 
\nonumber\\&& \times 
( \tr t^{a_l} ... t^{a_1} t^{d_m} ... t^{d_1} ) 
\nonumber\\
&&\hspace{4em}
\sim 
\left( {1\over2} \right)^2 
\left( {N\over2} \right)^{l+m-2} 
\hspace{-1.5em}
( \tr t^{b_1} ... t^{b_{n-1}} [t^{c_1},t^s] 
t^{c_1} ... t^{c_p} 
t^{c_p} ... t^{c_2} [t^{b_n},t^s] 
t^{b_n} ... t^{b_1} )
\hspace{-1em}
\nonumber\\&&\hspace{4em}
\sim
\left( {1\over2} \right)^2 
\left( {N\over2} \right)^{p+l+m+n-4} 
( \tr [t^{c_1},t^s] t^{c_1} [t^{b_n},t^s] t^{b_n} )
\nonumber\\&&\hspace{4em}
\sim
\left( {1\over2} \right) 
\left( {N\over2} \right)^{p+l+m+n} 
\end{eqnarray}
which give the leading large $N$ contributions to it. 
In the same fashion, the most significant terms 
in the correlators 
$\langle \OO \KK \OO_{\tiny \half} \rangle_{g^2}$ 
are 
\begin{eqnarray}
\label{OKH:leading-g^2}
&& 
( \tr t^{a_1} ... t^{a_l} t^{b_1} ... t^{b_{n-1}} [t^{c_1},t^s] ) 
( \tr t^{c_1} ... t^{c_p} ) 
~ 
( \tr t^{d_1} ... t^{d_m} t^{c_p} ... t^{c_2} [t^{b_n},t^s] 
t^{b_n} ... t^{b_1} ) 
\nonumber\\ && \times 
( \tr t^{a_l} ... t^{a_1} t^{d_m} ... t^{d_1} ) 
\nonumber\\
&&\hspace{4em}
\sim 
\left( {1\over2} \right)^2 
\left( {N\over2} \right)^{p+l+m+n-1} 
\sim 
\langle \KK \OO \OO_{\tiny \half} \rangle_{g^2} 
\hspace{11em}
\end{eqnarray}
while 
$\langle \OO \OO \OO_{\tiny \half} \rangle_{g^2}$ 
gets it leading $N$ behavior from terms like 
\begin{eqnarray}
\label{OOH:leading-g^2}
&& 
( \tr t^{a_1} ... t^{a_l} t^{b_1} ... t^{b_{n-1}} [t^{c_1},t^s] ) 
( \tr t^{c_1} ... t^{c_p} ) 
~ 
( \tr t^{d_1} ... t^{d_m} t^{c_p} ... ) ( \tr ... t^{c_2} [t^{b_n},t^s] 
t^{b_n} ... t^{b_1} ) 
\hspace{-2.5em}\nonumber\\ && \times 
( \tr t^{a_l} ... t^{a_1} t^{d_m} ... t^{d_1} ) 
\nonumber\\
&&\hspace{4em}\sim 
\left( {1\over2} \right)^3 
\left( {N\over2} \right)^{p+l+m+n-2} 
\hspace{15em}
\end{eqnarray}

\subsection{
$\langle \OO_{\tiny \quarter} \OO_{\tiny \quarter} \OO_{\tiny \half} \rangle$
correlators are protected
}

With just a little more work, we can find the ratios 
of the order $g^2$ corrections to the three-point functions 
$\langle \OO \OO \OO_{\tiny \half} \rangle_{g^2}$, 
$\langle \OO \KK \OO_{\tiny \half} \rangle_{g^2}$, 
$\langle \KK \OO \OO_{\tiny \half} \rangle_{g^2}$, 
and 
$\langle \KK \KK \OO_{\tiny \half} \rangle_{g^2}$. 
The argument proceeds along the same lines as in \cite{Ryzhov}. 
Given a term with generators in a particular order, 
contributing to $\langle \KK \KK \OO_{\tiny \half} \rangle_{g^2}$, 
such as the one shown in (\ref{KKH:leading-g^2}), 
we know that a term with the same order of generators 
also gives a leading contribution to 
$\langle \OO \KK \OO_{\tiny \half} \rangle_{g^2}$ 
as in 
(\ref{OKH:leading-g^2}). 
However, cyclic permutations 
within the two traces (of length $p$ and $p+q$) of $\OO$, 
contribute to 
$\langle \OO \KK \OO_{\tiny \half} \rangle_{g^2}$ 
in the same amount as the term (\ref{OKH:leading-g^2}). 
Therefore, 
\begin{equation}
\langle \OO \KK \OO_{\tiny \half} \rangle_{g^2}
/
\langle \KK \KK \OO_{\tiny \half} \rangle_{g^2} = 
{p(p+q) \over N} + \OO(N^{-3}) 
\equiv \beta 
\end{equation}  
In the same fashion 
\begin{equation}
\langle \KK \OO \OO_{\tiny \half} \rangle_{g^2}
/
\langle \KK \KK \OO_{\tiny \half} \rangle_{g^2} = 
{r(r+s) \over N} + \OO(N^{-3}) 
\equiv \beta' 
\end{equation}  
and 
\begin{equation}
\langle \OO \OO \OO_{\tiny \half} \rangle_{g^2}
/
\langle \KK \OO \OO_{\tiny \half} \rangle_{g^2} = 
{p(p+q) \over N} + \OO(N^{-3}). 
\end{equation}

Next consider the Born level correlators 
of Section \ref{section: Born level - large N}. 
When a pair of traces collapses 
completely 
(see equations \ref{OOH:leading-born}-\ref{OKH:leading-born}), 
we get 
\begin{eqnarray}
\label{QQH:free - collapse completely}
\langle \YY_{[p,q,p]} \YY_{[r,s,r]} 
\OO_{\tiny \half} \rangle_{\rm free} 
\sim 
\langle \OO \OO \OO_{\tiny \half} \rangle_{\rm free} \sim 
N^{p+l+m+n-1} 
\end{eqnarray}
Otherwise, the contributions add up to 
\begin{equation}
\label{QQH:free - do not collapse completely}
\langle \OO \OO \OO_{\tiny \half} \rangle_{\rm free} - 
\beta \langle \KK \OO \OO_{\tiny \half} \rangle_{\rm free} - 
\beta' \langle \OO \KK \OO_{\tiny \half} \rangle_{\rm free} + 
\beta \beta' \langle \KK \KK \OO_{\tiny \half} \rangle_{\rm free} 
\sim 
N^{p+l+m+n-3} 
\end{equation}
The terms in (\ref{QQH:free - do not collapse completely})
are all of the same order 
and do not cancel. 
The factors of $\beta$ and $\beta'$ discussed above are still present, 
but there are other complications. 
First, the string of $t^c$-s can be inserted 
anywhere in the third trace in (\ref{OOH:leading-born}), 
and cyclic permutations of the $t^b$-s in the same trace 
give terms of the same order in $N$. 
This results in an extra factor of $(r-p)^2$.
Second, different terms in the 
sum over antisymmetrizations 
(as in equation (\ref{def:OO}), for example)
contribute differently. 
The combinatorics is more involved, 
and we do not discuss this case in detail.

Bringing everything together, we see that 
the order $g^2$ corrections to the 
three-point function of the BPS operators 
in the large $N$ limit add up to 
\begin{eqnarray}
\label{QQH:leading-g^2}
\langle \YY_{[p,q,p]} \YY_{[r,s,r]} 
\OO_{\tiny \half} \rangle_{g^2} 
&\propto& 
\left( {1\over2} \right) 
\left( {N\over2} \right)^{p+l+m+n-1} 
\tilde B(x,y) N 
\nonumber\\&&\hspace{3em}\times
\left(
\matrix{
-\beta \cr 
1 
}\right)^t
\left( 
\matrix{
1 + \OO(N^{-2}) & \beta' \cr
\beta & \beta \beta' 
}
\right)
\left( 
\matrix{
-\beta' \cr
1 
}
\right)
\nonumber\\&=& 
\left( {1\over2} \right) 
\left( {N\over2} \right)^{p+l+m+n-1} 
\tilde B(x,y) N 
\times
\OO(N^{-4}) 
\quad\quad 
\end{eqnarray}
A comparison of (\ref{QQH:leading-g^2}) with 
(\ref{QQH:free - collapse completely}) 
or (\ref{QQH:free - do not collapse completely}) 
shows that order $g^2$ corrections to 
three-point functions 
of one \half-BPS operator with two \quarter-BPS operators 
vanish in the large $N$ limit, 
within working precision.

\section{Supergravity considerations}
\label{section:SUGRA}

In the AdS/CFT correspondence, there is a duality mapping 
single trace 
\half-BPS primary operators 
$\tr X^k$ 
of the SYM theory 
onto canonical supergravity fields, \cite{GKP}, \cite{Witten}. 
Given a set of such \half-BPS primary operators, 
one can compute their two- and three-point functions 
in SYM. 
The two-point functions define the normalization of 
operators, and the three-point functions probe the 
interactions between them. 
Independently, both the normalization 
of the SUGRA fields 
as well as their couplings, can be read off from the 
supergravity action (or supergravity equations of motion), \cite{LMRS}. 
So as a check of the AdS/CFT correspondence, 
one can compare the unambiguously defined 
three- and higher $n$-point functions of 
normalized \half-BPS operators in SYM, 
with the correlators of the corresponding 
elementary excitations in supergravity, 
\cite{LMRS}, \cite{AF}, \cite{ADS:2-and-3}, \cite{ADS:4-pt}.

We would like to proceed, in the same spirit, 
with the 
\quarter-BPS chiral primaries of 
the \NN=4 Super Yang Mills 
calculated in this paper and in \cite{Ryzhov}. 
We argued that these two- and three-point functions are 
independent of the SYM coupling constant 
(at least to order $g^2$), so it is reasonable 
to expect these correlators to agree with their dual AdS description. 
However, multiple trace operators 
do not correspond to any of the fields appearing in the 
supergravity action, so the discussion 
will be different than 
in the case of 
the previously studied operators $\tr X^k$.

\subsection{OPE definition of \quarter-BPS chiral primaries}
\label{section:OPE}

One of the ways to see \quarter-BPS chiral primaries 
is to consider higher $n$-point correlators of \half-BPS operators. 
For example, four-point functions of [0,2,0] operators 
reveal a pole corresponding to the exchange of 
a 
[2,0,2] operator 
with a protected dimension $\Delta=4$, \cite{BKRS}. 
In general, 
the \quarter-BPS primaries 
$\YY_{[p,q,p]}$ 
show up in the four-point functions 
\begin{equation}
\label{double trace:4pt of CPOs}
\langle \tr X^{(p+q)}(x) \tr X^p(x+\e) ~ 
\tr X^{(p+q)}(y) 
~ \tr X^p(w) 
\rangle
\end{equation} 
in the limit $\e\to0$, 
as the $[p,q,p]$ operators with 
the threshold value of scaling dimension 
$\Delta = 2p+q = {\rm dim} [\tr X^{(p+q)}] + {\rm dim} [\tr X^p]$. 
In other words, \quarter-BPS chiral primaries 
can be defined by the OPE-s 
of \half-BPS operators as 
\begin{equation}
\label{def:double trace - OPE}
\PP^{\Delta = 2p+q}_{[p,q,p]}
\left[ 
\lim_{\e\to0}
\tr X^{(p+q)}(x) \; \tr X^p(x+\e) 
\right] 
= 
\sum_i c_i \YY^{[p,q,p]}_i(x) 
\end{equation} 
Here, $\PP^{\Delta}_{[p,q,p]}$ projects onto 
the $[p,q,p]$ representation of $SU(4)$, 
and eliminates 
operators with scaling dimension other than $\Delta$ 
(e.g. the non-chiral descendants with the same 
$SU(4)$ quantum numbers). 
Singular terms normally subtracted from 
an OPE such as (\ref{def:double trace - OPE}), are 
automatically removed by applying $\PP^{\Delta = 2p+q}_{[p,q,p]}$. 

On the other hand, 
one can see by calculating three-point correlators that 
all \quarter-BPS primary operators $\YY^{[p,q,p]}_i$ 
are present in the OPE (\ref{def:double trace - OPE}). 
It appears that for general $N$, 
there is no canonical definition 
of the special $\YY_{[p,q,p]}$ 
that is a linear combination of 
the single and double-trace scalar composite $[p,q,p]$ operators only. 
However, this $\YY_{[p,q,p]}$ 
dominates in the $N \to \infty$ limit. 
For large $N$, 
all other terms in the right hand side of (\ref{def:double trace - OPE}) 
are suppressed by at least a factor of $1/N$, and 
the predominantly double-trace \quarter-BPS chiral primary operator 
\begin{equation}
\YY_{[p,q,p]} = 
\left(
\mbox{
\setlength{\unitlength}{1em}
\begin{picture}(7.5,1.6)
\put(0,.5){\framebox (1,1){}}
\put(1,.5){\framebox (2,1){\scriptsize $...$}}
\put(3,.5){\framebox (1,1){}}
\put(4,.5){\framebox (1,1){}}
\put(5,.5){\framebox (1,1){\scriptsize $...$}}
\put(6,.5){\framebox (1,1){}}
\put(0,-1){\framebox (1,1){}}
\put(1,-1){\framebox (2,1){\scriptsize $...$}}
\put(3,-1){\framebox (1,1){}}
\put(1.8,-1.7){\scriptsize $p$}
\put(5.4,-0.7){\scriptsize $q$}
\end{picture}}
\right) 
+ \OO(1/N)
\phantom{{}_\Big|}
\end{equation}
is uniquely defined by the OPE of \half-BPS primaries. 

When translated into the SUGRA language, 
the definition (\ref{def:double trace - OPE}) 
implies that 
\quarter-BPS primary operators of SYM 
should be thought of as 
threshold bound states of 
elementary SUGRA excitations. 
A threshold bound state is a state whose mass 
is precisely equal to the sum of the masses of all its 
constituents, and thus occurs at the lower end of the spectrum. 
Any bound state of BPS states which is itself BPS 
is automatically a threshold bound state. 
A familiar example is provided by an assembly of 
like sign charged Prasad-Sommerfield magnetic monopoles, 
whose classical static solution forms a 
threshold bound state of monopole constituents.

\subsection{
$\langle
\OO_{\tiny \quarter} \OO_{\tiny \quarter} 
\rangle$ 
and 
$\langle
\OO_{\tiny \half} \OO'_{\tiny \half} \OO_{\tiny \quarter} 
\rangle$ 
correlators
}

We are now going to 
illustrate the consistency of this dictionary. 
Specifically, 
we will look at two- and three-point functions 
involving \half- and \quarter-BPS operators 
in the large $N$ limit, 
that we calculated earlier in this paper 
and in \cite{Ryzhov}, in \NN=4 SYM. 
Then we will compare these correlators with 
their dual supergravity description.

The normalization of \half- and \quarter-BPS operators 
comes from their two-point functions, whose leading  
large $N$ behavior is \cite{LMRS}, \cite{Ryzhov} 
\begin{eqnarray}
\label{normalization:large N - half} 
\langle
\tr X^q (x) 
~ 
\tr X^q (y) 
\rangle
&\sim& 
{N^{q} \over (x-y)^{2 q} }
\\
\langle
\YY_{[p,q,p]} (x) 
\; 
\bar{\YY}^{[p,q,p]} (y) 
\rangle
&\sim& 
{N^{(2p+q)} \over (x-y)^{2(2p+q)} }
\label{normalization:large N - quarter} 
\end{eqnarray}
times some $N$-independent factors which we omit. 

The simplest three point functions 
involving \half- and \quarter-BPS operators 
are of the form 
$
\langle
\OO_{\tiny \half} (x) \OO'_{\tiny \half} (y) \OO_{\tiny \quarter} (w) 
\rangle
$. 
If the $SU(N)$ traces collapse completely 
(in which case 
$\langle \OO_{\tiny \half} \OO'_{\tiny \half} \OO_{\tiny \quarter} \rangle$ 
are extremal), 
the normalized three point-functions are then 
\begin{eqnarray}
\label{normalized:large N:HHQ-extremal} 
{1\over \sqrt{N^{(2p+q)+(p+q)+p}}}
\langle
\tr X^{(p+q)} (x) 
~ 
\tr X^{p} (y) 
~
\YY_{[p,q,p]} (w) 
\rangle 
\sim 
1
\end{eqnarray}
from a field theory calculation; 
the space-time coordinate dependence is fixed by 
conformal invariance, so we will not exhibit it anymore. 
If the traces do not collapse completely, the correlator 
is suppressed by $1/N^2$ 
(see the discussion around equations 
(\ref{OOH:leading-born:equal}-\ref{OOH:leading-born}) 
of Section \ref{section:large N}), 
and 
\begin{eqnarray}
\label{normalized:large N:HHQ} 
{1\over \sqrt{N^{(2p+q)+(k+l)+k}}}
\langle
\tr X^{(k+l)} (x) 
~ 
\tr X^{k} (y) 
~
\YY_{[p,q,p]} (w) 
\rangle 
\sim 
{1\over N^2}
\end{eqnarray} 
whenever $k \ne p$ of $l \ne q$. 
All this matches nicely 
with the corresponding supergravity diagrams: 
\begin{eqnarray}
\label {AdS:half-half-quarter}
&&\epsfig{height=1.4in, file=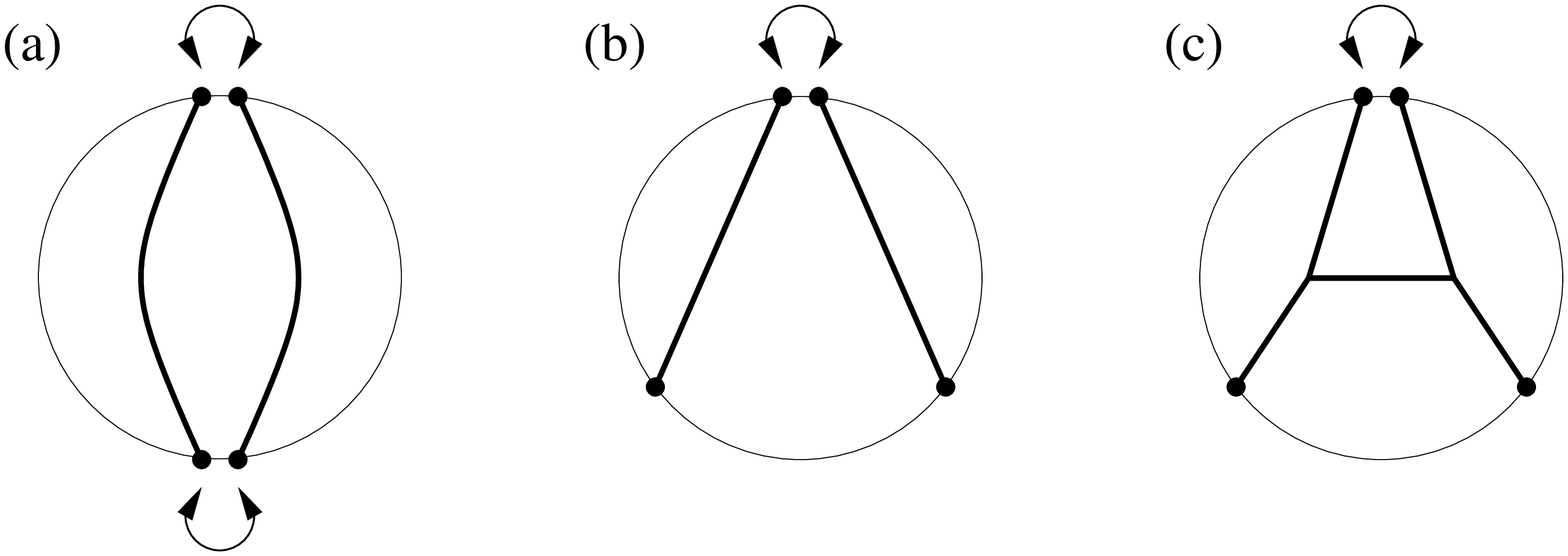, angle=0}
\hspace{-2em}
\\
&&\mbox{
Leading AdS diagrams for 
(a) equation (\ref{normalization:large N - quarter}); 
(b) equation (\ref{normalized:large N:HHQ-extremal}); 
}\hspace{0.6em}
\nonumber
\\
&&\mbox{
(c) equation (\ref{normalized:large N:HHQ}). 
Each cubic bulk interaction vertex goes like $1/N$. 
}
\nonumber
\end{eqnarray}
We denoted the \half-BPS operators by ``$\bullet$''; 
and the predominantly double trace \quarter-BPS 
primaries which arise from bringing two \half-BPS operators 
together by 
``$\hspace{0.2em}
\begin{picture}(1,1)
\put(0,-2){\epsfig{height=0.17in, file=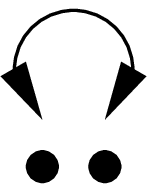, angle=0}}
\end{picture}
{\phantom{\Big|}}\hspace{0.85em}$''.

There are also AdS diagrams with 
quartic interactions in the bulk, 
which have the same large $N$ dependence as 
(\ref{AdS:half-half-quarter}c); 
we will not show these.

\subsection{
$\langle
\OO_{\tiny \quarter} \OO'_{\tiny \quarter} \OO_{\tiny \half} 
\rangle$ 
correlators
}

Other three point functions involving 
the \quarter-BPS as well as the \half-BPS operators 
can be analyzed similarly. 
Whenever traces of the SYM operators 
do not collapse completely, 
the supergravity counterparts of such correlators 
have extra bulk interaction vertices. 
The leading dependence 
of such correlators is then suppressed 
by the corresponding power of $1/N$. 
For example, correlators of the form 
$
\langle
\OO_{\tiny \quarter} (x) \OO'_{\tiny \quarter} (y) \OO_{\tiny \half} (w) 
\rangle
$, 
discussed in Section \ref{section:large N}, 
behave like 
\begin{equation}
\label{OOH:leading-born:normalized}
{1\over \sqrt{N^{(2p+q)+(2r+s)+k}}}
\langle 
\YY_{[p,q,p]} \YY_{[r,s,r]} \tr X^k \rangle 
\sim 
\left\{
\matrix{
1/N 
& 
\mbox{(a) if one pair of traces}
\cr &
\mbox{collapses completely}\hfill 
\cr &
\cr 
1/N^3 
& 
\mbox{(b) otherwise}\hfill
}
\right.
\end{equation}
From the AdS point of view, 
this difference is captured by the following diagrams 
\begin{eqnarray}
\label {AdS:quarter-quarter-half}
&&\epsfig{height=1.0in, file=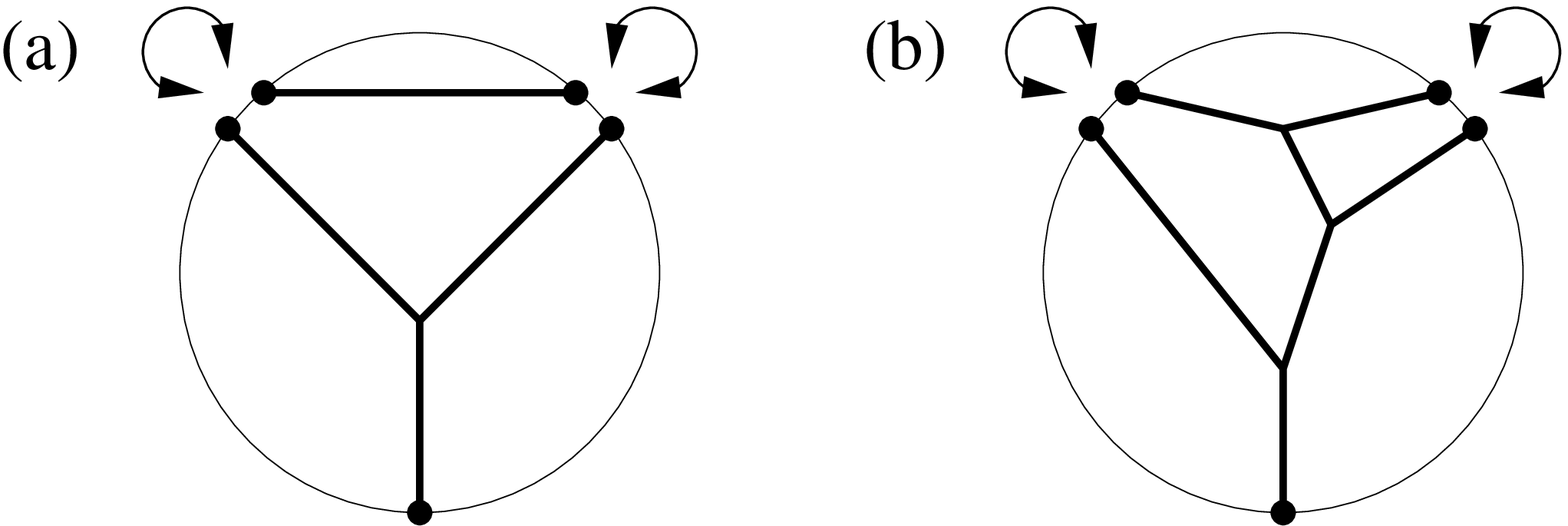, angle=0}
\hspace{2em}\phantom{\Bigg|}
\\
&&\hspace{4em}\mbox{
AdS description of 
equation (\ref{OOH:leading-born:normalized}). 
}
\nonumber
\end{eqnarray}

\subsection{
$\langle
\OO_{\tiny \quarter} \OO'_{\tiny \quarter} \OO''_{\tiny \quarter} 
\rangle$ 
correlators
}

Similar arguments show that when all 
operators are \quarter-BPS, the normalized 
three-point functions are 
\begin{equation}
\label{OOO:leading-born:normalized}
{1\over \sqrt{N^{(2p+q)+(2r+s)+(2l+k)}}}
\langle 
\YY_{[p,q,p]} \YY_{[r,s,r]} \YY_{[l,k,l]} \rangle 
\sim 
\left\{
\matrix{
1
&
\mbox{(a) if all traces}\hfill
\cr &
\mbox{collapse pairwise}\hfill 
\cr &
\cr
1/N^2 
&
\mbox{(b) if only one pair}\hfill
\cr &
\mbox{of traces collapses}\hfill
\cr &
\cr 
1/N^4 
&
\mbox{(c) otherwise}\hfill
}
\right.
\end{equation}
and the corresponding AdS diagrams 
\begin{eqnarray}
\label {AdS:quarter-quarter-quarter}
&&\epsfig{height=1.1in, file=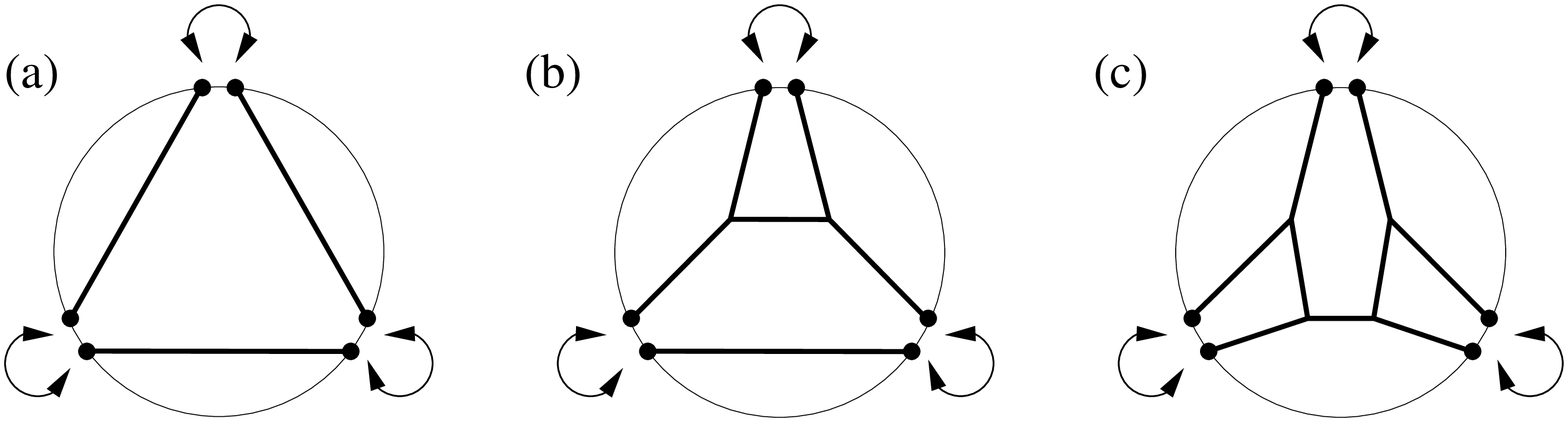, angle=0}
\hspace{1em}\phantom{\Bigg|}
\\
&&\hspace{7em}\mbox{
AdS description of 
equation (\ref{OOO:leading-born:normalized}). 
}
\nonumber
\end{eqnarray}
show the correct leading large $N$ behavior.

\subsection{
Detailed agreement between SYM and AdS
}

Unlike the \half-BPS calculations (e.g. \cite{LMRS}), 
this study does not provide a new independent check or application 
of the AdS/CFT correspondence. 
On the one hand, the definition of 
the predominantly double-trace \quarter-BPS 
operators in the SYM theory (in the large $N$ limit) 
is based on the OPE of \half-BPS primaries. 
On the other hand, AdS correlators of 
the duals of the \quarter-BPS operators 
(bound states of elementary SUGRA excitations)
are defined by the corresponding correlators of 
primary supergravity fields. 
Therefore, SYM correlators involving 
\quarter-BPS operators 
agree by construction 
with their SUGRA counterparts.

This is especially clear in the cases 
show in diagrams 
(\ref{AdS:half-half-quarter}a,b), 
(\ref{AdS:quarter-quarter-half}a), 
and (\ref{AdS:quarter-quarter-quarter}a). 
To leading order in $N$, these two- and three-point functions 
of \quarter-BPS scalar composite operators 
are expressed in terms of the previously studied 
two- and three-point functions of 
\half-BPS chiral primary operators.

\section{Conjectures} 

Let us summarize what has been done so far. 
First, \quarter-BPS primary operators were identified in \cite{Ryzhov}: 
for general representations $[p,q,p]$ in the large $N$ limit; 
and for general $N$ in the case when $2p+q \le 7$. 
Second, three-point functions 
involving \half-BPS operators as well as 
several infinite families of \quarter-BPS operators 
were considered in this paper, 
also for arbitrary $N$. 
It was found that there are no $\OO(g^2)$ corrections 
to such correlators. 
Next, all three-point functions involving the \quarter-BPS primaries 
with $2p+q \le 7$ 
were 
computed for 
general $N$, and were shown to be protected at order $g^2$. 
In the large $N$ approximation, 
three point functions involving 
two \quarter-BPS primaries and one \half-BPS primary 
were shown to receive no $\OO(g^2)$ corrections, 
for general representations of the operators involved. 
Finally, we presented AdS considerations which reproduced 
many features of the CFT two- and three-point functions 
in the large $N$ limit.

Collecting all the non-renormalization effects established above 
generates strong evidence for a number of natural conjectures, 
which we now state: 
\medskip

(1) 
We conjecture that on the CFT side, 
for every $[p,q,p]$ representation of $SU(4)$ 
and arbitrary $N$, 
there are \quarter-BPS chiral primaries. 
Within each $[p,q,p]$, one of these operators is a linear 
combination of double and single trace scalar 
composites only; 
the other \quarter-BPS chiral primaries in $[p,q,p]$ 
also involve operators with higher numbers of traces. 
\medskip

(2)
We speculate that two--point 
functions of \quarter-BPS operators, 
as well as three-point functions involving 
\half-BPS and \quarter-BPS operators, 
do not depend on the coupling $g^2$ of \NN=4 SYM. 
This non-renormalization also persist for all $N$, 
and is not just a large $N$ approximation. 
\medskip

(3)
One of the group theory arguments 
of Section \ref{section:simplifications}, 
and the analysis of Section \ref{section:large N}, 
generalize straightforwardly to 
extremal correlators, 
i.e. $(n+1)$-point functions of the form 
$\langle \OO_0 (x_0) \OO_1(x_1) ... \OO_n(x_n) \rangle$ 
with $\Delta_0 = \Delta_1 + ... + \Delta_n$. 
So do the AdS considerations of Section \ref{section:SUGRA}. 
Therefore, we conjecture that arbitrary extremal correlators 
of \half- and \quarter-BPS chiral primaries are 
also protected. 

\section{Acknowledgments}

We would like to thank 
Daniel Z. Freedman and Sergio Ferrara 
for useful conversations 
during various stages of this project.

This research is supported in part by NSF Grant 
No. PHY-98-19686.

\pagebreak

{\LARGE \bf \appendixname}
\appendix

\section{\quarter-BPS operators explicitly}
\label{operators explicitly}

Scalar composite gauge invariant operators 
$\OO \sim [(\phi^1)^{p+q} (\phi^2)^{p} ]$ 
corresponding to highest weights of representations $[p,q,p]$ of the 
$R$-symmetry group $SU(4) \sim SO(6)$ were 
studied in \cite{Ryzhov}. 
It was found that there are many such 
gauge invariant operators $\OO_i$ for a given set of fields. 
They depend on how we partition the products into traces. 

In general, none of the $\OO_i$-s are 
eigenstates of the dilatations operator. 
To construct operators with a well defined scaling 
dimension, one has to take particular linear combinations, 
$\YY_j = \sum_i C_j^i \OO_i$. 
Some of the $\YY_j$-s are \quarter-BPS: 
they are annihilated by a quarter of the Poincar\'e supercharges, 
they are not descendants of \eigth-BPS primaries, 
and at order $g^2$ they have a protected scaling dimension $\Delta = 2p+q$; 
their normalization also receives no $\OO(g^2)$ corrections.

In this Appendix we list some of the results of \cite{Ryzhov}. 
In particular, we define the operators $\OO_i$ 
for each representation $[p,q,p]$ with $2p+q \le 7$, 
and list their linear combinations that are \quarter-BPS. 
(We do not show the other pure operators here.) 
This is done for general $N$. 
We also write out the 
two special scalar composite operators which 
contribute to the \quarter-BPS primaries 
in the large $N$ limit, for all $[p,q,p]$ representations.

\subsection{[2,0,2]}

There are two linearly independent gauge invariant 
scalar composite operators with the highest [2,0,2] weight, 
\begin{eqnarray}
\label{202:operators}
\OO^{[2,0,2]}_1 &=& \tr z_1 z_1 z_2 z_2 - \tr z_1 z_2 z_1 z_2 
= 
- \half \tr [z_1 , z_2] [z_1 , z_2] 
\\ 
\OO^{[2,0,2]}_2 &=& 2 \left( \tr z_1 z_1 ~ \tr z_2 z_2 - 
\tr z_1 z_2 ~ \tr z_1 z_2 \right) 
\end{eqnarray}
The operator whose two-point functions with itself 
and other scalar composites receives no $\OO(g^2)$ corrections is 
\begin{eqnarray}
\YY_{[2,0,2]} &=& \OO^{[2,0,2]}_2 - {4\over N} \OO^{[2,0,2]}_1 
\label{202:protected}
\end{eqnarray}
Other \quarter-BPS operators in the same representation 
(the ones corresponding to different $SU(3)\times U(1)$ weights) 
have the same coefficients in front of 
double and single trace operators (provided they are normalized in the 
same way). 
For example, such are 
(\ref{202:different wts-1}-\ref{202:different wts-2}).

\subsection{[2,1,2]}

Here there are also two partitions of integer $2p+q$ 
which result in linearly independent scalar composite operators
\begin{eqnarray}
\label{212:operators}
\OO^{[2,1,2]}_1 &=& \tr z_1 z_1 z_1 z_2 z_2 - \tr z_1 z_1 z_2 z_1 z_2 
= 
- \half \tr [z_1 , z_2] [z_1^2 , z_2] 
\\ 
\OO^{[2,1,2]}_2 &=& \tr z_1 z_1 z_1 ~ \tr z_2 z_2 - 
2 \tr z_1 z_2 ~ \tr z_1 z_2 z_1 + \tr z_1 z_2 z_2 ~ \tr z_1 z_1 
\end{eqnarray}
The single \quarter-BPS operators is 
\begin{eqnarray}
\YY_{[2,1,2]} &=& \OO^{[2,1,2]}_2 - {6\over N} \OO^{[2,1,2]}_1 
\label{212:protected}
\end{eqnarray}

\subsection{[2,2,2]}

$2p+q = 6$ is the lowest dimension for which there 
are more than two linearly independent gauge invariant operators. 
Here we have five: 
\begin	{eqnarray}
\label{222:operators:begin}
\OO^{[2,2,2]}_1 &=& 
\tr z_1 z_1 z_1 z_1 z_2 z_2 
- \mbox{$2\over3$} \, \tr z_1 z_1 z_1 z_2 z_1 z_2 
- \mbox{$1\over3$} \, \tr z_1 z_1 z_2 z_1 z_1 z_2 
\quad\quad
\\
\OO^{[2,2,2]}_2 &=& 
\tr z_1 z_1 z_1 z_1 ~\tr z_2 z_2 
- 2 \, \tr z_1 z_1 z_1 z_2 ~\tr z_1 z_2 
\nonumber\\ && \quad\quad\quad\quad
+ \mbox{$1\over3$} 
\left( 
2 \, \tr z_1 z_1 z_2 z_2 + \tr z_1 z_2 z_1 z_2 
\right) ~\tr z_1 z_1 
\\ 
\OO^{[2,2,2]}_3 &=& 
\tr z_1 z_1 z_1  ~\tr z_1 z_2 z_2 
- \tr z_1 z_1 z_2 ~\tr z_1 z_1 z_2 
\\ 
\OO^{[2,2,2]}_4 &=& 
\half \tr z_1 z_1 z_1 z_1 ~\tr z_2 z_2 
- \tr z_1 z_1 z_1 z_2 ~\tr z_1 z_2 
\nonumber\\ && \quad\quad\quad\quad
+ \half
\left( 
4 \, \tr z_1 z_1 z_2 z_2 - 3 \, \tr z_1 z_2 z_1 z_2 
\right) ~\tr z_1 z_1 
\\
\OO^{[2,2,2]}_5 &=& 
\tr z_1 z_1 
\left( 
\tr z_1 z_1 ~\tr z_2 z_2 - 
\tr z_1 z_2 ~\tr z_1 z_2
\right) 
\label{222:operators:end}
\end	{eqnarray}
The two linear combinations of operators which satisfy 
$\langle \YY \bar{\OO}_i \rangle = 0$ for all $i$, are 
\begin	{equation}
\label	{222:protected-1}
\YY^{[2,2,2]}_1 = 
-{\frac{8 N}{\left( N^2 - 4 \right) }} \OO^{[2,2,2]}_1 + 
\OO^{[2,2,2]}_2 + 
{\frac{8}{3 \left( N^2 - 4 \right) }} 
\left( 2 \OO^{[2,2,2]}_3 + \OO^{[2,2,2]}_4 \right) 
\end	{equation}
and the one orthogonal to it 
(in the sense that $\langle \YY^{[2,2,2]}_1(x) 
\bar{\YY}^{[2,2,2]}_2(y) \rangle = 0$) 
\begin	{eqnarray}
\label	{222:protected-2}
\YY^{[2,2,2]}_2 &\!\!=\!\!& 
{144 \left( N^2 - 4 \right) \left( {N^2} -2 \right) 
\over 3 {N^6} - 47 {N^4} + 248 {N^2} -192 } \OO^{[2,2,2]}_1 - 
{3 N \left( {N^2} -7 \right) \left( 3 {N^2} + 8 \right) 
\over 3 {N^6} - 47 {N^4} + 248 {N^2} -192 } \OO^{[2,2,2]}_2 
\hspace{-3em}
\nonumber\\ &&-
{2 N \left( 3 {N^4} - 23 {N^2} + 104 \right) 
\over 3 {N^6} - 47 {N^4} + 248 {N^2} -192 } 
\left( 2 \OO^{[2,2,2]}_3 + \OO^{[2,2,2]}_4 \right)
+ 
\OO_5 
\end	{eqnarray}

\subsection{[3,1,3]}

Two $[p,q,p]$ representations have $2p+q=7$. 
These are [3,1,3] = ${\bf 960}$ and [2,3,2] = ${\bf 1470}$. 
In the first case, the 
scalar composite operators are 
\begin	{eqnarray}
\label{313:operators:begin}
\OO^{[3,1,3]}_1 &=& 
\third \tr z_1 z_1 z_1 z_1 z_2 z_2 z_2 
-\half \tr z_1 z_1 z_1 z_2 z_1 z_2 z_2 
-\half \tr z_1 z_1 z_1 z_2 z_2 z_1 z_2 
\hspace{-2em}\nonumber\\&&
+ 
\third \tr z_1 z_1 z_2 z_1 z_2 z_1 z_2 + 
\third \tr z_1 z_1 z_2 z_2 z_1 z_1 z_2 
\quad\quad
\\
\OO^{[3,1,3]}_2 &=& 
\tr z_1 z_1 z_1 z_1 ~\tr z_2 z_2 z_2 
- 3 \, \tr z_1 z_1 z_1 z_2 ~\tr z_1 z_2 z_2 
\\ && 
+ 
\left( 
2 \, \tr z_1 z_1 z_2 z_2 + \tr z_1 z_2 z_1 z_2 
\right) ~\tr z_1 z_1 z_2 
- \tr z_1 z_2 z_2 z_2 ~\tr z_1 z_1 z_1 
\nonumber\\ 
\OO^{[3,1,3]}_3 &=& 
- \left( 
\tr z_1 z_1 z_2 z_2 z_2 - \tr z_1 z_2 z_1 z_2 z_2 
\right) ~\tr z_1 z_1  
\nonumber\\ && 
+ 
\left( 
\tr z_1 z_1 z_1 z_2 z_2 - \tr z_1 z_1 z_2 z_1 z_2 
\right) ~\tr z_1 z_2 
\\
\OO^{[3,1,3]}_4 &=& 
\tr z_1 z_2 
\left( 
2 \, \tr z_1 z_2 ~\tr z_1 z_2 z_2 - 
\tr z_2 z_2 ~\tr z_1 z_1z_1 
- 3 \tr z_1 z_1 ~\tr z_1 z_2z_2 
\right) 
\hspace{-2em}\nonumber\\&&
+
\tr z_1 z_1 
\left( 
\tr z_2 z_2 ~\tr z_1 z_1 z_2 + 
\tr z_1 z_1 ~\tr z_2 z_2z_2
\right) 
\label{313:operators:end}
\end	{eqnarray}
The $\OO(g^2)$ protected operators work out to be 
\begin	{eqnarray}
\label	{313:protected-1}
\YY^{[3,1,3]}_1 &=& 
- {\frac{12 N}{{N^2}-2}} \OO^{[3,1,3]}_1 + \OO^{[3,1,3]}_2 
- {\frac{5}{{N^2}-2}} \OO^{[3,1,3]}_3
\\
\YY^{[3,1,3]}_2 &=& 
{\frac{96}{{N^2}-4}} \OO^{[3,1,3]}_1 - {\frac{4 N}{{N^2}-4}} \OO^{[3,1,3]}_2 
+ {\frac{10 N}{{N^2}-4}} \OO^{[3,1,3]}_3 + \OO^{[3,1,3]}_4
\label	{313:protected-2}
\hspace{2em}
\end	{eqnarray}

\subsection{[2,3,2]}

The other $[p,q,p]$ of $SU(4)$ with $2p+q=7$ is the [2,3,2]. 
Here we have seven linearly independent operators 
corresponding to the highest weight state: 
\begin	{eqnarray}
\label{232:operators-begin}
\OO^{[2,3,2]}_1 &=& 
2 \, \tr z_1 z_1 z_1 z_1 z_1 z_2 z_2 
- \tr z_1 z_1 z_1 z_1 z_2 z_1 z_2 
- \tr z_1 z_1 z_1 z_2 z_1 z_1 z_2 
\quad\quad
\\
\OO^{[2,3,2]}_2 &=& 
2 \, \tr z_1 z_1 z_1 z_1 z_1 ~\tr z_2 z_2 
- 4 \, \tr z_1 z_1 z_1 z_1 z_2 ~\tr z_1 z_2 
\\ && 
+ 
\left( 
\tr z_1 z_1 z_1 z_2 z_2 + \tr z_1 z_1 z_2 z_1 z_2 
\right) ~\tr z_1 z_1 
\nonumber\\ 
\OO^{[2,3,2]}_3 &=& 
\tr z_1 z_1 z_1 z_1 z_1 ~\tr z_2 z_2 
- 2 \, \tr z_1 z_1 z_1 z_1 z_2 ~\tr z_1 z_2 
\\ && 
+ 
\left( 
8 \, \tr z_1 z_1 z_1 z_2 z_2 - 7 \, \tr z_1 z_1 z_2 z_1 z_2 
\right) ~\tr z_1 z_1 
\nonumber\\ 
\OO^{[2,3,2]}_4 &=& 
3 \, \tr z_1 z_1 z_1 z_1 ~\tr z_1 z_2 z_2 
- 6 \, \tr z_1 z_1 z_1 z_2 ~\tr z_1 z_1 z_2 
\nonumber\\ && 
+ 
\left( 
2 \, \tr z_1 z_1 z_2 z_2 + \tr z_1 z_2 z_1 z_2 
\right) ~\tr z_1 z_1 z_1 
\\ 
\OO^{[2,3,2]}_5 &=& 
3 \, \tr z_1 z_1 z_1 z_1 ~\tr z_1 z_2 z_2 
- 6 \, \tr z_1 z_1 z_1 z_2 ~\tr z_1 z_1 z_2 
\nonumber\\ && 
+ 
\left( 
7 \, \tr z_1 z_1 z_2 z_2 - 4 \, \tr z_1 z_2 z_1 z_2 
\right) ~\tr z_1 z_1 z_1 
\\
\OO^{[2,3,2]}_6 &=& 
-8 \, \tr z_1 z_2 \tr z_1 z_2 \tr z_1 z_1 z_1 
-6 \, \tr z_1 z_1 \tr z_1 z_2 \tr z_1 z_1 z_2 
\hspace{-2em}\nonumber\\&&
+
\tr z_1 z_1 
\left( 
11 \, \tr z_2 z_2 ~\tr z_1 z_1 z_1 + 
 3 \, \tr z_1 z_1 ~\tr z_1 z_2 z_2 
\right) 
\\
\OO^{[2,3,2]}_7 &=& 
7 \, \tr z_1 z_2 \tr z_1 z_2 \tr z_1 z_1 z_1 
-6 \, \tr z_1 z_1 \tr z_1 z_2 \tr z_1 z_1 z_2 
\hspace{-2em}\nonumber\\&&
+
\tr z_1 z_1 
\left( 
-4 \, \tr z_2 z_2 ~\tr z_1 z_1 z_1 + 
 3 \, \tr z_1 z_1 ~\tr z_1 z_2 z_2 
\right) 
\label{232:operators-end}
\end	{eqnarray}
The \quarter-BPS operators are 
\begin	{eqnarray}
\label	{232:protected-1}
\YY^{[2,3,2]}_1 &\!\!=\!\!& 
-{\frac{10 N}{N^2-7}} \OO^{[2,3,2]}_1 + \OO^{[2,3,2]}_2 + 
{\frac{2}{N^2-7}} 
\left( \OO^{[2,3,2]}_3 + \OO^{[2,3,2]}_4 + \OO^{[2,3,2]}_5 \right) 
\nonumber\\
\\
\YY^{[2,3,2]}_2 &\!\!=\!\!& -20 \OO^{[2,3,2]}_1 
+ {\frac{2 \left( N^2+2 \right) }{N}} \OO^{[2,3,2]}_2 
-{\frac{2}{N}} \left( \OO^{[2,3,2]}_3 + \OO^{[2,3,2]}_4 \right) 
+ \OO^{[2,3,2]}_6 
\nonumber\\
\\
\YY^{[2,3,2]}_3 &\!\!=\!\!& 10 \OO^{[2,3,2]}_1 
-{\frac{\left( N^2-4 \right) }{N}} \OO^{[2,3,2]}_2 
-{\frac{2}{N}} \left( \OO^{[2,3,2]}_3 + \OO^{[2,3,2]}_4 \right) 
+ \OO^{[2,3,2]}_7
\nonumber\\
\label	{232:protected-2}
\end{eqnarray}
(The $\YY^{[2,3,2]}_{1,2,3}$ are not orthogonal. 
Although orthonormal linear combinations are easy to find, 
they look rather messy and we don't list them here.)

\subsection{$\OO_{[p,q,p]}$ and $\KK_{[p,q,p]}$ in the large $N$ limit}
\label{appendix:large N}

Recall that the $SO(6)$ Young tableau for the $[p,q,p]$ of $SU(4)$ 
consists of two rows 
(one of length $p+q$, and the other of length $p$). 
Among the possible partitions of the highest weight tableau, 
there are two special ones 
\begin{eqnarray}
\OO_{[p,q,p]} \sim 
\left(
\mbox{
\setlength{\unitlength}{1em}
\begin{picture}(7.5,1.6)
\put(0,.5){\framebox (1,1){\scriptsize $1$}}
\put(1,.5){\framebox (2,1){\scriptsize $...$}}
\put(3,.5){\framebox (1,1){\scriptsize $1$}}
\put(4,.5){\framebox (1,1){\scriptsize $1$}}
\put(5,.5){\framebox (1,1){\scriptsize $...$}}
\put(6,.5){\framebox (1,1){\scriptsize $1$}}
\put(0,-1){\framebox (1,1){\scriptsize $2$}}
\put(1,-1){\framebox (2,1){\scriptsize $...$}}
\put(3,-1){\framebox (1,1){\scriptsize $2$}}
\put(1.8,-1.7){\scriptsize $p$}
\put(5.4,-0.7){\scriptsize $q$}
\end{picture}}
\right) 
, \quad 
\KK_{[p,q,p]} \sim 
\left(
\mbox{
\setlength{\unitlength}{1em}
\begin{picture}(7.5,1.6)
\put(0,.2){\framebox (1,1){\scriptsize $1$}}
\put(1,.2){\framebox (2,1){\scriptsize $...$}}
\put(3,.2){\framebox (1,1){\scriptsize $1$}}
\put(4,.2){\framebox (1,1){\scriptsize $1$}}
\put(5,.2){\framebox (1,1){\scriptsize $...$}}
\put(6,.2){\framebox (1,1){\scriptsize $1$}}
\put(0,-.8){\framebox (1,1){\scriptsize $2$}}
\put(1,-.8){\framebox (2,1){\scriptsize $...$}}
\put(3,-.8){\framebox (1,1){\scriptsize $2$}}
\put(1.8,-1.5){\scriptsize $p$}
\put(5.4,-0.5){\scriptsize $q$}
\end{picture}}
\right) 
\end{eqnarray}
(each continuous group of boxes stands for 
a single trace). 
Explicitly, the corresponding operators are 
\begin{eqnarray}
\label{def:OO}
\OO_{[p,q,p]} &=& 
\sum_{k=0}^p {(-1)^k p! \over k! (p-k)!} ~
\tr \left( {z_1}^{p+q-k} {z_2}^{k} \right)_s ~ 
\tr \left( {z_1}^{k} {z_2}^{p-k} \right)_s 
\\
\label{def:KK}
\KK_{[p,q,p]} &=& 
\sum_{k=0}^p {(-1)^k p! \over k! (p-k)!} ~
\tr \left[ \left( {z_1}^{p+q-k} {z_2}^{k} \right)_s 
\left( {z_1}^{k} {z_2}^{p-k} \right)_s \right] 
\end{eqnarray}
after projecting $SU(4) \to SU(3) \times U(1)$ 
and keeping only 
the highest $U(1)$-charge pieces. 
Made of only $z_1$ and $z_2$, 
both types of operators 
are annihilated by four out of the sixteen 
Poincar\'e supersymmetry generators.

$\KK_{[p,q,p]}$ is special because it is 
the only single trace $[p,q,p]$ operator 
which can be constructed out of these fields. 
On the other hand, $\OO_{[p,q,p]}$ is 
``the most natural'' double trace composite operator 
in this representation. 
We also recognize it as the 
free theory chiral primary 
from the classification of \cite{AFSZ}.

With a slight abuse of notation, we will use the 
same name for 
operators with different $SU(4)$ weights; 
e.g. all $[p,q,p]$ single trace scalar composites 
will be referred to as $\KK_{[p,q,p]}$, etc.

\section{$[p,q,p] \otimes [r,s,r] \otimes [0,k,0]$ 
and BZ triangles}
\label{BZ-triangles}

Tensoring irreducible representations using Young tableaux 
can get quite tedious. 
Berenstein-Zelevinsky (BZ) triangles \cite{DiF} 
provide a powerful way to calculate the multiplicity of the 
scalar representation%
\footnote{
	It is conventional to choose 
	$\nu^*$ instead of $\nu$ for the third weight. 
	}
in $\lambda \otimes \mu \otimes \nu$.
We will discuss the construction for $SU(3)$ and $SU(4)$, 
the generalization to higher $SU(N)$ 
(but not to other Lie algebras, which is not currently known)
being straightforward.

For $SU(3)$, 
the triangles are constructed according to 
the following rules:
\begin{equation}
\label{BZ:su3}
\begin{tabular}{  c c c c c c c  }
         &          &          & $m_{13}$ &          &          &          \cr
         &          & $n_{12}$ &          & $l_{23}$ &          &          \cr
         & $m_{23}$ &          &          &          & $m_{12}$ &          \cr
$n_{13}$ &          & $l_{12}$ &          & $n_{23}$ &          & $l_{13}$ 
\end{tabular}
\end{equation}
where the nine non-negative integers 
$l_{ij}$, $m_{ij}$, $n_{ij}$ are related to the 
Dynkin labels 
$(\lambda_1,\lambda_2)$, $(\mu_1,\mu_2)$, $(\nu_1,\nu_2)$ 
of the highest weights of the three representations by 
\begin{equation}
\begin{tabular}{  c  c  c  }
$m_{13}+n_{12}=\lambda_1$ & $n_{13}+l_{12}=\mu_1$ & $l_{13}+m_{12}=\nu_1$ \cr
$m_{23}+n_{13}=\lambda_2$ & $n_{23}+l_{13}=\mu_2$ & $l_{23}+m_{13}=\nu_2$ 
\end{tabular}
\end{equation}
They must further satisfy the so-called hexagon conditions 
\begin{equation}
\label{hex:su3}
\begin{tabular}{  c  }
$n_{12}+m_{23}=m_{12}+n_{23}$ \cr
$l_{12}+m_{23}=m_{12}+l_{23}$ \cr
$l_{12}+n_{23}=n_{12}+l_{23}$ 
\end{tabular}
\end{equation}
This means that the length of opposite sides in the hexagon formed by 
$n_{12}$, $l_{23}$, $m_{12}$, $n_{23}$, $l_{12}$, and $m_{23}$ in 
(\ref{BZ:su3}) are equal, the length of a segment being the sum of its 
two vertices.

The number of such triangles gives the multiplicity ${\cal N}_{\lambda\mu\nu}$; 
if it is not possible to construct such a triangle, 
$\nu^*$ does not occur in the tensor product $\lambda \otimes \mu$. 

The integers in the BZ triangles have the following origin. 
Each pair of indices $ij$, $i<j$, on the labels of the triangle 
is related to a positive root of $SU(3)$. 
For $SU(N)$, positive roots can be written as 
$\e_i-\e_j$, $1 \le i \le j \le N$, 
in terms of orthonormal vectors $\e_i$ in ${\bf R}^N$. 

The triangle encodes three sums of positive roots:
\begin{equation}
\begin{tabular}{  c  }
$\mu + \nu - \lambda^* = \sum_{i<j} l_{ij} (\e_i-\e_j)$ \cr
$\nu + \lambda - \mu^* = \sum_{i<j} m_{ij} (\e_i-\e_j)$ \cr
$\lambda + \mu - \nu^* = \sum_{i<j} n_{ij} (\e_i-\e_j)$ 
\end{tabular}
\end{equation}
The hexagon relations (\ref{hex:su3}) can be seen as 
consistency conditions for these three expansions.

For $SU(4)$, the BZ triangles are defined in a similar way, 
in terms of 
\begin{equation}
\label{BZ:su4}
\begin{tabular}{  c c c c c c c c c c c  }
         &          &          &          &          & $m_{14}$ &          &          &          &          &          \cr
         &          &          &          & $n_{12}$ &          & $l_{34}$ &          &          &          &          \cr
         &          &          & $m_{24}$ &          &          &          & $m_{13}$ &          &          &          \cr
         &          & $n_{13}$ &          & $l_{23}$ &          & $n_{23}$ &          & $l_{24}$ &          &          \cr
         & $m_{34}$ &          &          &          & $m_{23}$ &          &          &          & $m_{12}$ &          \cr
$n_{14}$ &          & $l_{12}$ &          & $n_{24}$ &          & $l_{13}$ &          & $n_{34}$ &          & $l_{14}$ 
\end{tabular}
\end{equation}
eighteen non-negative integers, 
related to the Dynkin labels by 
\begin{equation}
\begin{tabular}{  c  c  c  }
$m_{14}+n_{12}=\lambda_1$ & $n_{14}+l_{12}=\mu_1$ & $l_{14}+m_{12}=\nu_1$ \cr
$m_{24}+n_{13}=\lambda_2$ & $n_{24}+l_{13}=\mu_2$ & $l_{24}+m_{13}=\nu_2$ \cr
$m_{34}+n_{14}=\lambda_3$ & $n_{34}+l_{14}=\mu_3$ & $l_{34}+m_{14}=\nu_3$ 
\end{tabular}
\end{equation}
Furthermore, an $SU(4)$ BZ triangle has three hexagons%
\footnote{
	The $SU(N)$ generalization is straightforward; 
	the BZ triangles are built out of three corner vertices
	and $\half (N-1)(N-2)$ hexagons. 
	}
\begin{equation}
\label{hex:su4}
\begin{tabular}{  c  c  c  }
$n_{12}+m_{24}=m_{13}+n_{23}$ & $n_{13}+l_{23}=l_{12}+n_{24}$ & $l_{24}+n_{23}=l_{13}+n_{34}$ \cr
$n_{12}+l_{34}=l_{23}+n_{23}$ & $n_{13}+m_{34}=n_{24}+m_{23}$ & $n_{23}+m_{23}=m_{12}+n_{34}$ \cr
$m_{24}+l_{23}=l_{34}+m_{13}$ & $m_{34}+l_{12}=l_{23}+m_{23}$ & $l_{13}+m_{23}=l_{24}+m_{12}$ 
\end{tabular}
\end{equation}

As an application, consider 
$\nu=[0,k,0] \subset [p,q,p] \otimes [r,s,r] = \lambda \otimes \mu$ 
of $SU(4)$; 
here all representations are self-conjugate. 
The restrictions on the $l_{ij}$, $m_{ij}$, $n_{ij}$ 
(these integers must all be non-negative) imply that 
the entries of the BZ triangle are actually 
\begin{eqnarray}
\label{BZ:ours}
m_{14} = l_{14} = m_{12} = l_{34} &=& 0 ,
\nonumber\\
n_{12} &=& p ,
\nonumber\\
n_{23} &=& n_{14} ,
\nonumber\\
n_{34} &=& r ,
\nonumber\\
l_{23} = m_{34} &=& p - n_{14} ,
\nonumber\\
l_{12} = m_{23} &=& r - n_{14} ,
\nonumber\\
l_{13} &=& \half (s + k - (2 p + q) + 2 n_{14}) , 
\nonumber\\
m_{24} &=& \half (q + k - (2 r + s) + 2 n_{14}) , 
\nonumber\\
n_{13} &=& \half (q - k + (2 r + s) - 2 n_{14}) , 
\nonumber\\
n_{24} &=& \half (s - k + (2 p + q) - 2 n_{14}) , 
\nonumber\\
m_{13} &=& \half ((2 p + q) + k - (2 r + s) ) , 
\nonumber\\
l_{24} &=& \half ((2 r + s) + k - (2 p + q) ) . 
\end{eqnarray}
All entries thus depend on a single parameter $n_{14}$ 
which is subject to restrictions 
$0 \le n_{14} \le p, r, \half(p+r-k)$; plus we get further constraints 
$k \ge | (2 p + q) - (2 r + s) |$, 
$p+q \ge r$, and $r+s \ge p$, etc.

Now, recall that $SO(6) \sim SU(4)$, and all our operators are 
in fact made of the scalars which are in the 
fundamental $\bf 6$ of $SO(6)$. In terms of the Young diagrams 
for $SO(6)$, the representations involved are partitioned as 
\begin{eqnarray}
\mbox{
\setlength{\unitlength}{1em}
\begin{picture}(15,3)
\put(0,2){\framebox (3,1){$n_{14}$}}
\put(3,2){\framebox (5,1){$m_{34}$}}
\put(8,2){\framebox (3,1){$m_{24}$}}
\put(11,2){\framebox (4,1){$n_{13}$}}
\put(0,1){\framebox (3,1){$n_{14}$}}
\put(3,1){\framebox (5,1){$m_{34}$}}
\put(4,0){$p$}
\put(11,1){$q$}
\end{picture}}
\quad
\mbox{
\setlength{\unitlength}{1em}
\begin{picture}(15,3)
\put(0,2){\framebox (3,1){$n_{14}$}}
\put(3,2){\framebox (5,1){$m_{23}$}}
\put(8,2){\framebox (3,1){$l_{13}$}}
\put(11,2){\framebox (4,1){$n_{24}$}}
\put(0,1){\framebox (3,1){$n_{14}$}}
\put(3,1){\framebox (5,1){$m_{23}$}}
\put(4,0){$r$}
\put(11,1){$s$}
\end{picture}}
\nonumber\\
\mbox{
\setlength{\unitlength}{1em}
\begin{picture}(16,2.5)
\put(0,1){\framebox (5,1){$m_{34}$}}
\put(5,1){\framebox (3,1){$m_{24}$}}
\put(8,1){\framebox (5,1){$m_{23}$}}
\put(13,1){\framebox (3,1){$l_{13}$}}
\put(8,0){$k$}
\end{picture}}
\hspace{8em}
\end{eqnarray}

An especially convenient decomposition is when the $[0,k,0]$ state 
is made up of say only 1-s and $\bar2$-s. 
In which case, by symmetry in the vertices, there will be no 
contributions proportional to $\tilde C$ provided 
only two flavors are involved in the diagram. 
Unfortunately, this can be achieved only when $s+q \ge k$.

Alternatively, we can take a 
$[p,q,p]$ state made up of 
$n_{14} + m_{34} + m_{24}$ 1-s and 
$n_{14} + m_{34} + n_{13}$ 2-s; 
$[r,s,r]$ state made up of 
$n_{14} + m_{23} + l_{13}$ $\bar1$-s and 
$n_{14} + m_{23} + n_{24}$ $\bar2$-s; 
$[0,k,0]$ state made up of 
$m_{34} + m_{24}$ $\bar1$-s and 
$m_{23} + l_{13}$ 1-s, 
minus contractions. 
Unlike the previous decomposition, 
this one works for any 
$[0,k,0] \otimes [p,q,p] \otimes [r,s,r]$ 
containing the singlet.

\section{Partitioning a tableau into 2 flavors}
\label{partition-into-two}

It is often convenient to choose the operators 
to have only two distinct flavors. Here we shall
see that it can always be done. 

Consider an operator in the $[p,q,p]$ 
of $Gl(6)$ made of 
$n_1$ 1-s and $n_2$ 2-s, to be concrete. 
We have the following constraints: 
$p \le n_2 , n_1 \le p+q$. 
This state can be assigned 
an $SU(4)$ weight 
$w = n_1 (0,1,0) + n_2 (1,-1,1) = (n_2,n_1-n_2,n_2)$. 

Next, we project this onto the $SU(3) \times u(1)$; 
for example, we can choose $1 \to 1 + \bar 1$, $2 \to 2 + \bar 2$. 
Then $w$ contains terms with 
$b$ 1-s, $n_1 - b$ $\bar 1$-s, 
$c$ 2-s, $n_2 - c$ $\bar 2$-s; 
these 
have weights 
$w'_{b,c} 
= (n_2-n_1 + 2(b-c) , 2c-n_2 )^{2(b+c)-(n_1+n_2)}$. 

To make an irrep of $Gl(6)$ into one of $so(6)$, 
we must subtract traces. Since traces 
have weight zero, contributions with $n$ contractions instead of 1
and $m$ instead of 2, 
are equivalent to $n_1'=n_1-2n$, $n_2'=n_2-2m$. 
They are projected onto 
$w'{}^{n,m}_{b,c} = 
(n_2-n_1 + 2(b+n)-2(c+m), 2(c+m)-n_2)^{2(b+n+c+m)-(n_1+n_2)}$. 
We see that for fixed $n_1$ and $n_2$, 
$w'{}^{n,m}_{b,c} = w'{}^{\tilde n,\tilde m}_{\tilde b,\tilde c}$ 
iff 
$b+n=\tilde b+\tilde n, c+m=\tilde c+\tilde m$. 

We are interested in having $b=n_1$, $c=n_2$, 
for example; then 
$w'_{n_1,n_2} = (n_1-n_2 , n_2 )^{n_1+n_2}$. 
In order for $w'{}^{n,m}_{b,c}$ to have the same weight we must have 
$n_1-n \ge b+n=n_1$, $n_2-m \ge c+m=n_2$, or $m=n=0$; 
likewise, traces also do not contribute to the projection 
onto $b=n_1$, $c=0$.

This means that $[p,q,p]$ states of $so(6) \sim SU(4)$ 
which consist of $n_1$ of any $\phi_a$ and $n_2$ of any other 
$\phi_b$ minus various contractions,
project onto {\it pure} states of $SU(3) \times u(1)$, ones  
containing $n_1$ $z_a$-s and $n_2$ $z_b$-s or 
$n_1$ $z_a$-s and $n_2$ $\bar z_b$-s, etc. 
{\it without having to subtract any traces}.

\section{
Details of 
$\langle
\OO_{\tiny \quarter} \OO_{\tiny \quarter} \OO_{\tiny \quarter} 
\rangle$
calculations}
\label{tedious details}

In Section \ref{section: all BPS} we needed to explicitly calculate 
several three-point functions 
$\langle
\OO_{[l,k,l]}(x) \OO_{[p,q,p]}(y) \OO_{[r,s,r]}(w) 
\rangle$
of \quarter-BPS operators. 
The flavor breakdown is discussed in Section \ref{section: all BPS}, 
see equations (\ref{QQQ:operators1}-\ref{QQQ:operators2}). 
For the five cases of (\ref{quarter-all: cases}), 
we list 
the values of combinatorial coefficients 
multiplying the Born diagram, 
as well as the ones in front of $\tilde B (x,w)$ and $\tilde B (w,y)$.

When operators $\OO_w$ are properly symmetrized, 
we can mark the $z_1$-s exchanged between $\OO_w$ and $\OO_x$ 
separately from the $z_1$-s exchanged between $\OO_w$ and $\OO_y$. 
As far as the combinatorial factors $\alpha_{\rm free}$, 
$\tilde \beta_{xw}$ and $\tilde \beta_{yw}$ 
are concerned, the difference is just a multiplicative factor. 
This would be equivalent to 
calculating the three-point functions with 
\begin{eqnarray}
\label{QQQ:operators1-new}
\OO_{[l,k,l]} (x) &\sim& [z_1^a z_2^b z_3^e] \\
\OO_{[p,q,p]} (y) &\sim& [\bar z_1^c \bar z_2^d \bar z_3^e] \\
\OO_{[r,s,r]} (w) &\sim& [\bar z_1^a z_1^c \bar z_2^b z_2^d] 
\label{QQQ:operators2-new}
\end{eqnarray}
rather than (\ref{QQQ:operators1}-\ref{QQQ:operators2})
instead,%
\footnote{
	As written in (\ref{QQQ:operators2-new}), $\OO_w$ is not even 
	a $[r,s,r]$ operator; we need to subtract $SO(6)$ traces. 
	But when calculating whether the three coefficients 
	$\alpha_{\rm free}$, 
	$\tilde \beta_{xw}$ and $\tilde \beta_{yw}$ 
	are zero or not, the answers are the same as if we had 
	done it properly.
	}
with the same 
$e \equiv \half [(2l+k) + (2p+q) - (2r+s)] \le l+k, p+q$, 
and integers $a$, $b$, $c$, $d$ partitioning 
$r+s = a+c$, $r = b+d$. 
This simplifies the calculations dramatically.

For operators in the [2,0,2] or [2,1,2] representations, 
we chose the \quarter-BPS operator from the beginning. 
In the other cases, several \quarter-BPS chiral primaries 
exist in each representation, so instead we choose the 
operators as $\OO_w = \sum_j C_w^j \OO_j$ for example 
(see Appendix \ref{operators explicitly} for the definitions). 
In all cases (\ref{202-202-222}-\ref{202-313-313}) listed 
above, Born level correlators are nonzero for general 
$N$, and so are the order $g^2$ contributions for a 
random set of coefficients $C^j$. 
But when we set the $C^j$ to their proper values 
(to make $\OO_y$ and $\OO_w$ \quarter-BPS), 
we recover correlators which are nonvanishing 
($\alpha_{\rm free} \ne 0$) 
and protected at order $g^2$ 
($\tilde \beta_{xw} = \tilde \beta_{yw} = 0$).%
\footnote{
	The representations involved are quite large, 
	so calculating directly Clebsch-Gordan coefficients 
	for the states involved in a particular 
	tensor product is difficult. 
	Instead, we compute the Born level 
	correlator and if it doesn't vanish, we 
	know the CG is not zero. 
	This is not necessarily the case: 
	for example if we chose the flavors as 
$
\langle 
[z_1 z_2^2 z_3]_x 
[\bar z_1^2 \bar z_2^2 \bar z_3]_y 
[z_1^2 z_2^2 \bar z_1 \bar z_2^2]_w 
\rangle 
$
	in (\ref{202-212-232}), the correlators would 
	vanish both at Born level and to order $g^2$. 
	}

With $\OO_x$, $\OO_y$, $\OO_w$ as in 
(\ref{QQQ:operators1-new}-\ref{QQQ:operators2-new}), 
we list the representations, 
choices of flavors, Born level and order $g^2$ results below. 
\begin{eqnarray}
\label{202-202-222}
&&\hspace{-3em} 
[2,0,2] \otimes [2,0,2] \otimes [2,2,2] : \quad 
\quad 
\langle 
[z_1 z_2^2 z_3]_x 
[\bar z_1 \bar z_2^2 \bar z_3]_y 
[z_1 z_2^2 \bar z_1 \bar z_2^2]_w 
\rangle 
\quad 
\\ \nonumber
\alpha_{\rm free} &\!\!=\!\!& 
{3 (N^2-1) \over 2048 N^3} 
       (12288 C_w^2 + 2048 C_w^3 + 6144 C_w^4 - 
\\ \nonumber && \hspace{5em} 
       5120 C_w^1 N - 4096 C_w^5 N - 12800 C_w^2 N^2 - 1280 C_w^3 N^2 - 
\\ \nonumber && \hspace{5em} 
       768 C_w^4 N^2 + 2688 C_w^1 N^3 - 2560 C_w^5 N^3 + 3120 C_w^2 N^4 + 
\\ \nonumber && \hspace{5em} 
       1096 C_w^3 N^4 - 168 C_w^4 N^4 - 176 C_w^1 N^5 + 1616 C_w^5 N^5 + 
\\ \nonumber && \hspace{5em} 
       8 C_w^2 N^6 - 46 C_w^3 N^6 - 6 C_w^4 N^6 + C_w^1 N^7) 
\\ \nonumber 
\tilde \beta_{xw} &\!\!=\!\!& 
{9 (N^2-1) (N^2-16) \over 4096 N^2} 
       (768 C_w^2 + 128 C_w^3 + 
\\ \nonumber && \hspace{5em} 
       384 C_w^4 + 64 C_w^1 N - 256 C_w^5 N - 464 C_w^2 N^2 - 
\\ \nonumber && \hspace{5em} 
       216 C_w^3 N^2 - 72 C_w^4 N^2 - 80 C_w^1 N^3 - 176 C_w^5 N^3 + 
\\ \nonumber && \hspace{5em} 
       8 C_w^2 N^4 - 14 C_w^3 N^4 - 6 C_w^4 N^4 + C_w^1 N^5) 
\\ \nonumber
\tilde \beta_{yw} &\!\!=\!\!& \tilde \beta_{xw} 
\end{eqnarray}
\begin{eqnarray}
\label{202-212-232}
&&\hspace{-3em} 
[2,0,2] \otimes [2,1,2] \otimes [2,3,2] : \quad 
\quad 
\langle 
[z_1 z_2^2 z_3]_x 
[\bar z_1 \bar z_2^3 \bar z_3]_y 
[z_1 z_2^3 \bar z_1 \bar z_2^2]_w 
\rangle 
\quad 
\\ \nonumber
\alpha_{\rm free} &\!\!=\!\!& 
{3 (N^2-1) (N^2-4) \over 10240 N^4}
       (34560 C_w^2 + 17280 C_w^3 + 
\\ \nonumber && \hspace{5em} 
       51840 C_w^4 + 51840 C_w^5 - 17280 C_w^1 N - 86400 C_w^6 N + 
\\ \nonumber && \hspace{5em} 
       17280 C_w^7 N - 24720 C_w^2 N^2 + 13560 C_w^3 N^2 - 
\\ \nonumber && \hspace{5em} 
       11160 C_w^4 N^2 + 6120 C_w^5 N^2 + 6528 C_w^1 N^3 + 
\\ \nonumber && \hspace{5em} 
       1320 C_w^6 N^3 - 3720 C_w^7 N^3 + 3840 C_w^2 N^4 + 
\\ \nonumber && \hspace{5em} 
       6240 C_w^3 N^4 + 2340 C_w^4 N^4 + 1860 C_w^5 N^4 - 254 C_w^1 N^5 + 
\\ \nonumber && \hspace{5em} 
       4140 C_w^6 N^5 + 240 C_w^7 N^5 + 10 C_w^2 N^6 - 160 C_w^3 N^6 - 
\\ \nonumber && \hspace{5em} 
       10 C_w^4 N^6 - 105 C_w^5 N^6 + C_w^1 N^7)
\\ \nonumber
\tilde \beta_{xw} &\!\!=\!\!& 
{9 (N^2-1) (N^2-4) (N^2-16) \over 20480 N^3 }
      (2160 C_w^2 + 1080 C_w^3 + 
\\ \nonumber && \hspace{5em} 
       3240 C_w^4 + 3240 C_w^5 - 5400 C_w^6 N + 1080 C_w^7 N - 
\\ \nonumber && \hspace{5em} 
       960 C_w^2 N^2 - 1920 C_w^3 N^2 - 540 C_w^4 N^2 - 1980 C_w^5 N^2 - 
\\ \nonumber && \hspace{5em} 
       126 C_w^1 N^3 - 660 C_w^6 N^3 + 240 C_w^7 N^3 + 10 C_w^2 N^4 - 
\\ \nonumber && \hspace{5em} 
       10 C_w^4 N^4 - 105 C_w^5 N^4 + C_w^1 N^5)        
\\ \nonumber
\tilde \beta_{yw} &\!\!=\!\!& 
{3 (N^2-1) (N^2-4) (N^2-36) \over 5120 N^3}
      (1920 C_w^2 + 960 C_w^3 + 2880 C_w^4 + 
\\ \nonumber && \hspace{5em} 
       2880 C_w^5 - 4800 C_w^6 N + 960 C_w^7 N - 840 C_w^2 N^2 - 
\\ \nonumber && \hspace{5em} 
       1380 C_w^3 N^2 - 780 C_w^4 N^2 - 1740 C_w^5 N^2 - 98 C_w^1 N^3 - 
\\ \nonumber && \hspace{5em} 
       540 C_w^6 N^3 - 120 C_w^7 N^3 + 10 C_w^2 N^4 - 100 C_w^3 N^4 - 
\\ \nonumber && \hspace{5em} 
       10 C_w^4 N^4 + 75 C_w^5 N^4 + C_w^1 N^5) 
\end{eqnarray}
\begin{eqnarray}
\label{202-212-313}
&&\hspace{-3em} 
[2,0,2] \otimes [2,1,2] \otimes [3,1,3] : \quad 
\quad 
\langle 
[z_1 z_2^2 z_3]_x 
[\bar z_1^2 \bar z_2^2 \bar z_3]_y 
[z_1^2 z_2^2 \bar z_1 \bar z_2^2]_w 
\rangle 
\quad 
\\ \nonumber
\alpha_{\rm free} &\!\!=\!\!& 
{(N^2-1) (N^2-4) \over 10240 N^2} 
      (-103680 C_w^2 + 11520 C_w^1 N - 34560 C_w^4 N + 
\\ \nonumber && \hspace{5em} 
       18000 C_w^2 N^2 - 9504 C_w^3 N^2 - 545 C_w^1 N^3 + 
\\ \nonumber && \hspace{5em} 
       8520 C_w^4 N^3 + 60 C_w^2 N^4 + 276 C_w^3 N^4 + 5 C_w^1 N^5) 
\\ \nonumber
\tilde \beta_{xw} &\!\!=\!\!& 
{3 (N^2-1) (N^2-4) (N^2-16) \over 20480 N} 
      (-2160 C_w^2 + 864 C_w^3 - 225 C_w^1 N - 
\\ \nonumber && \hspace{5em} 
       1080 C_w^4 N + 60 C_w^2 N^2 + 84 C_w^3 N^2 + 5 C_w^1 N^3)
\\ \nonumber
\tilde \beta_{yw} &\!\!=\!\!& 
{(N^2-1) (N^2-4) (N^2-36) \over 10240 N} 
      (-4320 C_w^2 + 1728 C_w^3 - 465 C_w^1 N - 
\\ \nonumber && \hspace{5em} 
       2520 C_w^4 N + 60 C_w^2 N^2 + 228 C_w^3 N^2 + 5 C_w^1 N^3)
\end{eqnarray}
\begin{eqnarray}
\label{202-313-232}
&&\hspace{-3em} 
[2,0,2] \otimes [3,1,3] \otimes [2,3,2] : \quad 
\quad 
\langle 
[z_1^2 z_3^2]_x 
[\bar z_1^2 \bar z_2^3 \bar z_3^2]_y 
[z_1^2 z_2^3 \bar z_1^2]_w 
\rangle 
\quad 
\\ \nonumber
\alpha_{\rm free} &\!\!=\!\!& 
{(N^2-1) (N^2-4) \over 57600 N^2} 
      (-129600 C^2_w C^2_y - 64800 C^2_y C^3_w - 194400 C^2_y C^4_w - 
\\ \nonumber && \hspace{5em} 
       194400 C^2_y C^5_w + 14400 C^1_y C^2_w N + 45900 C^1_w C^2_y N + 
\\ \nonumber && \hspace{5em} 
       7200 C^1_y C^3_w N - 8640 C^1_w C^3_y N + 21600 C^1_y C^4_w N -        
\\ \nonumber && \hspace{5em} 
       43200 C^2_w C^4_y N - 21600 C^3_w C^4_y N - 64800 C^4_w C^4_y N + 
\\ \nonumber && \hspace{5em} 
       21600 C^1_y C^5_w N - 64800 C^4_y C^5_w N + 324000 C^2_y C^6_w N - 
\\ \nonumber && \hspace{5em} 
       64800 C^2_y C^7_w N + 88200 C^2_w C^2_y N^2 - 3150 C^2_y C^3_w N^2 - 
\\ \nonumber && \hspace{5em} 
       14400 C^2_w C^3_y N^2 + 14400 C^3_w C^3_y N^2 + 67500 C^2_y C^4_w N^2 - 
\\ \nonumber && \hspace{5em} 
       12960 C^3_y C^4_w N^2 + 18900 C^1_w C^4_y N^2 + 43200 C^2_y C^5_w N^2 + 
\\ \nonumber && \hspace{5em} 
       4320 C^3_y C^5_w N^2 - 36000 C^1_y C^6_w N^2 + 108000 C^4_y C^6_w N^2 + 
\\ \nonumber && \hspace{5em} 
       7200 C^1_y C^7_w N^2 - 21600 C^4_y C^7_w N^2 - 6600 C^1_y C^2_w N^3 - 
\\ \nonumber && \hspace{5em} 
       13965 C^1_w C^2_y N^3 - 9675 C^1_y C^3_w N^3 + 3696 C^1_w C^3_y N^3 - 
\\ \nonumber && \hspace{5em} 
       7125 C^1_y C^4_w N^3 + 34200 C^2_w C^4_y N^3 - 7650 C^3_w C^4_y N^3 + 
\\ \nonumber && \hspace{5em} 
       20700 C^4_w C^4_y N^3 - 13500 C^1_y C^5_w N^3 + 5400 C^4_y C^5_w N^3 - 
\\ \nonumber && \hspace{5em} 
       4500 C^2_y C^6_w N^3 + 31680 C^3_y C^6_w N^3 + 11700 C^2_y C^7_w N^3 - 
\\ \nonumber && \hspace{5em} 
       6120 C^3_y C^7_w N^3 - 820 C^1_w C^1_y N^4 - 12750 C^2_w C^2_y N^4 - 
\\ \nonumber && \hspace{5em} 
       16050 C^2_y C^3_w N^4 + 6420 C^2_w C^3_y N^4 + 9930 C^3_w C^3_y N^4 - 
\\ \nonumber && \hspace{5em} 
       10500 C^2_y C^4_w N^4 + 5490 C^3_y C^4_w N^4 - 6750 C^1_w C^4_y N^4 - 
\\ \nonumber && \hspace{5em} 
       13500 C^2_y C^5_w N^4 + 8595 C^3_y C^5_w N^4 - 5100 C^1_y C^6_w N^4 - 
\\ \nonumber && \hspace{5em} 
       9900 C^4_y C^6_w N^4 - 300 C^1_y C^7_w N^4 + 5400 C^4_y C^7_w N^4 + 
\\ \nonumber && \hspace{5em} 
       50 C^1_y C^2_w N^5 + 210 C^1_w C^2_y N^5 - 925 C^1_y C^3_w N^5 + 
\\ \nonumber && \hspace{5em} 
       471 C^1_w C^3_y N^5 + 75 C^1_y C^4_w N^5 - 6300 C^2_w C^4_y N^5 - 
\\ \nonumber && \hspace{5em} 
       10350 C^3_w C^4_y N^5 - 4200 C^4_w C^4_y N^5 + 225 C^1_y C^5_w N^5 - 
\\ \nonumber && \hspace{5em} 
       5850 C^4_y C^5_w N^5 - 12600 C^2_y C^6_w N^5 + 5580 C^3_y C^6_w N^5 + 
\\ \nonumber && \hspace{5em} 
       180 C^3_y C^7_w N^5 + 5 C^1_w C^1_y N^6 - 30 C^2_w C^3_y N^6 + 
\\ \nonumber && \hspace{5em} 
       630 C^3_w C^3_y N^6 + 150 C^2_y C^4_w N^6 + 225 C^2_y C^5_w N^6 - 
\\ \nonumber && \hspace{5em} 
       6300 C^4_y C^6_w N^6) 
\\ \nonumber
\tilde \beta_{yw} &\!\!=\!\!& 
{(N^2-1) (N^2-4) \over 115200 N} 
      (1036800 C^2_w C^2_y + 518400 C^2_y C^3_w - 
\\ \nonumber && \hspace{5em} 
       414720 C^2_w C^3_y - 207360 C^3_w C^3_y + 1555200 C^2_y C^4_w - 
\\ \nonumber && \hspace{5em} 
       622080 C^3_y C^4_w + 1555200 C^2_y C^5_w - 622080 C^3_y C^5_w + 
\\ \nonumber && \hspace{5em} 
       115200 C^1_y C^2_w N + 
       57600 C^1_y C^3_w N + 172800 C^1_y C^4_w N + 
\\ \nonumber && \hspace{5em} 
       691200 C^2_w C^4_y N + 345600 C^3_w C^4_y N + 
       1036800 C^4_w C^4_y N + 
\\ \nonumber && \hspace{5em} 
       172800 C^1_y C^5_w N + 
       1036800 C^4_y C^5_w N - 2592000 C^2_y C^6_w N + 
\\ \nonumber && \hspace{5em} 
       1036800 C^3_y C^6_w N + 518400 C^2_y C^7_w N - 
       207360 C^3_y C^7_w N + 
\\ \nonumber && \hspace{5em} 
       750 C^1_w C^1_y N^2 - 
       482400 C^2_w C^2_y N^2 - 707400 C^2_y C^3_w N^2 + 
\\ \nonumber && \hspace{5em} 
       123840 C^2_w C^3_y N^2 + 248400 C^3_w C^3_y N^2 - 
       516600 C^2_y C^4_w N^2 + 
\\ \nonumber && \hspace{5em} 
       102960 C^3_y C^4_w N^2 + 
       18000 C^1_w C^4_y N^2 - 982800 C^2_y C^5_w N^2 + 
\\ \nonumber && \hspace{5em} 
       289440 C^3_y C^5_w N^2 - 
       270000 C^1_y C^6_w N^2 - 1296000 C^4_y C^6_w N^2 + 
\\ \nonumber && \hspace{5em} 
       53100 C^1_y C^7_w N^2 + 
       237600 C^4_y C^7_w N^2 - 52850 C^1_y C^2_w N^3 - 
\\ \nonumber && \hspace{5em} 
       60720 C^1_w C^2_y N^3 - 76800 C^1_y C^3_w N^3 + 
       22488 C^1_w C^3_y N^3 - 
\\ \nonumber && \hspace{5em} 
       56200 C^1_y C^4_w N^3 - 
       303600 C^2_w C^4_y N^3 - 428400 C^3_w C^4_y N^3 - 
\\ \nonumber && \hspace{5em} 
       315600 C^4_w C^4_y N^3 - 105825 C^1_y C^5_w N^3 - 
       574200 C^4_y C^5_w N^3 - 
\\ \nonumber && \hspace{5em} 
       392400 C^2_y C^6_w N^3 + 
       286560 C^3_y C^6_w N^3 - 18000 C^2_y C^7_w N^3 - 
\\ \nonumber && \hspace{5em} 
       16560 C^3_y C^7_w N^3 - 6755 C^1_w C^1_y N^4 + 
       3000 C^2_w C^2_y N^4 - 
\\ \nonumber && \hspace{5em} 
       70200 C^2_y C^3_w N^4 + 
       29160 C^2_w C^3_y N^4 + 70920 C^3_w C^3_y N^4 + 
\\ \nonumber && \hspace{5em} 
       4800 C^2_y C^4_w N^4 + 29640 C^3_y C^4_w N^4 - 40680 C^1_w C^4_y N^4 + 
\\ \nonumber && \hspace{5em} 
       13500 C^2_y C^5_w N^4 + 
       52020 C^3_y C^5_w N^4 - 45600 C^1_y C^6_w N^4 - 
\\ \nonumber && \hspace{5em} 
       309600 C^4_y C^6_w N^4 - 1500 C^1_y C^7_w N^4 + 
       250 C^1_y C^2_w N^5 + 
\\ \nonumber && \hspace{5em} 
       300 C^1_w C^2_y N^5 - 8000 C^1_y C^3_w N^5 + 3948 C^1_w C^3_y N^5 + 
\\ \nonumber && \hspace{5em} 
       400 C^1_y C^4_w N^5 - 51600 C^3_w C^4_y N^5 + 1125 C^1_y C^5_w N^5 + 
\\ \nonumber && \hspace{5em} 
       30960 C^3_y C^6_w N^5 + 
       25 C^1_w C^1_y N^6 + 5160 C^3_w C^3_y N^6)
\end{eqnarray}
\begin{eqnarray}
\label{202-313-313}
&&\hspace{-3em} 
[2,0,2] \otimes [3,1,3] \otimes [3,1,3] : \quad 
\quad 
\langle 
[z_1^2 z_3^2]_x 
[\bar z_1^2 \bar z_2^3 \bar z_3^2]_y 
[z_1^2 z_2^3 \bar z_1^2]_w 
\rangle 
\quad 
\\ \nonumber
\alpha_{\rm free} &\!\!=\!\!& 
{(N^2-1) (N^2-4) \over 1036800 N^2} 
      (2332800 C^2_w C^2_y - 2332800 C^2_y C^4_w N - 
\\ \nonumber && \hspace{5em} 
       2332800 C^2_w C^4_y N + 10800 C^1_w C^1_y N^2 + 324000 C^2_w C^2_y N^2 - 
\\ \nonumber && \hspace{5em} 
       492480 C^2_y C^3_w N^2 - 492480 C^2_w C^3_y N^2 + 114048 C^3_w C^3_y N^2 + 
\\ \nonumber && \hspace{5em} 
       259200 C^1_y C^4_w N^2 + 259200 C^1_w C^4_y N^2 - 777600 C^4_w C^4_y N^2 - 
\\ \nonumber && \hspace{5em} 
       44100 C^1_y C^2_w N^3 - 44100 C^1_w C^2_y N^3 - 8280 C^1_y C^3_w N^3 - 
\\ \nonumber && \hspace{5em} 
       8280 C^1_w C^3_y N^3 + 950400 C^2_y C^4_w N^3 - 250560 C^3_y C^4_w N^3 + 
\\ \nonumber && \hspace{5em} 
       950400 C^2_w C^4_y N^3 - 250560 C^3_w C^4_y N^3 - 5875 C^1_w C^1_y N^4 - 
\\ \nonumber && \hspace{5em} 
       133200 C^2_w C^2_y N^4 + 90720 C^2_y C^3_w N^4 + 90720 C^2_w C^3_y N^4 - 
\\ \nonumber && \hspace{5em} 
       22752 C^3_w C^3_y N^4 - 52800 C^1_y C^4_w N^4 - 52800 C^1_w C^4_y N^4 + 
\\ \nonumber && \hspace{5em} 
       396000 C^4_w C^4_y N^4 + 900 C^1_y C^2_w N^5 + 900 C^1_w C^2_y N^5 + 
\\ \nonumber && \hspace{5em} 
       4920 C^1_y C^3_w N^5 + 4920 C^1_w C^3_y N^5 - 100800 C^2_y C^4_w N^5 + 
\\ \nonumber && \hspace{5em} 
       51840 C^3_y C^4_w N^5 - 100800 C^2_w C^4_y N^5 + 51840 C^3_w C^4_y N^5 + 
\\ \nonumber && \hspace{5em} 
       75 C^1_w C^1_y N^6 + 3600 C^2_w C^2_y N^6 - 2880 C^3_w C^3_y N^6 - 
\\ \nonumber && \hspace{5em} 
       50400 C^4_w C^4_y N^6) 
\\ \nonumber
\tilde \beta_{yw} &\!\!=\!\!& 
{(N^2-1)(N^2-4) \over 518400 }
      (172800 C^1_y C^2_w + 172800 C^1_w C^2_y - 
\\ \nonumber && \hspace{5em} 
       69120 C^1_y C^3_w - 
       69120 C^1_w C^3_y + 4147200 C^2_y C^4_w - 
\\ \nonumber && \hspace{5em} 
       1658880 C^3_y C^4_w + 4147200 C^2_w C^4_y - 1658880 C^3_w C^4_y + 
\\ \nonumber && \hspace{5em} 
       33450 C^1_w C^1_y N - 
       837000 C^2_w C^2_y N - 79920 C^2_y C^3_w N - 
\\ \nonumber && \hspace{5em} 
       79920 C^2_w C^3_y N + 197856 C^3_w C^3_y N + 457200 C^1_y C^4_w N + 
\\ \nonumber && \hspace{5em} 
       457200 C^1_w C^4_y N + 
       2678400 C^4_w C^4_y N - 107700 C^1_y C^2_w N^2 - 
\\ \nonumber && \hspace{5em} 
       107700 C^1_w C^2_y N^2 - 2640 C^1_y C^3_w N^2 - 2640 C^1_w C^3_y N^2 - 
\\ \nonumber && \hspace{5em} 
       910800 C^2_y C^4_w N^2 + 
       96480 C^3_y C^4_w N^2 - 910800 C^2_w C^4_y N^2 + 
\\ \nonumber && \hspace{5em} 
       96480 C^3_w C^4_y N^2 - 13175 C^1_w C^1_y N^3 + 
       18000 C^2_w C^2_y N^3 + 
\\ \nonumber && \hspace{5em} 
       83880 C^2_y C^3_w N^3 + 
       83880 C^2_w C^3_y N^3 - 43200 C^3_w C^3_y N^3 - 
\\ \nonumber && \hspace{5em} 
       100800 C^1_y C^4_w N^3 - 100800 C^1_w C^4_y N^3 - 
       597600 C^4_w C^4_y N^3 + 
\\ \nonumber && \hspace{5em} 
       1500 C^1_y C^2_w N^4 + 
       1500 C^1_w C^2_y N^4 + 9480 C^1_y C^3_w N^4 + 
\\ \nonumber && \hspace{5em} 
       9480 C^1_w C^3_y N^4 + 59760 C^3_y C^4_w N^4 + 59760 C^3_w C^4_y N^4 + 
\\ \nonumber && \hspace{5em} 
       125 C^1_w C^1_y N^5 - 5976 C^3_w C^3_y N^5)
\end{eqnarray}

The last two sets of calculations are very formidable. 
Getting the coefficients $\tilde \beta_{yw}$ took 
2430 hrs for $[2,0,2] \otimes [3,1,3] \otimes [3,1,3]$
and 
2830 hrs for $[2,0,2] \otimes [3,1,3] \otimes [2,3,2]$ 
of a {\it Mathematica} computation on a Pentium-III with 1.4MHz speed.



\end{document}